\documentclass[aps,prx,reprint,superscriptaddress]{revtex4-2}

\pdfoutput=1
\usepackage{amsmath,amssymb}
\usepackage{mathrsfs,bm}
\usepackage{color,xcolor}
\usepackage{graphicx,epsfig}
\usepackage{appendix}
\usepackage{siunitx}
\usepackage{braket}
\usepackage{calc}
\usepackage{esint}
\usepackage{tabularx}
\usepackage{amsfonts,dsfont}
\usepackage{seqsplit}
\usepackage{tikz}
\usepackage[normalem]{ulem}
\usepackage{enumitem}
\newlength{\myl}%
\newcommand{\dif}{\mathrm{d}}%
\newcommand{\INT}[3]{\settowidth{\myl}{$\displaystyle\int_{#1}^{#2}$}{\int_{#1}^{#2}\;\;\:\hspace{-\the\myl}\dif #3}\,}
\newcommand{\ii}{\mathrm{i}}%

\usepackage{microtype}
\usepackage{hyperref}
\hypersetup{
  colorlinks=true,
  citecolor=teal,
  linkcolor=teal,
  urlcolor=blue,
  filecolor=blue,
  breaklinks=true,
}
\usepackage[capitalise]{cleveref}
\allowdisplaybreaks

\begin{document}

\title{Active pattern formation emergent from single-species nonreciprocity}
\author{Zhi-Feng Huang}
\thanks{These authors contributed equally.}
\affiliation{Department of Physics and Astronomy, Wayne State University, Detroit, Michigan 48201, USA}
\author{Michael te Vrugt}
\thanks{These authors contributed equally.}
\affiliation{DAMTP, Centre for Mathematical Sciences, University of Cambridge, Cambridge CB3 0WA, United Kingdom}
\affiliation{Institut f\"ur Physik, Johannes Gutenberg-Universit\"at Mainz, 55128 Mainz, Germany}
\author{Jonas Mayer Martins}
\affiliation{Institut f\"ur Informatik, Georg-August-Universit\"at G\"ottingen, 37077 G\"ottingen, Germany}
\author{Raphael Wittkowski}
\affiliation{Department of Physics, RWTH Aachen University, 52074 Aachen, Germany}
\affiliation{DWI -- Leibniz Institute for Interactive Materials, 52074 Aachen, Germany}
\author{Hartmut L\"{o}wen}
\email{hartmut.loewen@uni-duesseldorf.de}
\affiliation{Institut f\"{u}r Theoretische Physik II: Weiche Materie, Heinrich-Heine-Universit\"{a}t D\"{u}sseldorf, 40225 D\"{u}sseldorf, Germany}

\begin{abstract}
Nonreciprocal interactions violating Newton's third law are common in a plethora of nonequilibrium situations ranging from predator-prey systems to the swarming of birds and effective colloidal interactions under flow. While many recent studies have focused on two species with nonreciprocal coupling, less is examined for the basic single-component system breaking the \textit{actio} and \textit{reactio} equality of force within the same species. Here, we systematically derive a field theory for the case of single-species nonreciprocal interactions from the microscopic particle dynamics, leading to a generic continuum model termed \textit{Active Model N} (N denoting nonreciprocity). We explore the rich dynamics of pattern formation in this nonreciprocal system and the emergence of self-traveling states with persistent variation and flowing of active branched patterns. One particular new characteristic pattern is an interwoven self-knitting ``yarn'' structure with a unique feature of simultaneous development of micro- and bulk phase separations. The growth dynamics of a ``ball-of-wool'' active droplet towards these self-knitted yarn or branched states exhibits a crossover between different scaling behaviors. The mechanism underlying this distinct class of active phase separation is attributed to the interplay between nonreciprocity and competition of interparticle forces. Our predictions can be applied to various biological and artificial active matter systems controlled by single-species nonreciprocity.
\end{abstract}

\date{\today}

\maketitle

\section{Introduction}
Newton's famous third law states that the force ${\mathbf F}_{1 \rightarrow 2}$ which a body 1 exerts on another body 2 is, up to a sign, the same as the force $ \mathbf{F}_{2 \rightarrow 1}$ which body 2 exerts on body 1 such that there is equality of \textit{actio} and \textit{reactio}, i.e., $\mathbf{F}_{1 \rightarrow 2} = - \mathbf{F}_{2 \rightarrow 1}$. This law of reciprocity holds for all fundamental interactions and also for effective interactions in equilibrium situations. In systems out of equilibrium, however, reciprocity is typically broken such that $ \mathbf{F}_{1 \rightarrow 2} +  \mathbf{F}_{2 \rightarrow 1} \not= \mathbf{0}$ which implies that the center-of-mass of a two-body system is not force-free but experiences a driving force leading to spontaneous ``active'' motion of a pair. 

There are two fundamentally distinct typical examples of nonreciprocity: either the two bodies are of different species (such as a predator trying to catch an escaping prey) or they are of the same type (such as two birds flying together where only the following bird can see the leading one). In the former case of two-species nonreciprocity, typically only particles of different species couple nonreciprocally, while the latter case of single-species nonreciprocity concerns intrinsic nonreciprocal interactions within the same species. There are many examples of single-species nonreciprocity, both of living systems such as birds, fish, bacteria \cite{PetroffPRL15}, or starfish embryos \cite{TanNature22}, and artificial robotic and colloidal systems with vision-cone-like interactions \cite{LavergneWBB2019}. Various aspects of single-species nonreciprocity stemming from vision-cone effects \cite{BarberisPRL16,Bastien2020,NegiWG2022,NegiWG2024,NegiIG2024,LoosKM2023,RouzaireLP2024}, transverse interactions \cite{PetroffPRL15,TanNature22,BililignNatPhys22} or the general cases of parity-symmetric and parity-antisymmetric forces \cite{PoncetPRL22} have been considered in theories. 
Systems governed by nonreciprocal interactions have shown potential technological importance, as demonstrated in various contexts ranging from robotic metamaterials \cite{VeenstraGGSMC2024,BrandenbourgerLLC2019} to drug design \cite{VosseldGG2023}.

In the field-theoretical description of active matter, most of the recent works on pattern formation induced by nonreciprocal interactions treat coupling-nonreciprocity of a binary system while ignoring nonreciprocity between the same species \cite{IvlevEtAl2015,BartnickEtAl2016,YouPNAS20,SahaAG2020,FruchartNature21,KreienkampK2022}. These studies have provided crucial insights into the dynamics of multicomponent systems \cite{FrohoffT2023,SuchanekKL2023,FrohoffTP2023,BraunsPRX24} and a wealth of novel dynamical behavior has been discovered, including symmetry-broken nonreciprocal phase transitions \cite{FruchartNature21}, traveling waves with interface undulations and spatiotemporal chaos \cite{BraunsPRX24}, as well as traveling bands \cite{YouPNAS20} and transverse ripples \cite{SahaAG2020}. The case of single-species vision-cone nonreciprocity was examined from a field-theoretical perspective in Ref.\ \cite{BarberisPRL16} which, like most studies employing particle-based simulations \cite{BarberisPRL16,Bastien2020,NegiWG2022,NegiWG2024,NegiIG2024}, considered self-propelled particles whose self-propulsion is modified by a vision-cone mechanism. 

Here we investigate the rich behavior of pattern formation in a single-species system that becomes active not via self-propulsion of individual particles, but via force nonreciprocity. We start from the microscopic equations of motion for particles interacting via nonreciprocal interaction forces ${\mathbf F}_{2 \rightarrow 1}$ and systematically derive a fundamental underlying continuum field theory based on microscopic dynamics, establishing a new model which we term \textit{Active Model N} (where N denotes the aspect of nonreciprocity). A crucial advantage of this type of continuum field theories \cite{Cates2019b,teVrugtBW2022,DoostmohammadiIYS2018,BickmannW2020}, in addition to the fact that they can reveal the complex mechanisms involved, is the accessibility of much larger spatial and temporal scales for a greater amount of particles compared to discrete particle-based simulations. Unlike previous active field theories such as Active Model B \cite{WittkowskiTSAMC2014} and its extensions \cite{TiribocchiPRL15,TjhungNC2018,teVrugtFHHTW2023} as well as active phase field crystal models \cite{MenzelPRL13,MenzelPRE14,HuangPRL20,HuangCP22,NestlerJPCM24}, this model is based on a generic framework to study single-species force nonreciprocity. Through the study of vision-cone-like interactions, it builds an inherent connection between nonreciprocal interaction forces and self-propelled active matter. We predict new types of self-traveling chiral patterns for which both spatial mirror symmetry and time-reversal symmetry are spontaneously broken. In particular, we discover a unique interwoven pattern reminiscent of self-knitting threads that we term ``active yarn'', which self-organizes through simultaneous evolution of micro- and bulk phase separations over multiple length and time scales. 
This new class of active phase separation is induced by an intrinsic mechanism showing as the interplay between interparticle force nonreciprocity and the strength and range of interactions, leading to the simultaneous occurrence of two distinct classes of collective behavior incorporating phase-separation-induced assembly and orientationally aligning, persistent collective self-migration. Notably, the size of the phase-separating yarn region is controllable via, e.g., the vision-cone angle. This mechanism has the potential for interesting applications in the realization of programmable matter, particularly in the form of controlled phase separation.

\begin{figure}[tb]
  \centering 
  \includegraphics[trim=30px 3px 30px 30px,clip,width=\linewidth]{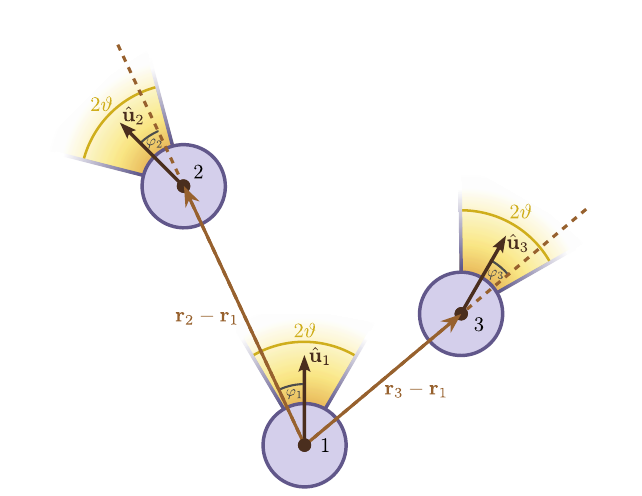}
  \caption{Schematic of nonreciprocal single-species vision-cone model. An agent~1 with orientation $\mathbf{\hat{u}_1}$ sees and interacts with other agents within a vision cone of full opening angle $2\vartheta$. Agent~2 is included in the perception zone of agent~1, such that there is a resulting nonreciprocal force acting on agent~1 which is directed along $\mathbf{r} = \mathbf{r}_2 - \mathbf{r}_1$. There is also a torque turning its orientation to be aligned with the orientation $\mathbf{\hat{u}_2}$ of the second agent. In this example, agent~2 with its own vision cone does not see agent~1 and thus there is no interaction between them that influences the motion of agent~2. Likewise, agent~3 is not visible for agent~1 such that there is no mutual interaction between agents~1 and 3 that influences the motion of either of them.}
  \label{fig:schematic}
\end{figure}

\section{Active Model N}

\subsection{Microscopic basis}

We start from a microscopic picture of particles or agents interacting through vision cones, as illustrated in Fig.~\ref{fig:schematic}. Consider a general form of pairwise nonreciprocal force between single-component particles with an orientational degree of freedom on a plane, i.e.,
\begin{equation}
  \mathbf{F}_{2 \rightarrow 1} = \hat{\mathbf r} f(|\mathbf{r}_2 - \mathbf{r}_1|, \hat{\mathbf u}_1 \cdot \hat{\mathbf r}, \hat{\mathbf u}_2 \cdot \hat{\mathbf r}),
  \label{eq:F21}
\end{equation}
which satisfies both global translational symmetry via the dependence on $\mathbf{r} = \mathbf{r}_2 - \mathbf{r}_1$ and global rotational symmetry via the dependence on $\hat{\mathbf u}_i \cdot \hat{\mathbf r} = \cos \varphi_i$, where $\hat{\mathbf u}_i$ denotes the unit vector of orientation of particle $i$ and $\hat{\mathbf r} = (\mathbf{r}_2 - \mathbf{r}_1) / |\mathbf{r}_2 - \mathbf{r}_1| = (\mathbf{r}_2 - \mathbf{r}_1) / r$. The nonreciprocity of interaction is given by
\begin{equation}
  \mathbf{F}_{2 \rightarrow 1} \neq - \mathbf{F}_{1 \rightarrow 2} = -(-\hat{\mathbf r}) f(r, -\hat{\mathbf u}_2 \cdot \hat{\mathbf r}, -\hat{\mathbf u}_1 \cdot \hat{\mathbf r}),
\end{equation}
such that
\begin{equation}
  f(r, \cos\varphi_1, \cos\varphi_2) \neq f(r, -\cos\varphi_2, -\cos\varphi_1).
\end{equation}
When particles interact within a finite vision cone of full opening angle $2\vartheta$ with $\vartheta \in (0, \pi]$ (Fig.~\ref{fig:schematic}), the above nonreciprocal condition can be achieved via 
\begin{equation}
  f(r, \cos\varphi_1, \cos\varphi_2) = h(r) \Theta ( -\cos\vartheta + \cos\varphi_1 ), \label{eq:f}
\end{equation}
giving $\mathbf{F}_{2 \rightarrow 1} = \hat{\mathbf r} h(r) \Theta(-\cos\vartheta + \cos\varphi_1)$ and $\mathbf{F}_{1 \rightarrow 2} = -\hat{\mathbf r} h(r) \Theta(-\cos\vartheta - \cos\varphi_2)$, with $h(r) > 0$ for interparticle attraction and $h(r)<0$ for repulsion. The step function $\Theta$ is equal to one when the first particle can see the second one and zero otherwise, with nonreciprocity arising for $\vartheta\in (0,\pi)$.

Similarly, a nonreciprocal torque aligning the particle's orientations can be set up as 
\begin{equation}
  M(r, \hat{\mathbf u}_1, \hat{\mathbf u}_2) = 
    \tau(r) \Delta \varphi_{21} \Theta ( -\cos\vartheta + \cos\varphi_1), \label{eq:M}
\end{equation}  
where $\Delta \varphi_{21} = \varphi_2 - \varphi_1$ if $\varphi_1 - \varphi_2 \in [0, \pi]$ and $\Delta \varphi_{21} = \varphi_2 - \varphi_1 - 2\pi$ if $\varphi_1 - \varphi_2 \in (\pi, 2\pi)$, representing different directions of alignment rotation (a particle rotates in the direction closer to the desired orientation). The function forms of $h(r)$ and $\tau(r)$ depend on the specific types of attractive and repulsive forces and aligning torque. 

Our main focus here is on establishing the corresponding continuum field theory for single-species nonreciprocity, and predicting the emergence of new patterns and dynamics particularly those with structural chirality and parity-time (PT) symmetry breaking. To derive the continuum theory, we start from the microscopic Langevin equations containing the interparticle forces and torques. The dynamics of the full one-particle density field $\varrho(\mathbf{r}, \hat{\mathbf u}, t)$ is then derived by integrating out the stochastically equivalent Smoluchowski equation. By a systematic expansion in terms of a particle density field $\rho(\mathbf{r},t)$ and a polarization density field $\mathbf{P}(\mathbf{r},t)$, the coarse-grained continuum field equations can be obtained. 

\subsection{Derivation of generic continuum field theory}
\label{amnderivation}

In \cref{micro} we construct a general nonreciprocal continuum field theory based on this approach, before applying the specific forms of nonreciprocal forces and torques given above to derive Active Model N. In the following we explain the main idea of the derivation. Similar strategies have been successfully applied to other physical systems, including the different case for self-propelled particles \cite{teVrugtBW2022}.

We consider a system of two-dimensional (2D) Brownian particles. The position $\mathbf{r}_i$ and orientation $\phi_i$ of the $i$th particle obey the Langevin equations
\begin{align}
\dot{\mathbf{r}}_i(t)&=\beta D_\mathrm{T}\mathbf{F}(\{\mathbf{r}_j(t),\phi_j(t)\}) + \sqrt{2D_\mathrm{T}}\boldsymbol{\xi}_i(t),\label{eq:langevin1}\\
\dot{\phi}_i(t)&=\beta D_\mathrm{R}M(\{\mathbf{r}_j(t),\phi_j(t)\}) + \sqrt{2D_\mathrm{R}}\chi_i(t),\label{eq:langevin2}
\end{align}
where $\beta = 1/(k_\mathrm{B} T)$ with Boltzmann constant $k_\mathrm{B}$ and temperature $T$, $D_\mathrm{T}$ and $D_\mathrm{R}$ represent the translational and rotational diffusion coefficients, $\mathbf{F}$ is the interaction force, $M$ is the interaction torque, and $\boldsymbol{\xi}_i$ and $\chi_i$ are white noises with zero mean and unit variance. Here the dot denotes a time derivative. This microscopic model is generic, since no restrictions have been made on force and torque apart from the assumption that they are pairwise (without three- or higher-body interactions). We deliberately do \textit{not} include any self-propulsion terms to make sure that any nonequilibrium effects observed are just a consequence of nonreciprocal interactions. This allows us to study and understand the effect of nonreciprocity in isolation and to build a continuum model that incorporates single-species nonreciprocity in a minimal way (similar to the nonreciprocal Cahn-Hilliard equation \cite{SahaAG2020,FrohoffT2023,BraunsPRX24,SuchanekKL2023b} which constitutes a minimal model for the case of multispecies nonreciprocity). 

We then derive the corresponding Smoluchowski equation which is stochastically equivalent to Eqs.~\eqref{eq:langevin1} and \eqref{eq:langevin2}, integrate over the degrees of freedom of all particles except for one. This gives the dynamic equation for the orientation-dependent one-body density $\varrho$, i.e.,
\begin{align}
    &\frac{\partial}{\partial t} \varrho(\mathbf{r},\hat{\mathbf{u}},t) 
    = D_\mathrm{T}  \bm{\nabla}^2 \varrho(\mathbf{r},\hat{\mathbf{u}},t)
        + D_\mathrm{R} \partial_\phi^2\varrho(\mathbf{r},\hat{\mathbf{u}},t)
        \notag\\& \qquad
        + \beta D_\mathrm{T}  \bm{\nabla}\cdot
            \bigg\{
                \varrho(\mathbf{r},\hat{\mathbf{u}},t) \INT{}{}{^2 r'} \INT{}{}{^2 u'}
                \notag\\& \qquad
                \Big[
                    -\mathbf{F}(\mathbf{r},\mathbf{r}',\hat{\mathbf{u}},\hat{\mathbf{u}}',t)g(\mathbf{r},\mathbf{r}',\hat{\mathbf{u}},\hat{\mathbf{u}}',t) \varrho(\mathbf{r}',\hat{\mathbf{u}}',t)
                \Big]
            \bigg\}
        \notag\\& \qquad
        + \beta D_\mathrm{R} \partial_\phi
            \bigg\{
                \varrho(\mathbf{r},\hat{\mathbf{u}},t) \INT{}{}{^2 r'} \INT{}{}{^2 u'}
                \notag\\& \qquad
                \Big[
                    -M(\mathbf{r},\mathbf{r}',\hat{\mathbf{u}},\hat{\mathbf{u}}',t)g(\mathbf{r},\mathbf{r}',\hat{\mathbf{u}},\hat{\mathbf{u}}',t) \varrho(\mathbf{r}',\hat{\mathbf{u}}',t)
                \Big]
            \bigg\}, \label{eq:DDFT_raw_exact}
\end{align}
where $\mathbf{F}$ and $M$ are the force and torque that a particle at position $\mathbf{r}'$ with orientation $\hat{\mathbf{u}}'$ exerts on a particle at position $\mathbf{r}$ with orientation $\hat{\mathbf{u}}$, and $g$ is the pair-distribution function \cite{HansenMD2009}, which measures how the probability of finding a particle at position $\mathbf{r}'$ with orientation $\hat{\mathbf{u}}'$ is modified by the presence of a particle at position $\mathbf{r}$ with orientation $\hat{\mathbf{u}}$.

A general difficulty in deriving field theories for interacting particles is the requirement of an expression for the pair-distribution function $g$. This problem can be solved in the passive case by approximating it via relations known from equilibrium density functional theory \cite{teVrugtLW2020} and in the active case by exploiting the fact that $g$ is known from simulation results \cite{JeggleSW2020,BrokertVJSW2024}. However, neither of these is possible here since the system is considerably out of equilibrium and since $g$ has never been studied for systems of this type governed by nonreciprocity. Hence, we use the mean-field approximation, by simply setting $g=1$. This is better justified for soft interaction potentials (as used in this study; see Sec.~\ref{sec:AMN}), for which in the passive case the mean-field approximation is known to be very accurate \cite{LouisBG2000}. Equation \eqref{eq:DDFT_raw_exact} is then reduced to
\begin{align}
    \frac{\partial}{\partial t} \varrho(\mathbf{r},\hat{\mathbf{u}},t) 
    =\,& D_\mathrm{T}  \bm{\nabla}^2 \varrho(\mathbf{r},\hat{\mathbf{u}},t)
        + D_\mathrm{R} \partial_\phi^2\varrho(\mathbf{r},\hat{\mathbf{u}},t)
        \notag\\&\!
        + \beta D_\mathrm{T}  \bm{\nabla}\cdot
            \bigg\{
                \varrho(\mathbf{r},\hat{\mathbf{u}},t) \INT{}{}{^2 r'} \INT{}{}{^2 u'}
                \notag\\&\quad
                \Big[
                    -\mathbf{F}(\mathbf{r},\mathbf{r}',\hat{\mathbf{u}},\hat{\mathbf{u}}',t) \varrho(\mathbf{r}',\hat{\mathbf{u}}',t)
                \Big]
            \bigg\}
        \notag\\&\!
        + \beta D_\mathrm{R} \partial_\phi
            \bigg\{
                \varrho(\mathbf{r},\hat{\mathbf{u}},t) \INT{}{}{^2 r'} \INT{}{}{^2 u'}
                \notag\\&\quad
                \Big[
                    -M(\mathbf{r},\mathbf{r}',\hat{\mathbf{u}},\hat{\mathbf{u}}',t) \varrho(\mathbf{r}',\hat{\mathbf{u}}',t)
                \Big]
            \bigg\}. \label{eq:DDFT_raw}
\end{align}
Note that \cref{eq:DDFT_raw_exact} actually coincides with \cref{eq:DDFT_raw} if we define a modified interaction force $\Tilde{\mathbf{F}}=g\mathbf{F}$ and a modified torque $\Tilde{M}=gM$. Thus, a derivation without the mean-field approximation would essentially correspond to using the modified form of interactions, and since our results do not depend on specific details of the interactions, they are expected to be fairly general. (This would not be the case for, e.g., motility-induced phase separation in active spheres, since there $g$ has a different angular dependence than $\mathbf{F}$ and using $\Tilde{\mathbf{F}}$ rather than $\mathbf{F}$ changes the underlying symmetries. In the setup here, $\mathbf{F}$ and $M$ already have the most general angular dependence compatible with the symmetries of the system, implying that our result will also be general.) 

Given global translational and rotational invariance, $\mathbf{F}$ and $M$ can be written as $\mathbf{F} = f(r,\varphi_1,\varphi_2)\mathbf{\hat{u}}_\mathrm{F}$ and $M=M(r,\varphi_1,\varphi_2)$ with orientation angles $\varphi_1 = \phi_{\mathrm{R}}-\phi$, $\varphi_2 = \phi' -\phi$, and $\mathbf{r}'-\mathbf{r} = r\mathbf{\hat{u}}(\phi_{\mathrm{R}})$. The scalar functions $f$ and $M$ depend only on rotationally invariant quantities, whereas the direction of $\mathbf{F}$, given by the unit vector $\mathbf{\hat{u}}_\mathrm{F}$, is obviously not rotationally invariant. We now assume $\mathbf{\hat{u}}_\mathrm{F} = \mathbf{\hat{u}}(\phi_\mathrm{R})$ (cf. \cref{eq:F21}), as is the case for many pairwise interaction forces usually considered.

The rest of the derivation is rather lengthy and discussed in detail in Appendix \ref{micro}. Here, we list the key steps of the derivation as an overview:
(i) A truncated Fourier expansion is performed for the dependence of $f$ and $M$ on angles $\varphi_1$ and $\varphi_2$.
(ii) A Cartesian orientational expansion \cite{ReinkenKBH2018,teVrugtW2020b} is performed for the dependence of $\varrho$ on $\mathbf{\hat{u}}$, which allows for the replacement
\begin{equation}
\varrho(\mathbf{r},\mathbf{\hat{u}},t) = \rho(\mathbf{r},t) + \mathbf{P}(\mathbf{r},t)\cdot\mathbf{\hat{u}} + \underline{\mathbf{Q}}(\mathbf{r},t):(\mathbf{\hat{u}}\otimes \mathbf{\hat{u}}),
\end{equation}
where ``:'' is a double tensor contraction.
(iii) A truncated gradient expansion \cite{ArcherRRS2019,teVrugtBW2022} is performed for the spatially nonlocal terms in \cref{eq:DDFT_raw}, which allows to express them as a sum of local terms.
(iv) We evaluate the angular integrals, which can be done analytically in closed form, to obtain coupled dynamical equations for the fields $\rho$, $\mathbf{P}$, and $\underline{\mathbf{Q}}$.
(v) We then make a quasi-stationary approximation \cite{teVrugtBW2022} $\partial \underline{\mathbf{Q}} / \partial t =\underline{\mathbf{0}}$, solve the resulting equation for $\underline{\mathbf{Q}}$, and insert the result into the dynamical equations for $\rho$ and $\mathbf{P}$ to obtain closed equations of motion for these two fields.

This results in a general field-theoretical model (given by Eqs.~\eqref{dotpreduced} and \eqref{dotrhoreduced} in Appendix \ref{micro}) which is valid for any reciprocal or nonreciprocal interaction forces and torques. Note that the derivation is predictive, i.e., it provides explicit microscopic expressions for all model coefficients.

\subsection{Nonreciprocal field theory: Active Model N}
\label{sec:AMN}

We can derive Active Model N from the general theoretical model (Eqs.~\eqref{dotpreduced} and \eqref{dotrhoreduced}) simply by explicitly calculating the model coefficients using the nonreciprocal force and torque equations \eqref{eq:f} and \eqref{eq:M}. For this choice of force and torque, most of the coefficients in the general model are equal to zero, such that the theory considerably simplifies. The model is further simplified by dropping the terms of second order contribution (see Appendix \ref{micro}). 
Finally, the resulting model is nondimensionalized via a length scale $a$ representing the interaction range of force and a time scale $a^2/D_\mathrm{T}$, with density fields $\rho$, $\mathbf{P}$, and $\underline{\mathbf{Q}}$ rescaled by $a^2$.
We then reach a new continuum density-field theory, i.e., Active Model N,
\begin{align}
  \frac{\partial \rho}{\partial t} & = \nabla^2 \rho + \bm{\nabla} \cdot \left [ B_1 \rho \mathbf{P}
    + B_2 \rho \bm{\nabla} \rho + B_3 \mathbf{P} \nabla^2 \rho \right. \nonumber\\
    &\quad\, + \left. 2B_3 \left (\mathbf{P} \cdot \bm{\nabla} \right ) \bm{\nabla} \rho
    + B_4 \rho \bm{\nabla} \nabla^2 \rho \right ], \label{eq:rho}\\
  \frac{\partial \mathbf{P}}{\partial t} & = \nabla^2 \mathbf{P} - \tilde{D}_\mathrm{R} \mathbf{P}
  + 2B_1 \bm{\nabla} \rho^2 + B_1 \bm{\nabla} \cdot \left ( \rho \underline{\mathbf{Q}} \right ) \nonumber\\
  &\quad\, - \tilde{D}_\mathrm{R}B_5 \left ( 2\rho \mathbf{P} - \underline{\mathbf{Q}} \cdot \mathbf{P} \right ), \label{eq:P}
\end{align}
where the nematic order parameter tensor $\underline{\mathbf{Q}}$ is given by 
\begin{align}
  Q_{ij} &= B_5 \Big( -P_iP_j + \frac{1}{2} \delta_{ij} P_k^2 \Big) \nonumber\\
  &\quad\, + \frac{B_1}{4\tilde{D}_\mathrm{R}} \left [ \partial_i \left ( \rho P_j \right ) + \partial_j \left ( \rho P_i \right )
    - \delta_{ij} \partial_k \left ( \rho P_k \right ) \right ] \label{eq:Qij}
\end{align}
with coupling coefficients $B_i$ ($i=1,...,5$) and the nondimensionalized rotational diffusion constant $\tilde{D}_\mathrm{R} = a^2 D_\mathrm{R} / D_\mathrm{T}$. Active Model N developed here constitutes a minimal microscopic theory for single-species nonreciprocal interactions since its microscopic derivation (Appendix \ref{micro}) involves the minimal number of orders in the Fourier expansion \cite{teVrugtW2020b} (namely one) required for obtaining a model that is not passive \cite{teVrugtBW2022}, and the minimal number of orders in the gradient expansion \cite{ArcherRRS2019} (namely three) that are required for capturing the dynamics of active phase separation \cite{Cates2019b}. As shown in Appendix \ref{relationamb}, this model contains the previously developed Active Model B+ \cite{TjhungNC2018} as a limiting case, which is notable since Active Model B+ has been developed as a field theory for self-propelling particles \cite{teVrugtBW2022}. This is in line with similar observations made recently for reaction-diffusion systems \cite{RobinsonMS2024} and shows that theories like Active Model B+ are significantly more general than what specific microscopic derivations might suggest.

The coupling coefficients $B_i$ can be explicitly calculated as moments of the microscopic nonreciprocal forces and torques. They vary with the vision-cone opening angle $\vartheta$ and the force and torque parameters, as shown in Appendix \ref{micro}. In this study we choose the specific form of force and torque functions as
\begin{equation}
  h(r) = -F_0 e^{-{r^2}/{a^2}} + F_1 e^{-{r}/(\alpha a)}, \quad
  \tau(r) = b e^{-r/c}, \label{eq:g_tau}
\end{equation}
which describes a force with a short-ranged repulsive contribution of strength $F_0$ and a long-ranged attractive contribution of strength $F_1$, and a torque of strength $b$. The length scales $a$ and $c$ determine the range on which the forces and torques act. After rescaling over the length scale $a$, the nondimensionalized coefficients $B_i$ in Eqs.~\eqref{eq:rho} and \eqref{eq:P} are expressed as
\begin{align}
  B_1 & = \pi \beta a \left ( F_0 - 2\alpha^2 F_1 \right ) \sin\vartheta, \nonumber\\
  B_2 & = \frac{\pi}{2} \beta a \left ( \sqrt{\pi} F_0 - 8\alpha^3 F_1 \right ) \vartheta, \nonumber\\
  B_3 & = \frac{\pi}{8} \beta a \left ( F_0 - 12\alpha^4 F_1 \right ) \sin\vartheta, \nonumber\\
  B_4 & = \frac{3}{32} \pi \beta a \left ( \sqrt{\pi} F_0 - 64\alpha^5 F_1 \right ) \vartheta, \nonumber\\
  B_5 & = -2\pi \beta b \left ( \frac{c}{a} \right )^2 \vartheta.
\end{align}
In the following, we use this choice of soft interaction and set a weaker attractive force (e.g., $F_1/F_0 = 10^{-3}$) and torque ($b/(F_0a) = 0.3$) relative to the repulsive force, but with a longer interaction range ($\alpha = c/a =2$). The corresponding expressions of rescaled model parameters are summarized in Appendix \ref{sec:parameters}. Note that the nonreciprocal active field theory given by Eqs.~\eqref{eq:rho} and \eqref{eq:P} is generic and the forms of model terms do not depend on the specific choices of force and torque functions (see Appendix \ref{micro}). The combination of a short-ranged repulsive and a long-ranged attractive force holds for many interactions usually considered.

It is also important to note that in the model constructed here, only nonreciprocal interactions are incorporated (with the use of Eqs.~\eqref{eq:f} and \eqref{eq:M}), such that all the observed effects are solely a consequence of single-species nonreciprocity. Of course, actual physical or biological living systems are governed by both nonreciprocal and reciprocal interactions (such as reciprocal steric repulsion that comes into play at very short distances) as well as self-propulsion. Thus, the effects and phenomena observed in our model will not directly map onto those observed in, e.g., real living systems. Nevertheless, they provide fundamental insights into novel physical phenomena that, in a more complex form, are relevant for living systems; also importantly, the resulting predictions can be examined in experimentally controllable artificial systems such as specifically-designed colloidal particles \cite{LavergneWBB2019}. Moreover, it is quite common in biological systems that activity takes the form not of self-propulsion of individual agents but of active relative motion, and the study of such phenomena can benefit substantially from microscopically derived active field theories, as demonstrated recently for microtubule-motor mixtures interacting via sliding and alignment \cite{DeLucaMF2024}.

\section{Bifurcations and phase diagram of active patterns}
\label{sec:phase_diagram}

Two steady-state fixed points can be identified from the model equations (\ref{eq:rho}) and (\ref{eq:P}), for a fully disordered state with constant particle density $\rho = \bar{\rho}$ and zero polarization density $\mathbf{P} = \bar{\mathbf P} = \mathbf{0}$, and a flocking state still with constant $\rho = \bar{\rho}$ but having nonzero fixed magnitude of polarization
\begin{equation}
  |\mathbf{P}|^2 = |\bar{\mathbf P}|^2 = -\frac{4}{B_5} \left ( \bar{\rho} + \frac{1}{2B_5} \right ),
  \label{eq:Pbar}
\end{equation}
where $B_5$ is given by \cref{eq:b5}, when
\begin{equation}
  1 + 2B_5\bar{\rho} < 0, \quad {\rm i.e.,} \quad \bar{\rho} \vartheta > \frac{1}{4\pi \beta b c^2}.
  \label{eq:sig1_conditionmain}
\end{equation}
A primary bifurcation analysis can be conducted for the fully disordered state (Appendix \ref{sec:bifurcation}). For the soft interaction forms of \cref{eq:g_tau} chosen here, it yields a flocking transition (without phase separation) when \cref{eq:sig1_conditionmain} is satisfied at large enough average particle density $\bar{\rho}$ or opening angle $\vartheta$, with lower transition threshold for larger torque strength $b$ and range $c$, consistent with the torque aligning effect. The resulting homogeneous flocking phase is characterized by uniform distribution of particle density ($\rho = \bar{\rho}$) with constant polarization alignment ($\mathbf{P} = \bar{\mathbf P} \neq \mathbf{0}$), i.e., a uniform aligned phase where the orientational symmetry is spontaneously broken. More interesting transitions occur for the subsequent secondary bifurcation from this orientationally symmetry-breaking state, with the linearized dynamical equation for Fourier components of perturbations $\widehat{\rho}_{\mathbf{q}}$ and $\widehat{\mathbf P}_{\mathbf{q}}$ given by
\begin{equation}
  \frac{\partial}{\partial t} \left (
  \begin{array}{c}
    \widehat{\rho}_{\mathbf{q}} \\ \widehat{P}_{x\mathbf{q}} \\ \widehat{P}_{y\mathbf{q}}
  \end{array}
  \right )
  = \mathcal{L}(\mathbf{q}, \bar{\rho}, \bar{\mathbf P}, \vartheta) \left (
  \begin{array}{c}
    \widehat{\rho}_{\mathbf{q}} \\ \widehat{P}_{x\mathbf{q}} \\ \widehat{P}_{y\mathbf{q}}
  \end{array}
  \right ), 
\end{equation}
where details of the dynamical matrix $\mathcal{L}$ are presented in \cref{eq:dynamicalmatrix} of Appendix \ref{sec:bifurcation}. When $\vartheta \neq \pi$ (with a limited vision perception $2\vartheta < 2\pi$), $\mathcal{L}^\dagger \neq \mathcal{L}$, i.e., $\mathcal{L}$ is non-Hermitian, leading to nonreciprocal, PT symmetry breaking phase transitions \cite{YouPNAS20,FruchartNature21}. Note that here the system non-Hermiticity originates from vision-cone nonreciprocal interactions within single species, fundamentally different from the previous cases featured by asymmetric coupling between different populations (e.g., interspecies cross diffusivities) \cite{SahaAG2020,YouPNAS20,KreienkampK2022,FruchartNature21} or different harmonic modes \cite{FruchartNature21}.

\begin{figure*}[tb]
\includegraphics[width=\textwidth]{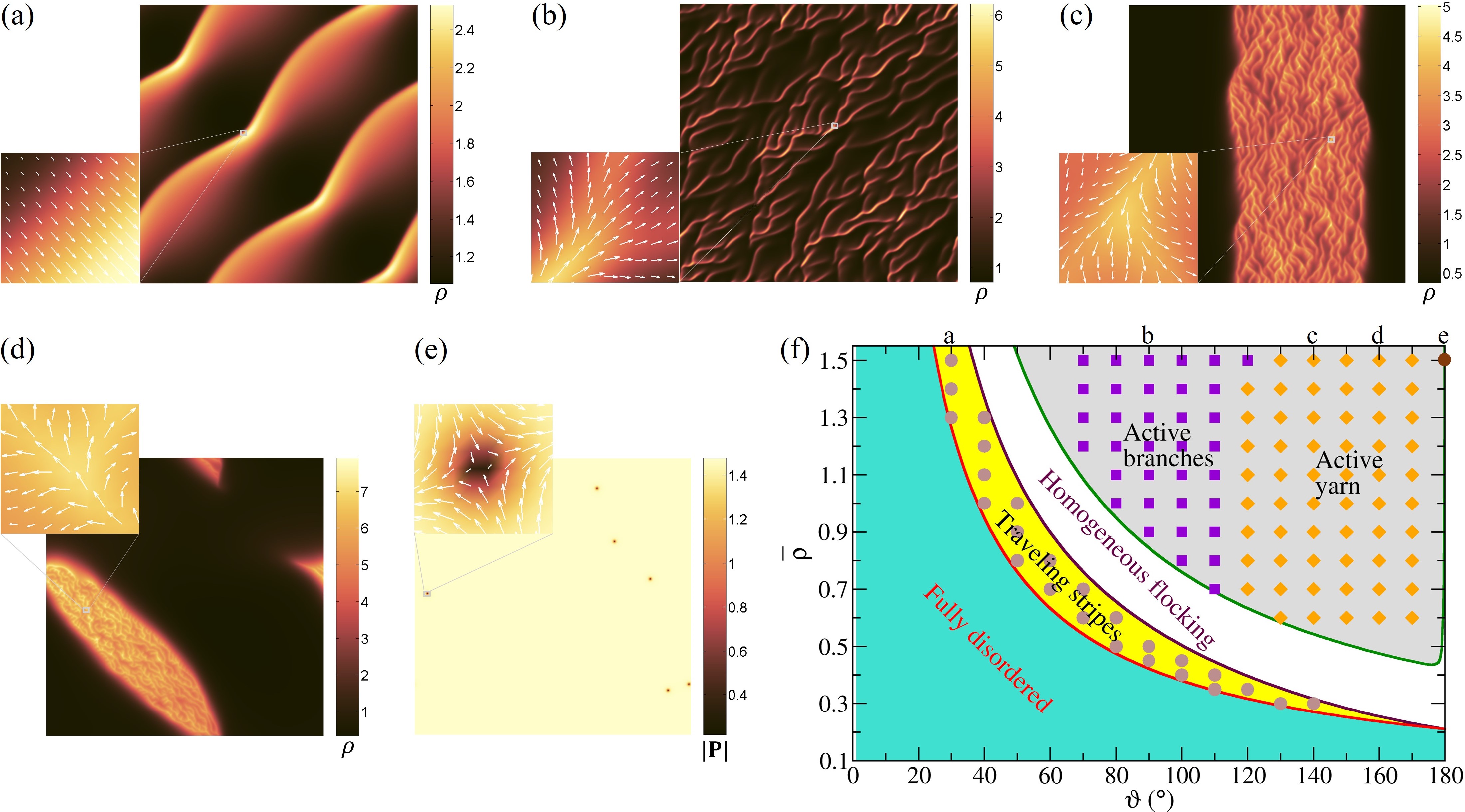}
\caption{
(a)--(e) Simulation snapshots for various active patterns that emerge at average particle density $\bar{\rho}=1.5$, including active stripes at $\vartheta=30^\circ$ ((a) and Supplemental Video 1), active branches at $\vartheta=90^\circ$ ((b) and Supplemental Video 2), active yarn at $\vartheta=140^\circ$ ((c) and Supplemental Video 3) and $160^\circ$ ((d) and Supplemental Video 4), and a homogeneous flocking phase with vortex defects of polarization field $\mathbf{P}$ at the reciprocal limit $\vartheta=180^\circ$ ((e) and Supplemental Video 5). In (a)--(d) with patterns induced by nonreciprocity, the spatial profiles of particle density $\rho$ are presented, and the local distribution of vector field $\mathbf{P}$ are indicated as arrows in the insets which are enlarged portions of the boxed regions. The patterns self-travel opposite to the average direction of $\mathbf{P}$ (Supplemental Videos 1-5). In (e) the spatial profile of polarization magnitude $|\mathbf{P}|$ is shown. (f) The phase diagram of $\bar{\rho}$ vs $\vartheta$, with solid curves evaluated from the analytical results of bifurcation analysis and the symbols identified via outcomes of numerical simulations (those giving disordered or homogeneous phases are not shown).}
\label{fig:phase_diagram}
\end{figure*}

We have calculated the corresponding $\bar{\rho}$ vs $\vartheta$ phase diagram, as shown in Fig.~\ref{fig:phase_diagram}, by evaluating the analytical results of bifurcation analyses (Appendix \ref{sec:bifurcation}), with phase boundaries well agreeing with numerical simulations of the full model equations (\ref{eq:rho})--(\ref{eq:Qij}). A noteworthy phenomenon is the occurrence of phase transitions with the increase of $\vartheta$ or $\bar{\rho}$, i.e., from (i) the disordered state of both particle and polarization densities, to (ii) a traveling striped state of low- vs high-density microphase separation and local flocking alignment (Fig.~\ref{fig:phase_diagram}(a) and Supplemental Video 1), then to (iii) homogeneous flocking with again disordered particle density, and back to (iv) a phase separated, flocking state with emergent dynamical chiral patterns, i.e., patterns for which spatial mirror symmetry is broken, including active branches (Fig.~\ref{fig:phase_diagram}(b) and Supplemental Video 2) and active yarn pattern which shows as a phase consisting of traveling bands containing microphase-separated fine texture that are embedded in a homogeneous medium (see Figs.~\ref{fig:phase_diagram}(c), \ref{fig:phase_diagram}(d) and Supplemental Videos 3, 4), before (v) a reentrance to the homogeneous flocking phase at $\vartheta=\pi$. In the limit of $\vartheta=\pi$ with complete perception, the system becomes reciprocal and the dynamics of $\widehat{\rho}_{\mathbf{q}}$ and $\widehat{\mathbf P}_{\mathbf{q}}$ are decoupled, resulting in the stability of uniform flocking with homogeneous $\rho$ and the usual vector field dynamics for polarization $\mathbf{P}$ with motion and annihilation of vortex and antivortex defects (Fig.~\ref{fig:phase_diagram} (e) and Supplemental Video 5). The phase separation effects observed here are thus induced by vision-cone-type force nonreciprocity.

The active branches and active yarn patterns are distinguished through two main characteristics, (i) microphase separation with the formation of spatially dispersed fibers for active branches vs the simultaneous occurrence of micro and bulk phase separations for the dense banding of active yarn, and (ii) the appearance of different fine textures that are dominated by bifurcated microstructures in active branches but multi-branching in active yarn, as will be further detailed in the next section and Fig.~\ref{fig:pattern_evolution}. Both types of patterns vary persistently with time and are the dynamical results of nonlinear evolution far from equilibrium; thus the phase boundary between them (see Fig.~\ref{fig:phase_diagram}(f)) cannot be determined by linear stability or bifurcation analysis but was estimated through simulations. With the increase of opening angle $\vartheta$, around the boundary of transition the patterns start to develop large gaps of homogeneous medium separating clusters/bands of fibers (instead of spreading throughout the whole system), with the mixing of both bifurcated and multi-branched fiber textures.

\begin{figure*}[tb]
\includegraphics[width=\textwidth]{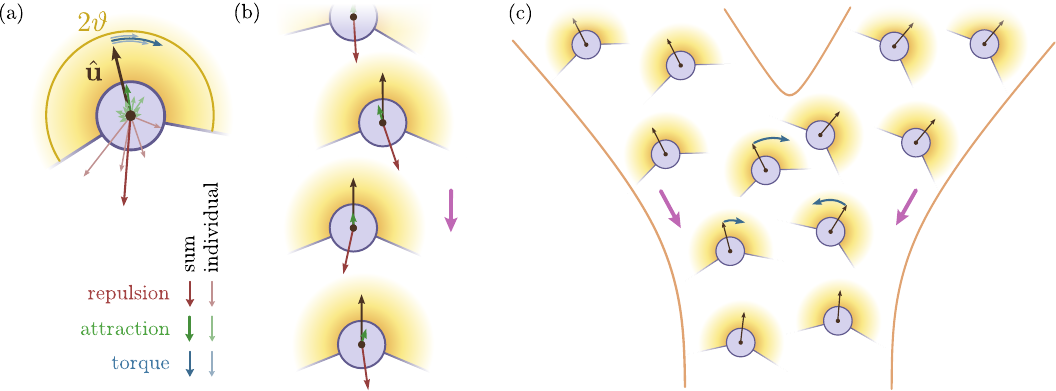}
\caption{(a) Visualization of a single particle with orientation $\hat{\mathbf{u}}$ and vision-cone opening angle $2\vartheta$. Via interacting with other nearby particles within its field of vision, it experiences strong short-range repulsive (red) and weak long-range attractive (green) forces, as well as aligning torques (blue). The individual forces and torques (light arrows) result in a net force and a net torque (bold arrows). (b) On a microscopic level, spontaneous motion originates from the net repulsive forces in the bulk. Each particle experiences a net repulsion by the particles in front of it (within its vision cone) but no force from those behind, leading to collective motion in the direction of the bold violet arrow on the right. (c) Visualization of the microscopic mechanism for branch merging. If two branches cross, the particles belonging to different branches start seeing each other and, due to the aligning torques, acquire the same orientation. As a result, they travel within one merged branch.}
\label{fig:pattern_mechanism}
\end{figure*}

Although no explicit self-driving factor has been incorporated in the model, all phase-separated patterns self-travel unidirectionally, which leads to the breaking of time-reversal symmetry. The direction of pattern traveling is opposite to that of the average polarization field $\langle \mathbf{P} \rangle$, with average migration velocity $\bar{\mathbf{v}} \propto -\langle \mathbf{P} \rangle$. This is consistent with the microscopic picture: Due to nonreciprocity, each particle can only interact with the neighboring particles within its own vision cone centered around its orientation (Fig.~\ref{fig:schematic}). Since in the setup here the short-range repulsion is much stronger than the long-range attraction, the particle feels the net effect of repulsion from the particles in front of it (within the vision cone) but not from those behind, such that every particle ends up being pushed towards the direction opposite to its orientation, as illustrated in Figs.~\ref{fig:pattern_mechanism}(a) and \ref{fig:pattern_mechanism}(b). Given that $\mathbf{P}$ measures the particle orientation, collective particle migration occurs if \cref{eq:sig1_conditionmain} is satisfied with strong enough torque alignment. As particles are still being attracted to next neighbors, such a competition between short- and long-range interactions leads to phase separation at large enough particle density or opening angle. This phenomenon of nonreciprocity-induced phase separation with local orientational alignment then incorporates both types of collective behavior of active particles \cite{ZhangNatPhys21,DauchotNatPhys21}, assembly via phase separation, and aligning collective self-migration, with mechanisms distinct from the previously known motility-induced \cite{CatesT2015} or nonreciprocal-torque-based \cite{ZhangNatPhys21} phase separation.

The unidirectional migration can be understood more quantitatively from the nonzero locally averaged force for a particle surrounded by its neighboring bath, i.e.,
\begin{align}
  \langle \mathbf{F}_{2 \rightarrow 1} \rangle_{\rm local} & \simeq \bar{\rho}_{\rm local} \INT{}{}{\mathbf{r}_2} \dif\hat{\mathbf{u}}_2 \mathbf{F}_{2 \rightarrow 1} (\mathbf{r}_2 - \mathbf{r}_1, \hat{\mathbf u}_1 \cdot \hat{\mathbf r}, \hat{\mathbf u}_2 \cdot \hat{\mathbf r}) \nonumber\\
  & = 4\pi \bar{\rho}_{\rm local} \left [\INT{0}{\infty}{r} r h(r) \right ] \sin(\vartheta) \hat{\mathbf u}_1 \neq \mathbf{0},
  \label{eq:F_avg}
\end{align}
where $\bar{\rho}_{\rm local}$ represents local particle density and Eqs.~(\ref{eq:F21}) and (\ref{eq:f}) have been used. Using the choice of \cref{eq:g_tau}, we have $\int_0^\infty \dif r\, r h(r) = -a^2 F_0 (1 - 2\alpha^2 F_1/F_0)/2 < 0$ for weak enough attraction as considered here, leading to $\langle \mathbf{F}_{2 \rightarrow 1} \rangle_{\rm local} \propto -\hat{\mathbf u}_1$ for finite perception $0 < \vartheta < \pi$, consistent with the simulation result of reverse traveling and the microscopic mechanism described above. (If considering the opposite case of strong enough attraction such that $\int_0^\infty \dif r\, r h(r) > 0$, from \cref{eq:F_avg} and similar microscopic picture the forward traveling of particles would be expected.) Thus, our results demonstrate that single-species nonreciprocity induces spontaneous self-propulsion on the many-body level.

It is instructive to compare the traveling fibers or strands observed here with the wormlike structure observed in Refs.~\cite{BarberisPRL16,NegiWG2022,NegiWG2024}, where the system under consideration consists of self-propelling particles that have a tendency to move towards particles within their vision cones. This gives rise to a moving worm phase where particles, in a conga-like behavior, follow a leader that is itself unaffected by the particles behind as it cannot see them. While this phenomenon of chain-like winding structure is somewhat related to the fiber formation in the active branched patterns emerging here (both originating from vision-cone-like interactions), a crucial difference is that in our model particles are not self-propelled. Consequently, our results demonstrate that self-traveling structures can occur in systems of particles with vision-cone nonreciprocal interactions even when there is no self-propulsion on the level of individual particles.

It is also noted that the rich behavior of self-traveling patterns identified here is a result of coupling between particle density and polarization vector fields in this single-species nonreciprocal system. A recent work showed that two nonreciprocally coupled, conserved scalar fields in multicomponent systems can also generate rich dynamics of traveling wave patterns \cite{BraunsPRX24}, although with different mechanisms via nonreciprocal Cahn-Hilliard equations. Interesting dynamics of traveling undulations occurs along the interfaces of wave bands or droplets in those systems governed by purely scalar fields, while here the coupling to the polarization vector field (with vision cone orientation and nonreciprocity) allows for more complex nonequilibrium evolution of the pattern that is not limited at interfaces, showing as perpetual variations and flow of chiral patterns and their branched or twisted/undulated micro-textures in active branches and active yarn, as detailed below.

\section{Active branches and active yarn}
\label{sec:active_branches_yarn}

\begin{figure*}[tb]
\includegraphics[width=\textwidth]{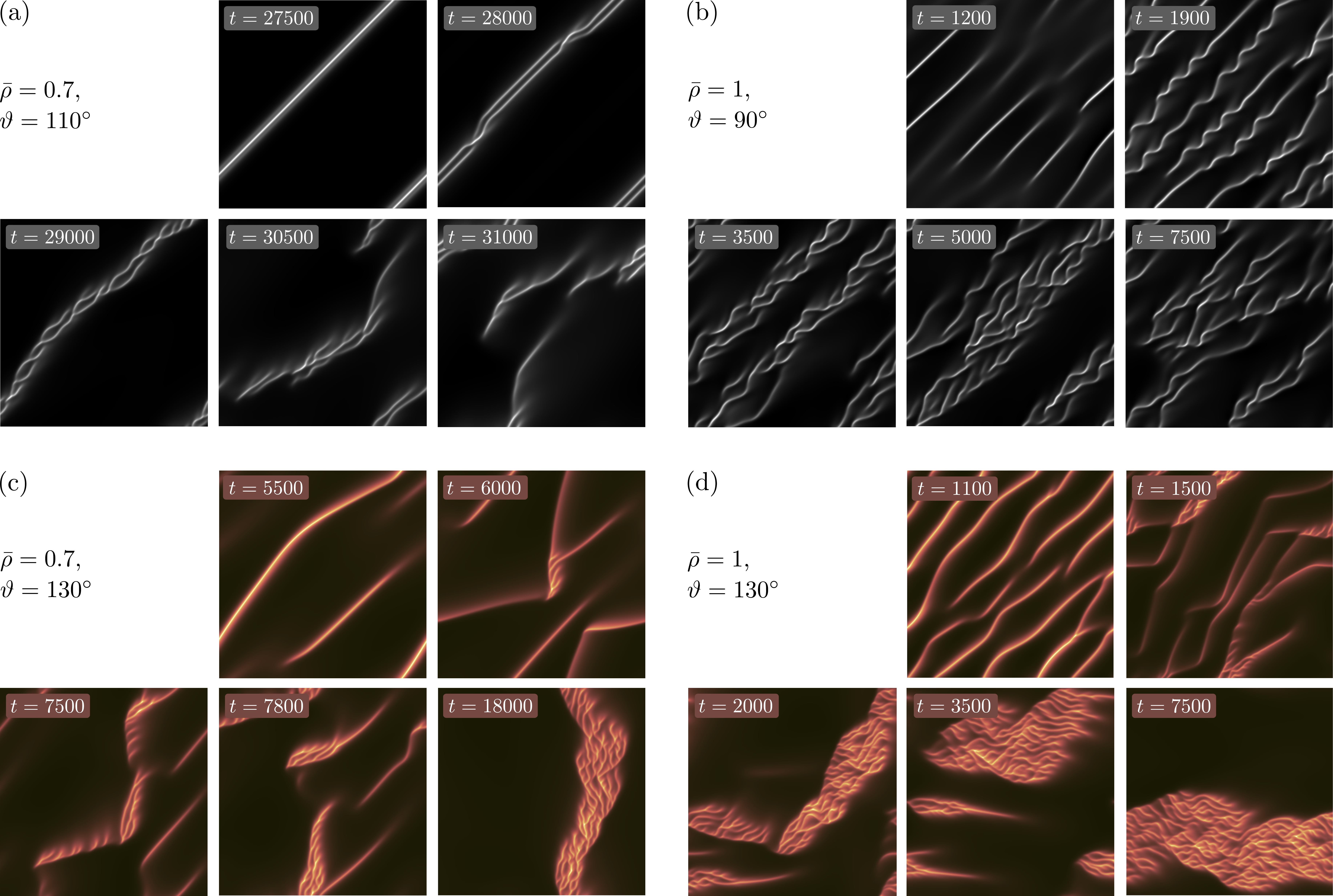}
\caption{Dynamical processes of active pattern formation. Snapshots of density profile at different time stages of system evolution for two sample average densities $\bar{\rho}=0.7$ ((a), (c)) and $\bar{\rho}=1$ ((b), (d)), illustrating the rich dynamics of phase-separated fibers/strands which leads to the persistent variation of active branches ((a), (b) with $\vartheta=110^\circ$, $90^\circ$; Supplemental Videos 6 and 7) or the self-knitting of active yarn ((c), (d) with $\vartheta=130^\circ$; Supplemental Video 8).}
\label{fig:pattern_evolution}
\end{figure*}

One of the key findings here is the emergence of self-migrating chiral patterns induced by single-species nonreciprocity, in the form of active branches and active yarn, at high enough opening angles or average particle densities. These chiral patterns break time-reversal symmetry (traveling unidirectionally) and 2D parity symmetry (lacking mirror symmetry axes in space; see Figs.~\ref{fig:phase_diagram}(b), \ref{fig:phase_diagram}(c), and \ref{fig:phase_diagram}(d)) for both particle and polarization density fields. The formation and nonlinear evolution of these exotic patterns are well beyond the stage of bifurcation and instability development, as illustrated in Fig.~\ref{fig:pattern_evolution}.

Figure \ref{fig:pattern_evolution} shows that two categories of pattern-forming dynamics can be identified. For intermediate vision-cone opening angles, we observe persistently varying active branches, while for large opening angles, the banding of active yarn occurs. For the dynamics of active branches there are two sub-types: At low enough average density (e.g., $\bar{\rho}=0.7$ for $\vartheta = 110^\circ$; Supplemental Video 6), a microphase-separated stripe developed via instability sharpens to an individual fiber or thread, which splits and self-interweaves to form twisted or 2D double-helix type configurations. The strands then deform and bifurcate/branch, before breaking apart and constantly dispersing or regrouping. This double-twisting behavior does not occur at larger $\bar{\rho}$ (e.g., $\bar{\rho}=1$ for $\vartheta = 90^\circ$; Supplemental Video 7), where one instead finds the curving and undulating of individual sharpened fibers during the secondary stage of phase separation. This is followed by their local self-braiding and bundling to form loose, varying clusters in the subsequent tertiary stage of evolution. For both sub-types the resulting branched traveling texture is highly dynamical, aggregating and dispersing incessantly.

\begin{figure*}[htb]
\centerline{\includegraphics[width=\textwidth]{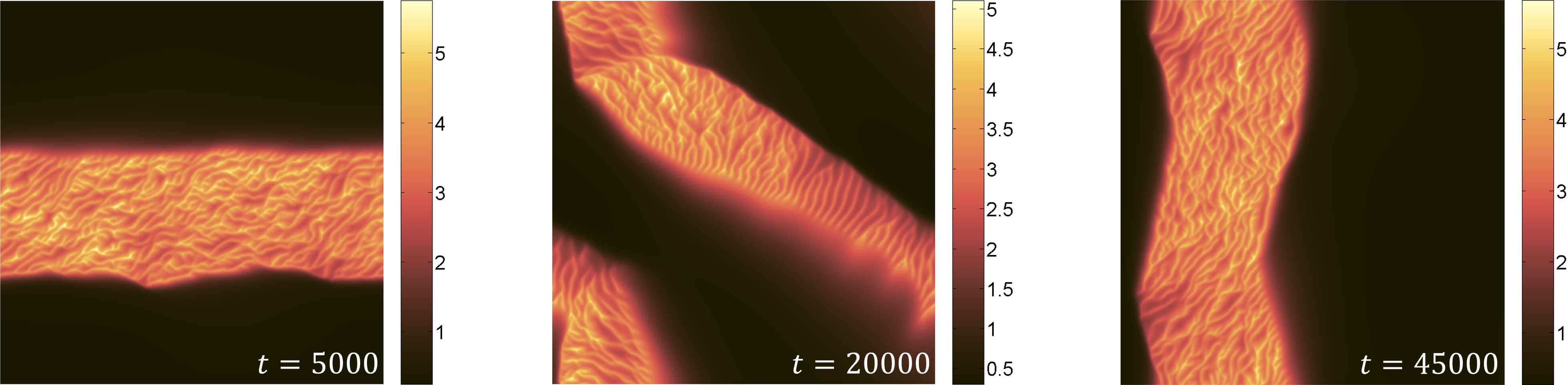}}
  \caption{Time variation of active yarn pattern. During the nonreciprocity-induced self migration, the orientation and morphology of active yarn can vary with time, as illustrated in this sample simulation at $\bar{\rho}=1.5$ and $\vartheta=150^\circ$ (see also Supplemental Video 9).}
  \label{fig:activeyarn_varying}
\end{figure*}

For the second category -- formation of active yarn at large opening angles -- the branching occurs earlier and directly from the bent sharpening fibers. The multi-branched threads aggregate and merge to compact bands with fluctuating interfaces that separate from the homogeneous melt, yielding a self-knitting process of a strongly condensed, active yarn pattern (see Fig.~\ref{fig:pattern_evolution} at $\vartheta = 130^\circ$ and Supplemental Video 8). The pattern not only shows persistent variations of micro-textures within each band, but could also evolve via an alternating process of aggregation/breaking/rebanding with the change of traveling band orientation, while the overall pattern morphology still remains without coarsening with time (see Fig.~\ref{fig:activeyarn_varying} and Supplemental Video 9).

\begin{figure}[htb]
  \centerline{\includegraphics[width=0.5\textwidth]{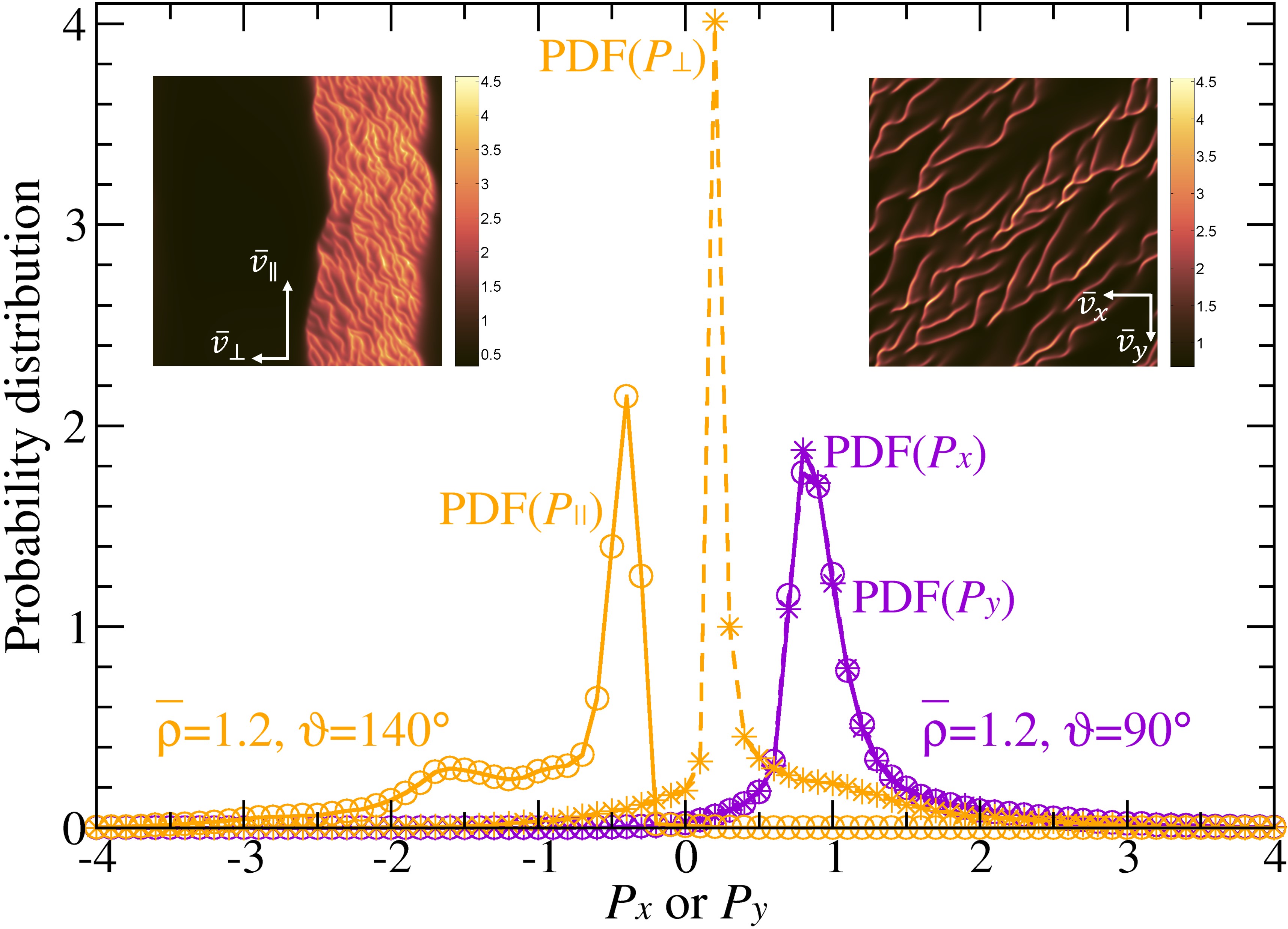}}
  \caption{Probability distribution of polarization and velocity. The probability density function (PDF) for the polarization component $P_x$ or $P_y$ is calculated from simulations at $\bar{\rho}=1.2$ and opening angle $\vartheta=90^\circ$ (active branches at $t=50000$) or $140^\circ$ (active yarn at $t=37500$, with $P_x=P_\perp$ and $P_y=P_{\parallel}$ in this example). It corresponds to the estimated probability distribution of locally coarse-grained particle velocity. The average migration velocity $\bar{\mathbf{v}} = (\bar{v}_x, \bar{v}_y)$, which is proportional to the reverse average polarization $-\langle \mathbf{P} \rangle$, is indicated in the inset images.}
  \label{fig:PDF_PxPy}
\end{figure}

\begin{figure}[htb]
\centerline{\includegraphics[width=0.5\textwidth]{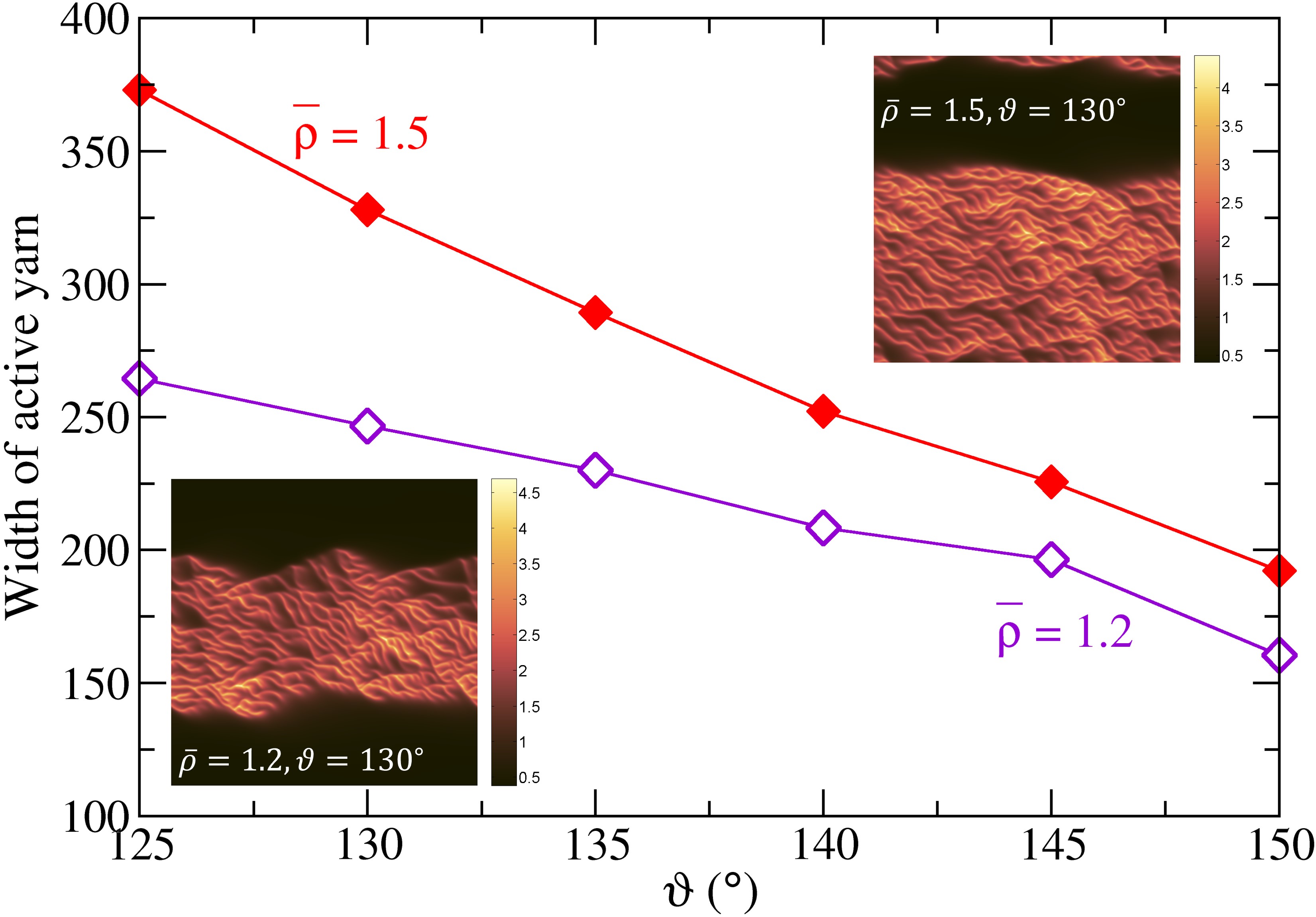}}
\caption{\label{fig:activeyarn_width} Width of active yarn as a function of opening angle $\vartheta$. Results are estimated from simulations at average densities $\bar{\rho}=1.2$ and $1.5$ with system size $512 \times 512$, showing the decrease of active yarn banding width with the increase of $\vartheta$ and the decrease of $\bar{\rho}$.}
\end{figure}

Both active branches and active yarn patterns evolve at two different length and two separate time scales. The fine branched textures evolve rapidly and perpetually on short timescales, while the flow and variation of the entire pattern structure occur on longer timescales (see, e.g., Supplemental Videos 2-4, 9). As seen also in Fig.~\ref{fig:PDF_PxPy} for the estimated probability distribution of velocity, the average migration velocity of the pattern, with value close to the sharp peak of the skewed distribution observed, corresponds to the overall timescale of pattern flow, while the extended range of the distribution, particularly for large magnitudes of velocity, reflects the faster variations of fine textures. Spatially, a prominent property is given by the dense banding of active yarn, which represent a large-scale phase separation between a surrounding homogeneous state and bundles or bands with fine-scale microphase-separated branched structures inside. Our simulations show that the larger length scale of bulk separation depends on the size of the simulation domain, suggesting that this is related to phase coexistence. The small length scale, on the other hand, depends on the physical parameters of the system and is about an order of magnitude smaller than the characteristic width of active yarn which is estimated in Fig.~\ref{fig:activeyarn_width}. It is important to note that the formation of active yarn is beyond the traditional phase coexistence and is characteristic of a new dynamical pattern, since it involves the simultaneous development of small-scale multi-branching and larger-scale local bundling or banding (Fig.~\ref{fig:pattern_evolution} and Supplemental Video 8), belonging to a different category of formation process as compared to active branches (noting also the difference of fine textures between active yarn with multi-branched strands within the highly condensed band and active branches consisting of mostly bifurcated fibers). Thus our results demonstrate the realizability of this unique type of active yarn pattern which self-organizes and evolves at multiple-scales through simultaneous microphase and bulk phase separations.

Microscopically, the banding phenomenon, particularly during the active yarn formation, can be understood by noting that, since the interaction combines a strong and short-ranged repulsive force with a weaker long-ranged attractive force (\cref{eq:g_tau}), the interaction of a particle with its nearest neighbors is dominated by repulsion, whereas the interaction with its next-nearest neighbors is dominated by attraction. At very large vision-cone angle (e.g., $2\vartheta \geq 240^\circ$) with narrow blind spot, the net attraction from a broad range of next neighbors would be strong enough to forge and confine the particle, then bundling or banding together with other nearby particles while traveling concurrently. Thus narrower, denser bands can be expected for larger $\vartheta$ and smaller $\bar{\rho}$, consistent with the numerical simulation results shown in Fig.~\ref{fig:activeyarn_width}. If the perception is further increased to $2\vartheta = 2\pi$, there would be no blind spot and the interactions are reciprocal; the strong net repulsion would then push away particles from all directions ($\langle \mathbf{F}_{2 \rightarrow 1} \rangle_{\rm local} = \mathbf{0}$ from \cref{eq:F_avg}), making the density distribution disordered without any phase separation.

A microscopic mechanism leading to the lightning-like branched structure found in active branches and active yarn is illustrated in Fig.~\ref{fig:pattern_mechanism}(c). If two traveling branches cross, the particles in these two differently oriented branches start to see each other and consequently interact and align their orientations. Afterwards, they travel together as one merged branch driven by the net repulsion (as described in Sec.~\ref{sec:phase_diagram}). This microscopic picture is supported by the simulation result that the patterns generally travel towards the direction of the branching points rather than away from them (see, e.g., Supplemental Videos 2 and 3). Consequently, it is likely that the lightning pattern arises from branches merging rather than from branches bifurcating.

\section{Active droplet growth dynamics}

To further examine the chiral pattern dynamics, we simulate the growth process of an active droplet developing branched structures when embedded in a homogeneous flocking medium. Details of the simulation setup are given in Appendix \ref{sec:parameters}. As shown in Fig.~\ref{fig:droplet_growth}, the droplet domain grows anisotropically, extending along the droplet self-migration direction. At the early stage with the development of linear instability to form weak stripes within the self-traveling droplet, the domain grows slowly. The growth accelerates at an intermediate time stage when microphase and/or bulk phase separations develop inside the droplet during the secondary and tertiary processes of nonlinear evolution of either active branches or active yarn (see also Sec.~\ref{sec:active_branches_yarn} and Fig.~\ref{fig:pattern_evolution}); i.e., the stripes sharpen to become fibers which then curve/twist and branch, together with their local aggregation and bundling, and particularly in the case of active yarn, segregate at a larger scale to the two internal ends of droplet, overall manifesting as a growing and moving ``ball of wool''. 

\begin{figure}[tb]
  \centerline{\includegraphics[width=0.5\textwidth]{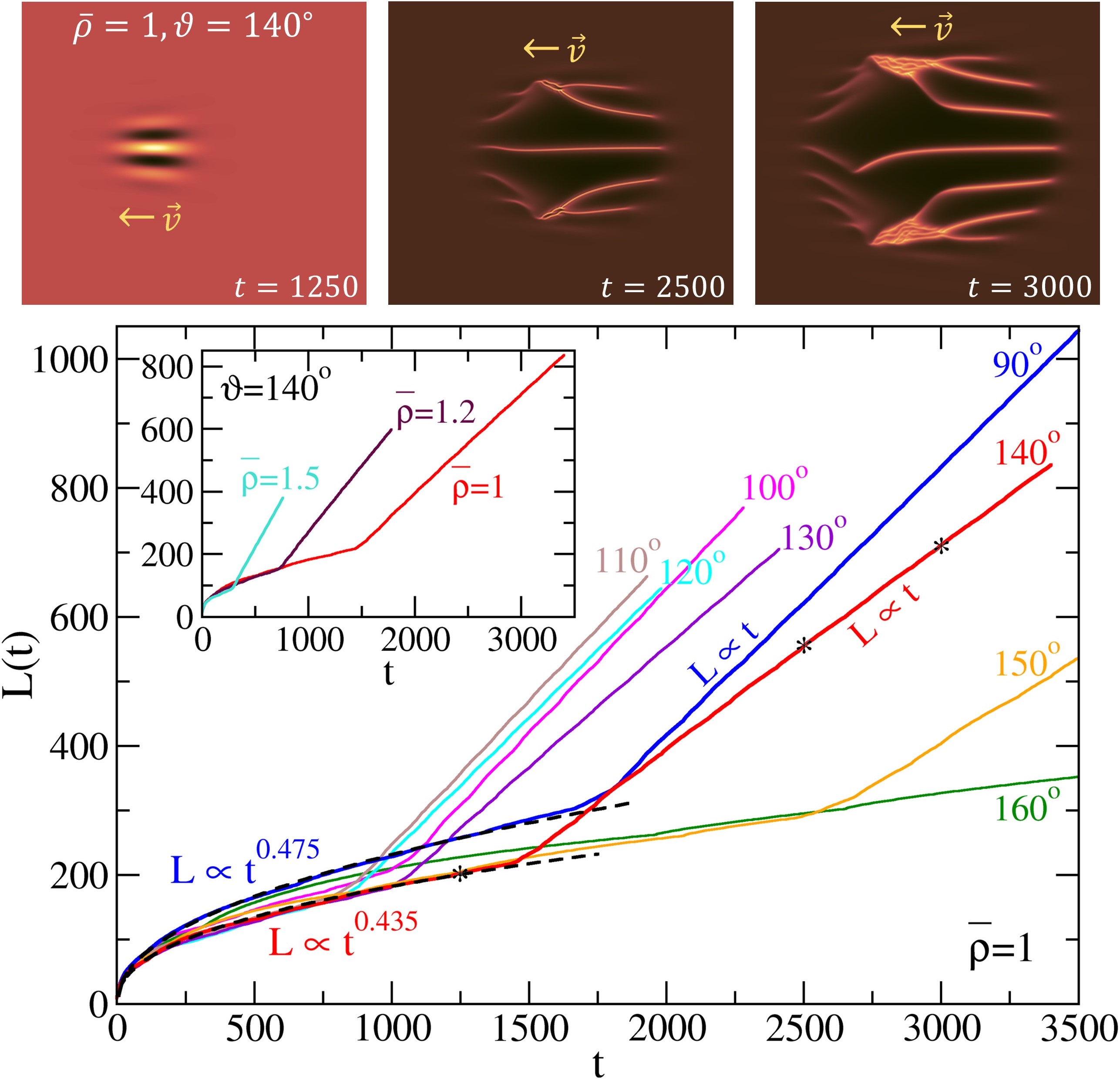}}
  \caption{Active droplet growth. The droplet size $L(t)$ as a function of time $t$, at $\bar{\rho}=1$ and $\vartheta$ ranging from $90^\circ$ to $160^\circ$ for active branches or active yarn patterns (inset: $\bar{\rho}=1, 1.2, 1.5$ at $\vartheta=140^\circ$). The time-evolving snapshots, which are small portions of the large system simulated, correspond to the stars indicated on the $\vartheta=140^\circ$ curve. Also shown are the fittings to a power law scaling $L \propto t^n$ at early time (dashed lines), with $n = 0.475 \pm 0.001$ at $\vartheta=90^\circ$ and $n = 0.435 \pm 0.002$ at $\vartheta=140^\circ$, before the crossover to a linear regime of $L \propto t$.}
  \label{fig:droplet_growth}
\end{figure}

A much faster growth takes place at late time stages, where the phase-separated region spreads both vertically and horizontally out of the droplet domain and expands up to the simulation boundaries. The droplet shape thus cannot be maintained, and the resulting much more rapid domain growth is caused by simulation boundary effects and is not physical. Thus, in the plots of Fig.~\ref{fig:droplet_growth} for the time evolution of characteristic droplet size $L(t)$, only the data points at the early and intermediate time stages are shown, without those unphysical results at late stage. All these results are generic and consistent for different opening angle $\vartheta$ and average density $\bar{\rho}$ yielding active branched and yarn states.

It has been known \cite{BrayAP94} that the growth of droplet or domain size $L(t)$ obeys a power law behavior, $L \sim t^n$, with the scaling exponent $n=1/3$ for the diffusion-controlled mechanism (Lifshitz-Slyozov-Wagner law), $n=1/2$ for the curvature-driven growth, and $n=1$, i.e., linear growth, when the hydrodynamic advection effect dominates, while activity could possibly arrest phase separation and growth in ``wet'' systems \cite{TiribocchiPRL15,CaballeroPRL22}. Our simulation results summarized in Fig.~\ref{fig:droplet_growth} indicate a crossover between two scaling regimes at early and intermediate time stages, before the influence of boundaries takes place at late times as described above. At the early stage during the development of linear instability (with the emergence of very weak and diffuse stripes inside droplet), a scaling exponent $n$ of value between $1/3$ and $1/2$ is found, which can be attributed to a combination of diffusion-driven and curvature-driven effects. As time increases, a transition to a linear growth regime with $L \sim t$ occurs, when the system is governed by the formation and nonlinear evolution of active branches or active yarn pattern within the droplet, with the growth mechanism dominated by those behaviors of phase separations induced by nonreciprocity. Around the crossover between these two scaling regimes, sharpening fibers or strands start to form, before further undulating, twisting, or banding when getting into the linear growth regime.  It is noted that in contrast to the linear growth previously found in ``wet'' systems \cite{TiribocchiPRL15,CaballeroPRL22}, both advective effect and activity arrest are absent in the nonreciprocal ``dry'' system studied here.

\section{Conclusions}

We have constructed a nonreciprocal continuum field theory, Active Model N, which incorporates the many-body effects induced by nonreciprocity and constitutes a minimal model for collective dynamics in a general class of single-species systems characterized by nonreciprocal forces. It allows us to reveal the intriguing results arising from the mechanism of competition between repulsive and attractive interactions of different strength and range and their combination with interparticle nonreciprocity and local alignment. Counterintuitively, in the absence of self-propulsion in the equations of motion for individual particles, our results show that unidirectional self-traveling of particles and patterns occurs spontaneously, demonstrating a fundamental and intrinsic connection between nonreciprocal matter and self-propelling active matter. This leads to the emergent phenomenon of single-species active pattern formation with PT symmetry breaking, a hallmark of nonreciprocal phase transitions. The resulting out-of-equilibrium chiral branched patterns, particularly the self-knitted active yarn, originate from a new class of active phase separation which simultaneously incorporates nonreciprocity-induced micro- and bulk phase separations, with local structural details being highly dynamical and varying persistently with time. 

An important factor responsible for these phenomena is the interplay of vision-cone interactions and aligning torques. Given how ubiquitous orientational alignment is in active matter -- it can arise spontaneously in the absence of explicit alignment interactions \cite{CapriniMP2020} and even in the presence of turn-away torques \cite{Das2024} -- the effects described in this work are expected to be fairly common in systems with nonreciprocal interactions.

A similar behavior of banding and self-traveling of single-species agents with vision-cone interactions can be found in various real biological systems, such as the migration of caterpillars or ants \cite{HeckenthalerPRXLife23} and the marching of penguins \cite{ZitterbartWBF2011} or pedestrians \cite{SchadschneiderS2011}.  Although for the specific terms of Active Model N derived here, only pure effects of nonreciprocal forces and torques are incorporated as our focus is on isolating and identifying the nonreciprocity-induced mechanisms, the field theory we have developed here (in Sec.~\ref{amnderivation} and Appendix \ref{micro}) is generic and can be readily extended by incorporating various forms of reciprocal and nonreciprocal interactions governing real material or living systems. In addition, our results of active pattern formation and dynamical mechanisms originating from single-species nonreciprocity are expected to be applicable and extendable to a range of artificial nonreciprocal systems, particularly Janus colloids \cite{LavergneWBB2019}, that can be realized and controlled in experiments. Moreover, the active yarn phase, where a rather chaotic but patterned state is embedded within a homogeneous medium, has interesting potential applications in the realization of programmable materials. For example, if the particles are immersed in a fluid, this effect allows to achieve controlled mixing of this fluid in one region (the size of which can be controlled by tuning $\vartheta$ or $\bar{\rho}$) while leaving it unmixed elsewhere.
If one additionally incorporates a mechanism in which the particles can adapt the value of $\vartheta$ in response to external stimuli, it would also allow for the realization of adaptive or even intelligent matter \cite{KasparRvdWWP2021,Walther2020}.

\begin{acknowledgments}
We thank Filippo De Luca, Ken Elder, Helder Hugo, and Simiso Mkhonta for helpful discussions. Z.-F.H. acknowledges support from the National Science Foundation under Grant No.\ DMR-2006446. M.t.V., R.W., and H.L.\ are funded by the Deutsche Forschungsgemeinschaft (DFG, German Research Foundation) -- Project-IDs 525063330, 433682494 -- SFB 1459, and SPP 2265 -- LO418/25. J.M.M.\ thanks the Studienstiftung des deutschen Volkes for financial support.
\end{acknowledgments}

\appendix

\section{Microscopic derivation of field theory}
\label{micro}

Here, we present the microscopic derivation of active model N, given by Eqs.~\eqref{eq:rho}--\eqref{eq:Qij}, in more detail. We also present some models for general forms of nonreciprocal interactions.

\subsection{General model}

Our starting point is \cref{eq:DDFT_raw} given in Sec.~\ref{amnderivation} with force $\mathbf{F} = f(r,\varphi_1,\varphi_2)\mathbf{\hat{u}}_F$ and torque $M=M(r,\varphi_1,\varphi_2)$, where $\varphi_1$ and $\varphi_2$ are the orientation angles of particles 1 and 2 respectively. We set $\varphi_1 = \phi_{\mathrm{R}}-\phi$, $\varphi_2 = \phi' -\phi$, and $\mathbf{r}'-\mathbf{r} = r\mathbf{\hat{u}}(\phi_{\mathrm{R}})$ as given in Sec.~\ref{amnderivation}. For ease of notation, we drop any possible time dependence of $\mathbf{F}$ and $M$ since such a time dependence would not affect the calculations.

First, we make the truncated Fourier expansions 
\begin{align}
f&= -\sum_{|n_1|+|n_2| \leq 2}^{} f_{n_1 n_2}(r)\exp[\ii(n_1\varphi_1+n_2\varphi_2)],\label{fourier1}\\
M&= \sum_{|n_1|,|n_2| \leq 1}^{} M_{n_1 n_2}(r)\exp[\ii(n_1\varphi_1+n_2\varphi_2)], \label{fourier2}
\end{align}
with the expansion coefficients \cite{teVrugtW2020b}
\begin{align}
f_{n_1 n_2}(r)&= -\frac{1}{(2\pi)^2}\int_{0}^{2\pi}\!\dif \varphi_1\int_{0}^{2\pi}\!\dif \varphi_2 \notag\\
& \qquad\qquad \times f(r, \varphi_1, \varphi_2)e^{-\ii(n_1\varphi_1+n_2\varphi_2)},\\
M_{n_1 n_2}(r)&=\frac{1}{(2\pi)^2}\int_{0}^{2\pi}\!\dif \varphi_1\int_{0}^{2\pi}\!\dif \varphi_2 \notag\\
& \qquad\qquad \times M(r, \varphi_1, \varphi_2)e^{-\ii(n_1\varphi_1+n_2\varphi_2)}.
\end{align}
The minus sign in \cref{fourier1} ensures that the force terms appear with a plus sign in the dynamic equations. We will mostly restrict ourselves to Fourier modes of first order in the derivation, but use one second-order mode in the expansion of $f$ to allow for an explicit coupling to the nematic tensor $Q_{ij}$ in the equation of $\partial\rho / \partial t$ for the interaction force considered here. The fact that $f$ and $M$ are real implies
\begin{align}
f_{n_1 n_2} &= f^*_{-n_1 -n_2},\\  
M_{n_1 n_2} &= M^*_{-n_1 -n_2},
\end{align}
where ``$*$'' denotes complex conjugation.  We make the substitution $\mathbf{r}'\to\mathbf{r}+\mathbf{r}'$ and use the gradient expansion \cite{teVrugtBW2022,ArcherRRS2019}
\begin{equation}
\varrho(\mathbf{r}+\mathbf{r}',\mathbf{\hat{u}}')= \sum_{l=0}^{\infty}\frac{r^l}{l!}(u_j(\phi_{\mathrm{R}})\partial_j)^l\varrho(\mathbf{r},\mathbf{\hat{u}}').\label{gradientexpansion}
\end{equation}

Substituting \cref{fourier1,fourier2,gradientexpansion} into \cref{eq:DDFT_raw} leads to
\begin{widetext}
\begin{align}
    &\frac{\partial}{\partial t} \varrho(\mathbf{r},\mathbf{\hat{u}},t)
    = D_\mathrm{T} \partial_j \varrho(\mathbf{r},\mathbf{\hat{u}},t)
        + D_\mathrm{R} \partial_\phi^2\varrho(\mathbf{r},\mathbf{\hat{u}},t)\notag
    \\&\qquad
        +   \beta D_\mathrm{T} \partial_j
            \bigg\{
                \varrho(\mathbf{r},\mathbf{\hat{u}},t) \INT{0}{\infty}{r}\INT{0}{2\pi}{\phi_\mathrm{R}} \INT{0}{2\pi}{\phi'}
                \sum_{l=0}^{\infty}\sum_{|n_1|+|n_2| \leq 2}^{} \frac{1}{l!}r^{l+1}\Big[
                    u_j(\phi_{\mathrm{R}})f_{n_1 n_2}(r)e^{\ii(n_1\varphi_1+n_2\varphi_2)}(u_k(\phi_{\mathrm{R}})\partial_k)^l\varrho(\mathbf{r},\mathbf{\hat{u}}')
                \Big]
            \bigg\}\notag
        \\&\qquad
        -\beta D_\mathrm{R} \partial_\phi
            \bigg\{
                \varrho(\mathbf{r},\mathbf{\hat{u}},t) \INT{0}{\infty}{r}\INT{0}{2\pi}{\phi_\mathrm{R}} \INT{0}{2\pi}{\phi'}
                 \sum_{l=0}^{\infty}\sum_{|n_1|,|n_2| \leq 1}^{} \frac{1}{l!}r^{l+1}\Big[
                     M_{n_1 n_2}(r)e^{\ii(n_1\varphi_1+n_2\varphi_2)}(u_k(\phi_{\mathrm{R}})\partial_k)^l\varrho(\mathbf{r},\mathbf{\hat{u}}')
                \Big]
            \bigg\}. \label{eq:DDFT_raw2}
\end{align}
\end{widetext}
For the one-particle density $\varrho$, we make the Cartesian orientational expansion \cite{teVrugtW2020b}
\begin{equation}
    \varrho(\mathbf{r},\mathbf{\hat{u}},t)=\rho(\mathbf{r},t)+P_i(\mathbf{r},t)u_i + Q_{ij}(\mathbf{r},t)u_iu_j,
\label{eq:orientation_expansion}
\end{equation}
with the mean particle density 
\begin{equation}
    \rho(\mathbf{r},t) = \frac{1}{2\pi} \INT{0}{2\pi }{\phi} \varrho(\mathbf{r},\mathbf{\hat{u}},t),
\end{equation}
the local polarization
\begin{equation}
    P_i(\mathbf{r},t) = \frac{1}{\pi} \INT{0}{2\pi}{\phi} \varrho(\mathbf{r},\phi ,t)u_i,
\end{equation}
and the nematic order parameter tensor
\begin{equation}
    Q_{ij}(\mathbf{r},t) = \frac{2}{\pi} \INT{0}{2\pi}{\phi} \varrho(\mathbf{r},\phi ,t)\bigg(u_iu_j-\frac{1}{2}\delta_{ij}\bigg).
\end{equation}
To simplify the notation, for the remainder of this section we do not write the dependence of variables on space and time and express the time derivative of fields with an overdot.

We insert \cref{eq:orientation_expansion} into \cref{eq:DDFT_raw2}, truncate the gradient expansion at $l=3$, evaluate the integrals over $\phi'$ and $\phi_{\mathrm{R}}$, and integrate also over $\phi$ to find
\begin{align}
\dot{\rho}&=D_\mathrm{T}\partial_i^2 \rho + D_\mathrm{T}\partial_i\Big\{ \left [(A_{1}+A_{2})\delta_{ij}+(A_{3}+A_{4})\epsilon_{ij} \right ]P_j\rho \notag\\
&+A_{5} Q_{ij}P_j + A_{6} Q_{ij}\epsilon_{jk}P_k +A_{7} \rho \partial_i \rho + A_{8} P_j \partial_i P_j \notag\\
& + A_{9} P_j \partial_i \epsilon_{jk}P_k  +2A_{10}\rho \partial_i \partial_j P_j + A_{10}\rho \partial_j^2 P_i \notag\\
&+2A_{11}\rho \partial_i \partial_j \epsilon_{jk}P_k + A_{11}\rho \partial_j^2 \epsilon_{ik}P_k
+2A_{12} P_j  \partial_i \partial_j\rho \notag\\
& + A_{12} P_i\partial_j^2 \rho +2A_{13} \epsilon_{jk}P_k\partial_i \partial_j  \rho+ A_{13} \epsilon_{ik}P_k \partial_j^2\rho \notag\\
& + A_{14} \rho\partial_i \partial_j^2\rho+A_{15}Q_{ij}\partial_j \rho+A_{16}Q_{ij}\epsilon_{jk}\partial_k \rho \notag\\
& +A_{17}Q_{lm}\partial_iQ_{lm} +A_{18}Q_{lm}\epsilon_{mk}\partial_i Q_{lk}\Big\},
\label{dotrho}
\end{align}
with the 2D Levi-Civita symbol
\begin{equation}
\mathbf{\epsilon}=
\begin{pmatrix}
 0 &  1\\
 -1 & 0
\end{pmatrix}.
\end{equation}
For simplicity, we have dropped all terms involving $Q_{ij}$ that are of higher than first order in gradients, all terms involving $P_i$ that are of higher than third order in gradients, and all terms of higher than second order in gradients that involve Fourier modes exceeding first order.
The coefficients are given by
\begin{align}
A_{1}&= 2\pi^2 \beta\INT{0}{\infty}{r}r\Re(f_{10}(r)),\label{eq:b1}\\
A_{2}&= 2\pi^2 \beta\INT{0}{\infty}{r}r\Re(f_{-11}(r)),\label{eq:b2}\\
A_{3}&= 2\pi^2 \beta\INT{0}{\infty}{r}r\Im(f_{10}(r)),\label{eq:b3}\\
A_{4}&= -2\pi^2 \beta\INT{0}{\infty}{r}r\Im(f_{-11}(r)),\label{eq:b4}\\
A_{5}&= \pi^2 \beta\INT{0}{\infty}{r}r\Re(f_{11}(r)),\label{eq:b5}\\ 
A_{6}&= -\pi^2 \beta\INT{0}{\infty}{r}r\Im(f_{11}(r)),\label{eq:b6}\\ 
A_{7}&= 2\pi^2 \beta\INT{0}{\infty}{r}r^2 f_{00}(r),\label{eq:b7}\\
A_{8}&= \pi^2 \beta\INT{0}{\infty}{r}r^2\Re(f_{01}(r)),\label{eq:b8}\\ 
A_{9}&= -\pi^2 \beta\INT{0}{\infty}{r}r^2\Im(f_{01}(r)),\label{eq:b9}\\
A_{10}&= \frac{\pi^2}{4} \beta\INT{0}{\infty}{r}r^3\Re(f_{-11}(r)),\label{eq:b10}\\
A_{11}&= -\frac{\pi^2}{4} \beta\INT{0}{\infty}{r}r^3\Im(f_{-11}(r)),\label{eq:b11}\\
A_{12}&= \frac{\pi^2}{4} \beta\INT{0}{\infty}{r}r^3\Re(f_{10}(r)),\label{eq:b12}\\
A_{13}&= \frac{\pi^2}{4} \beta\INT{0}{\infty}{r}r^3\Im(f_{10}(r)),\label{eq:b13}\\
A_{14}&= \frac{\pi^2}{4} \beta\INT{0}{\infty}{r}r^4 f_{00}(r),\label{eq:b14}\\
A_{15}&= \pi^2 \beta\INT{0}{\infty}{r}r^2\Re(f_{20}(r)),\label{eq:b15}\\
A_{16}&= -\pi^2 \beta\INT{0}{\infty}{r}r^2\Im(f_{20}(r)),\label{eq:b16}\\
A_{17}&= \frac{\pi^2}{2} \beta\INT{0}{\infty}{r}r^2\Re(f_{01}(r)),\label{eq:b17}\\
A_{18}&= -\frac{\pi^2}{2} \beta\INT{0}{\infty}{r}r^2\Im(f_{01}(r)).\label{eq:b18}
\end{align}
Similarly, multiplying \cref{eq:DDFT_raw2} by $u_i(\phi)$, inserting \cref{eq:orientation_expansion}, truncating at $l=0$, and dropping Fourier modes of higher than first order, after integrating we get
\begin{align}
\dot{P}_i=&~D_\mathrm{T}\partial_j^2 P_i - D_\mathrm{R}P_i \notag\\
&~ + D_\mathrm{T}\partial_j \Big[2(A_{1}\delta_{ij} +A_{3}\epsilon_{ij})\rho^2 + (A_{5}\delta_{ij} + A_{6}\epsilon_{ij}) P_k^2 \notag\\
&~ + (A_{2}\delta_{jk}+ A_{4}\epsilon_{jk}) P_i P_k + (A_{1}\delta_{jk} + A_{3}\epsilon_{jk})Q_{ik}\rho\Big] \notag\\
&~ -D_\mathrm{R}\Big[A_{19}\epsilon_{ij}P_j\rho  +2(A_{20}\delta_{ij}+A_{21}\epsilon_{ij})P_j\rho \notag\\
&~ -(A_{20}\delta_{jk}+A_{21}\epsilon_{jk})Q_{ij}P_k\Big],
\label{dotp}
\end{align}
with the coefficients
\begin{align}
A_{19}&= 4\pi^2 \beta\INT{0}{\infty}{r}rM_{00}(r),\label{eq:b19}\\
A_{20}&= 2\pi^2 \beta\INT{0}{\infty}{r}r\Im(M_{01}(r)),\label{eq:b20}\\
A_{21}&= 2\pi^2 \beta\INT{0}{\infty}{r}r\Re(M_{01}(r)).\label{eq:b21}
\end{align}
Finally, multiplying \cref{eq:DDFT_raw2} by $u_i(\phi)u_j(\phi)-\delta_{ij}/2$, inserting \cref{eq:orientation_expansion}, truncating at $l=1$, dropping Fourier modes of higher than first order, and integrating give
\begin{align}
\dot{Q}_{ij}=&~D_\mathrm{T}\partial_k^2 Q_{ij} - 4D_{\mathrm{R}}Q_{ij} \nonumber\\
&+D_\mathrm{T} \partial_k\Big\{ \left [ (2A_{5} +A_{1})\delta_{ml}+(2A_{6}-A_{3})\epsilon_{ml} \right ] \nonumber\\
& \qquad\qquad \times (\delta_{ik}\delta_{jm}+\delta_{im}\delta_{jk}-\delta_{ij}\delta_{km})P_l\rho \nonumber\\
& \qquad\qquad + (A_{2}\delta_{kl}+ A_{4}\epsilon_{kl})Q_{ij}P_l\Big\} \nonumber\\
&-D_{\mathrm{R}} \Big[ A_{19}(\epsilon_{il}\delta_{jm}+\epsilon_{im}\delta_{jl})Q_{lm}\rho \nonumber\\
& ~\qquad + (2A_{20}\delta_{ml}+2A_{21}\epsilon_{ml}) \nonumber\\
& ~\qquad \times (\delta_{ik}\delta_{jm}+\delta_{im}\delta_{jk}-\delta_{ij}\delta_{km})P_lP_k \Big].
\label{dotq}
\end{align}
Equations \eqref{dotrho}, \eqref{dotp}, and \eqref{dotq} together constitute a general continuum field theory based on microscopic dynamics for particles with arbitrary two-body interactions. What is notable here in particular are the terms involving $\epsilon_{ij}$, which represent chiral contributions. These are not present in active model N as the corresponding coefficients vanish for the nonreciprocal interaction force and torque considered in this work. However, they can in general be present for other types of interactions. Given the recent interest in and the rich phenomenology of chiral active matter \cite{DunajovaEtAl2023,KoleARM2021,LiebchenL2022b}, a detailed investigation of the more general model presented here would be an interesting perspective for future work.

\subsection{Simplified models}
\label{sec:simplified}

The next step is to simplify the above general model by reducing the number of dynamic order parameter fields. To eliminate the nematic tensor field $Q_{ij}$, we make the quasi-stationary approximation \cite{teVrugtBW2022}
\begin{equation}
\dot{Q}_{ij}=0.
\end{equation}
From \cref{dotq}, we thereby find
\begin{align}
Q_{ij}=&~\frac{D_\mathrm{T}}{4D_\mathrm{R}} \Big \{ \partial_k^2 Q_{ij} +\partial_k\big\{(A_{2}\delta_{kl}+ A_{4}\epsilon_{kl})Q_{ij}P_l \nonumber\\
&\qquad\qquad+[(2A_{5} +A_{1})\delta_{ml}+(2A_{6}-A_{3})\epsilon_{ml}] \nonumber\\
&\qquad\qquad\times (\delta_{ik}\delta_{jm}+\delta_{im}\delta_{jk}-\delta_{ij}\delta_{km})P_l\rho \big\} \Big\} \nonumber\\
&-\frac{1}{4}\big[A_{19}(\epsilon_{il}\delta_{jm}+\epsilon_{im}\delta_{jl})Q_{lm}\rho \nonumber\\
&\qquad + (2A_{20}\delta_{ml}+2A_{21}\epsilon_{ml}) \nonumber\\
&\qquad \times (\delta_{ik}\delta_{jm}+\delta_{im}\delta_{jk}-\delta_{ij}\delta_{km})P_lP_k\big].
\label{dotqgleichnull}
\end{align}
A closed expression for $Q_{ij}$ can be obtained from a recursive procedure, by repeatedly substituting its expression back into  \cref{dotqgleichnull} and truncating the result at some order in fields and gradients. For simplicity, here we truncate at first order in gradients and second order in fields, which gives
\begin{align}
Q_{ij}=&~-\frac{1}{2}\big[(A_{20}\delta_{ml}+A_{21}\epsilon_{ml}) \nonumber\\
&\qquad \times (\delta_{ik}\delta_{jm}+\delta_{im}\delta_{jk}-\delta_{ij}\delta_{km})P_lP_k\big] \nonumber\\
&~+\frac{D_\mathrm{T}}{4D_\mathrm{R}} \partial_k\big\{[(2A_{5} +A_{1})\delta_{ml}+(2A_{6}-A_{3})\epsilon_{ml}] \nonumber\\
&\qquad \times (\delta_{ik}\delta_{jm}+\delta_{im}\delta_{jk}-\delta_{ij}\delta_{km})P_l\rho\big\}.
\label{qij2}
\end{align}
Equations \eqref{dotrho}, \eqref{dotp}, and \eqref{qij2} together constitute a closed and general field theory for the order parameter fields $\rho$ and $\mathbf{P}$, since \cref{qij2} allows to express $Q_{ij}$ via $\rho$ and $P_i$. To see this explicitly, we insert \cref{qij2} into \cref{dotp} to find
\begin{widetext}
\begin{align}
\dot{P}_i=&~D_\mathrm{T}\partial_j^2 P_i - D_\mathrm{R}P_i \nonumber\\
&+ D_\mathrm{T}\partial_j \bigg(2(A_{1}\delta_{ij} +A_{3}\epsilon_{ij})\rho^2 + (A_{5}\delta_{ij} + A_{6}\epsilon_{ij}) P_k^2 + (A_{2}\delta_{jk}+ A_{4}\epsilon_{jk}) P_i P_k \nonumber\\
&- \frac{1}{2}(A_{1}\delta_{jk} + A_{3}\epsilon_{jk})((A_{20}\delta_{ml}+A_{21}\epsilon_{ml})(\delta_{in}\delta_{km}+\delta_{im}\delta_{kn}-\delta_{ik}\delta_{nm})P_lP_n)\rho
\nonumber\\
&+\frac{D_\mathrm{T}}{4D_\mathrm{R}}(A_{1}\delta_{jk} + A_{3}\epsilon_{jk})\rho\partial_n(((2A_{5} +A_{1})\delta_{ml}+(2A_{6}-A_{3})\epsilon_{ml})(\delta_{in}\delta_{km}+\delta_{im}\delta_{kn}-\delta_{ik}\delta_{nm})P_l\rho)\bigg) \nonumber\\
&-D_\mathrm{R}\bigg(A_{19}\epsilon_{ij}P_j\rho  +2(A_{20}\delta_{ij}+A_{21}\epsilon_{ij})P_j\rho \nonumber\\
&+\frac{1}{2}(A_{20}\delta_{jk}+A_{21}\epsilon_{jk})((A_{20}\delta_{ml}+A_{21}\epsilon_{ml})(\delta_{in}\delta_{jm}+\delta_{im}\delta_{jn}-\delta_{ij}\delta_{nm})P_lP_n)P_k \nonumber\\
&-\frac{D_\mathrm{T}}{4D_\mathrm{R}}(A_{20}\delta_{jk}+A_{21}\epsilon_{jk})P_k\partial_n(((2A_{5} +A_{1})\delta_{ml}+(2A_{6}-A_{3})\epsilon_{ml})(\delta_{in}\delta_{jm}+\delta_{im}\delta_{jn}-\delta_{ij}\delta_{nm})P_l\rho)\bigg).
\label{dotpreduced}
\end{align} 
Substituting \cref{qij2} into \cref{dotrho} gives
\begin{align}
\dot{\rho}=&~D_\mathrm{T}\partial_i^2 \rho + D_\mathrm{T}\partial_i\bigg(((A_{1}+A_{2})\delta_{ij}+(A_{3}+A_{4})\epsilon_{ij})P_j\rho-\frac{1}{2}A_{5} ((A_{20}\delta_{ml}+A_{21}\epsilon_{ml}) (\delta_{ik}\delta_{jm}+\delta_{im}\delta_{jk}-\delta_{ij}\delta_{km})P_lP_k)P_j
\nonumber\\
&+\frac{D_\mathrm{T}}{4D_\mathrm{R}} A_5 P_j \partial_k(((2A_{5} +A_{1})\delta_{ml}+(2A_{6}-A_{3})\epsilon_{ml})(\delta_{ik}\delta_{jm}+\delta_{im}\delta_{jk}-\delta_{ij}\delta_{km})P_l\rho)
\nonumber\\
&+\frac{D_\mathrm{T}}{4D_\mathrm{R}} A_6 \epsilon_{jn} P_n \partial_k(((2A_{5} +A_{1})\delta_{ml}+(2A_{6}-A_{3})\epsilon_{ml})(\delta_{ik}\delta_{jm}+\delta_{im}\delta_{jk}-\delta_{ij}\delta_{km})P_l\rho)
\nonumber\\
&- \frac{1}{2}A_{6} ((A_{20}\delta_{ml}+A_{21}\epsilon_{ml})(\delta_{ik}\delta_{jm}+\delta_{im}\delta_{jk}-\delta_{ij}\delta_{km})P_lP_k)\epsilon_{jn}P_n +A_{7} \rho \partial_i \rho + A_{8} P_j \partial_i P_j + A_{9} P_j \partial_i \epsilon_{jk}P_k \nonumber\\
& +2A_{10}\rho \partial_i \partial_j P_j + A_{10}\rho \partial_j^2 P_i +2A_{11}\rho \partial_i \partial_j \epsilon_{jk}P_k + A_{11}\rho \partial_j^2 \epsilon_{ik}P_k +2A_{12} P_j  \partial_i \partial_j\rho + A_{12} P_i\partial_j^2  \rho \nonumber\\
&+2A_{13} \epsilon_{jk}P_k\partial_i \partial_j  \rho+ A_{13} \epsilon_{ik}P_k \partial_j^2\rho+ A_{14} \rho\partial_i \partial_j^2\rho-\frac{A_{15}}{2}((A_{20}\delta_{ml}+A_{21}\epsilon_{ml}) (\delta_{ik}\delta_{jm}+\delta_{im}\delta_{jk}-\delta_{ij}\delta_{km})P_lP_k)\partial_j\rho \nonumber\\
&+\frac{D_\mathrm{T}}{4D_\mathrm{R}} A_{15} (\partial_j\rho) \partial_k(((2A_{5} +A_{1})\delta_{ml}+(2A_{6}-A_{3})\epsilon_{ml})(\delta_{ik}\delta_{jm}+\delta_{im}\delta_{jk}-\delta_{ij}\delta_{km})P_l\rho)
\nonumber\\
&-\frac{A_{16}}{2}((A_{20}\delta_{ml}+A_{21}\epsilon_{ml}) (\delta_{ik}\delta_{jm}+\delta_{im}\delta_{jk}-\delta_{ij}\delta_{km})P_lP_k)\partial_j\rho \nonumber\\
&+\frac{D_\mathrm{T}}{4D_\mathrm{R}} A_{16} (\partial_j\rho) \partial_k(((2A_{5} +A_{1})\delta_{ml}+(2A_{6}-A_{3})\epsilon_{ml}) (\delta_{ik}\delta_{jm}+\delta_{im}\delta_{jk}-\delta_{ij}\delta_{km})P_l\rho)\epsilon_{jn}\partial_n \rho \nonumber\\
&+A_{17}\bigg(-\frac{1}{2}((A_{20}\delta_{no}+A_{21}\epsilon_{no}) (\delta_{lk}\delta_{mn}+\delta_{ln}\delta_{mk}-\delta_{lm}\delta_{kn})P_oP_k) \nonumber\\
&\quad+\frac{D_\mathrm{T}}{4D_\mathrm{R}} \partial_k(((2A_{5} +A_{1})\delta_{no}+(2A_{6}-A_{3})\epsilon_{no}) (\delta_{lk}\delta_{mn}+\delta_{ln}\delta_{mk}-\delta_{lm}\delta_{kn})P_o\rho)\bigg) \nonumber\\
&\partial_i\bigg(-\frac{1}{2}((A_{20}\delta_{no}+A_{21}\epsilon_{no})(\delta_{lk}\delta_{mn}+\delta_{ln}\delta_{mk}-\delta_{lm}\delta_{kn})P_oP_k) \nonumber\\
&\quad +\frac{D_\mathrm{T}}{4D_\mathrm{R}} \partial_k(((2A_{5} +A_{1})\delta_{no}+(2A_{6}-A_{3})\epsilon_{no}) (\delta_{lk}\delta_{mn}+\delta_{ln}\delta_{mk}-\delta_{lm}\delta_{kn})P_o\rho)\bigg) \nonumber\\
&+A_{18}\bigg(-\frac{1}{2}((A_{20}\delta_{no}+A_{21}\epsilon_{no}) (\delta_{lp}\delta_{mn}+\delta_{ln}\delta_{mp}-\delta_{lm}\delta_{pn})P_oP_p) \nonumber\\
&\quad +\frac{D_\mathrm{T}}{4D_\mathrm{R}} \partial_p(((2A_{5} +A_{1})\delta_{no}+(2A_{6}-A_{3})\epsilon_{no})(\delta_{lp}\delta_{mn}+\delta_{ln}\delta_{mp}-\delta_{lm}\delta_{pn})P_o\rho)\bigg)\epsilon_{mk} \nonumber\\
&\partial_i \bigg(-\frac{1}{2}((A_{20}\delta_{no}+A_{21}\epsilon_{no}) (\delta_{lp}\delta_{kn}+\delta_{ln}\delta_{kp}-\delta_{lk}\delta_{pn})P_oP_p) \nonumber\\
&\quad +\frac{D_\mathrm{T}}{4D_\mathrm{R}} \partial_p(((2A_{5} +A_{1})\delta_{no}+(2A_{6}-A_{3})\epsilon_{no})(\delta_{lp}\delta_{kn}+\delta_{ln}\delta_{kp}-\delta_{lk}\delta_{pn})P_o\rho)\bigg)\bigg).
\label{dotrhoreduced}
\end{align}
\end{widetext}

This is all that is required for deriving Active Model N. In certain contexts, however, it would be useful to have a model involving only particle density $\rho$. For this purpose, we make the further quasi-stationary approximation
\begin{equation}
\dot{P}_i = 0,
\end{equation}
and apply a recursive procedure to \cref{dotpreduced} to obtain a closed expression of $P_i$ by dropping terms of higher than first order in gradients $\partial_i$ and of higher than second order in fields. This leads to
\begin{equation}
\begin{split}
P_i&= \frac{2D_{\mathrm{T}}}{D_{\mathrm{R}}}\partial_j(A_{1}\delta_{ij} +A_{3}\epsilon_{ij})\rho^2.
\end{split}
\label{pi2}
\end{equation}
Finally, we substitute \cref{pi2} into \cref{dotrhoreduced} and drop terms of higher than third order in $\rho$ and terms resulting from second-order Fourier modes, giving
\begin{widetext}
\begin{align}
\dot{\rho}=&~D_\mathrm{T}\partial_i^2 \rho + D_\mathrm{T}\partial_i\bigg(\frac{2D_{\mathrm{T}}}{D_{\mathrm{R}}}((A_{1}+A_{2})\delta_{ij}+(A_{3}+A_{4})\epsilon_{ij})\rho\partial_k(A_{1}\delta_{jk} +A_{3}\epsilon_{jk})\rho^2 \nonumber\\
& +A_{7} \rho \partial_i \rho  +\frac{2D_{\mathrm{T}}}{D_{\mathrm{R}}}(2A_{10}\rho \partial_i \partial_j \partial_k(A_{1}\delta_{jk} +A_{3}\epsilon_{jk})\rho^2+ A_{10}\rho \partial_j^2 \partial_k(A_{1}\delta_{ik} +A_{3}\epsilon_{ik})\rho^2 \nonumber\\
&+2A_{11}\rho \partial_i \partial_j \epsilon_{jk}(\partial_n(A_{1}\delta_{kn} +A_{3}\epsilon_{kn})\rho^2)+ A_{11}\rho \partial_j^2 \epsilon_{ik}(\partial_n(A_{1}\delta_{kn} +A_{3}\epsilon_{kn})\rho^2) \nonumber\\
& +2A_{12} (\partial_k(A_{1}\delta_{jk} +A_{3}\epsilon_{jk})\rho^2)  \partial_i \partial_j\rho + A_{12} (\partial_k(A_{1}\delta_{ik} +A_{3}\epsilon_{ik})\rho^2)\partial_j^2 \rho \nonumber\\
&+2A_{13} \epsilon_{jk} (\partial_n(A_{1}\delta_{kn} +A_{3}\epsilon_{kn})\rho^2)\partial_i \partial_j \rho+ A_{13} \epsilon_{ik}(\partial_n(A_{1}\delta_{kn} +A_{3}\epsilon_{kn})\rho^2) \partial_j^2\rho) + A_{14} \rho\partial_i \partial_j^2\rho\bigg),
\end{align}
which can be simplified as
\begin{align}
\dot{\rho}=&~ \frac{2D_{\mathrm{T}}^2}{D_{\mathrm{R}}}\partial_i (\rho(\tilde{T} \partial_i \ln(\rho)+ (A_{1}(A_{1}+A_{2})- A_{3}(A_{3}+A_{4}))\partial_i\rho^2 + (A_{1}(A_{3}+A_{4})+ A_{3}(A_{1}+A_{2}))\epsilon_{ij}\partial_j\rho^2 \nonumber\\
& +\tilde{B}_7 \partial_i \rho  +(3A_{10}A_{1} -3A_{11}A_{3})\partial_i \partial_j^2 \rho^2 + (A_{10}A_{3}+A_{11}A_{1}) \epsilon_{ik}\partial_k \partial_j^2\rho^2 \nonumber\\
& +(4A_{12} A_{1}-4A_{13}A_{3})(\partial_j \rho) \partial_i \partial_j\rho+(4A_{12} A_{3}+4A_{13}A_{1})\epsilon_{jk}(\partial_k \rho) \partial_i \partial_j\rho \nonumber\\
&+( 2A_{12} A_{1}- 2A_{13}A_{3}  ) (\partial_i\rho)\partial_j^2  \rho+ (2A_{12} A_{3}+ 2A_{13} A_{1})\epsilon_{ik}(\partial_k\rho)\partial_j^2  \rho+ \tilde{B}_{14} \partial_i \partial_j^2\rho)),
\label{dotrhoclosed}
\end{align}
where $\tilde{T}=D_\mathrm{R}/(2D_\mathrm{T})$, $\tilde{B}_7 = D_\mathrm{R}A_{7}/(2D_\mathrm{T})$, $\tilde{B}_{14} = D_\mathrm{R}A_{14}/(2D_\mathrm{T})$, and we have exploited the relation $\epsilon_{ij}\epsilon_{jk} = - \delta_{ik}$.

\end{widetext}

\subsection{Derivation of Active Model N}

The discussion so far has been concerned with general interactions. To derive Active Model N, we now assume the interaction force and torque to be specified by the nonreciprocal forms of Eqs.~\eqref{eq:F21}, \eqref{eq:f}, \eqref{eq:M}, and \eqref{eq:g_tau}. Combining these equations with \cref{eq:b1,eq:b2,eq:b3,eq:b4,eq:b5,eq:b6,eq:b7,eq:b8,eq:b9,eq:b10,eq:b11,eq:b12,eq:b13,eq:b14,eq:b15,eq:b16,eq:b17,eq:b18,eq:b19,eq:b20,eq:b21} allows us to compute the expansion coefficients, which require the integrals
\begin{align}
\INT{0}{\infty}{r}h(r)r &= -\frac{a^2}{2}(F_0 - 2\alpha^2 F_1),\\  \INT{0}{\infty}{r}h(r)r^2 &= -\frac{a^3}{4}(\sqrt{\pi}F_0-8\alpha^3F_1),\\
\INT{0}{\infty}{r}h(r)r^3 &= -\frac{a^4}{2}(F_0 - 12 \alpha^4 F_1),\\
\INT{0}{\infty}{r}h(r)r^4 &= -\frac{3a^5}{8}(\sqrt{\pi}F_0 - 64\alpha^5 F_1),\\  
\INT{0}{\infty}{r}\tau(r)r &= bc^2,
\end{align}
and the non-vanishing Fourier modes of $f$ and $M$ that are given by (noting $\vartheta\in [0,\pi]$)
\begin{align}
f_{00}&=-\frac{h(r)}{\pi}\vartheta,\\ 
\Re(f_{10})&=-\frac{h(r)}{\pi}\sin\vartheta,\\ 
\Im(M_{01})&=-\frac{b}{\pi}e^{-\frac{r}{c}}\vartheta,\\
\Re(f_{20})&=-\frac{h(r)}{2\pi}\sin 2\vartheta.
\end{align}
All other Fourier modes are zero, i.e., $\Im(f_{10})=\Re(f_{01})=\Im(f_{01})=\Re(f_{11})=\Im(f_{11})=\Re(f_{-11})=\Im(f_{-11})=\Im(f_{20})=\Re(f_{02})=\Im(f_{02})=M_{00}=\Re(M_{01})=0$. We can then find the coefficients
\begin{align}
&A_{1}=a^2\pi\beta(F_0-2\alpha^2 F_1)\sin\vartheta,\label{b1}\\
&A_{2}=A_{3}=A_{4}=A_{5}=A_{6}= 0,\label{b2-6}\\
&A_{7}= \frac{a^3}{2}\pi\beta(\sqrt{\pi}F_0-8\alpha^3F_1)\vartheta,\label{b7}\\
&A_{8}=A_{9}=A_{10}=A_{11}= 0,\label{b8-11}\\
&A_{12}= \frac{a^4}{8}\pi\beta(F_0-12 \alpha^4 F_1)\sin\vartheta,\label{b12}\\
&A_{13}=0,\label{b13}\\
&A_{14}=\frac{3a^5}{32}\pi\beta(\sqrt{\pi}F_0-64\alpha^5 F_1)\vartheta,\label{b14}\\
&A_{15}= \frac{a^3}{8}\pi\beta(\sqrt{\pi}F_0-8\alpha^3F_1)\sin 2\vartheta,\label{b15}\\
&A_{16}=A_{17}=A_{18}=A_{19}=0,\label{b16-19}\\
&A_{20}=-2\pi bc^2\beta\vartheta,\label{b20}\\
&A_{21}=0.\label{b21}
\end{align}
Substituting Eqs.~\eqref{b1}--\eqref{b21} into \cref{dotrho,dotp} gives
\begin{align}
\dot{\rho}=&~D_\mathrm{T}\partial_i^2 \rho + D_\mathrm{T}\partial_i\bigg(A_{1}P_i\rho +A_{7} \rho \partial_i \rho + 2A_{12} P_j  \partial_i \partial_j\rho \nonumber\\
& + A_{12} P_i\partial_j^2  \rho+ A_{14} \rho\partial_i \partial_j^2\rho +A_{15}Q_{ij}\partial_j\rho\bigg),
 \label{modelwithp1}\\
\dot{P}_i=&~D_\mathrm{T}\partial_j^2 P_i - D_\mathrm{R}P_i + 2D_\mathrm{T}A_{1}\partial_i\rho^2  \nonumber\\
& + D_\mathrm{T}A_{1}\partial_j Q_{ij} \rho-D_\mathrm{R}A_{20}P_j(2\delta_{ij} \rho-Q_{ij}).
\label{modelwithp2}  
\end{align}
Here we have assumed $Q_{ij}$ to relax at a much faster time scale and thus be determined by the following function of $\rho$ and $P_i$ (as obtained from substituting Eqs.~\eqref{b1}--\eqref{b21} into \cref{qij2})
\begin{align}
Q_{ij}=&-\frac{1}{2}A_{20}(2P_i P_j - \delta_{ij}P_k^2) \nonumber\\
& + \frac{D_\mathrm{T}A_1}{4D_\mathrm{R}}\left [\partial_i(P_j\rho) +\partial_j (P_i\rho) - \delta_{ij}\partial_k (P_k\rho) \right ].
\label{qij2here}
\end{align}
Thereby, we have a closed field theory for $\rho$ and $P_i$ for the nonreciprocal interactions considered in this work. If further setting $A_{15}=0$ and thereby considering only up to first Fourier mode, we obtain Active Model N (Eqs.~\eqref{eq:rho}--\eqref{eq:Qij}). This approximation is motivated by the interest in identifying a minimal model. We have also performed both bifurcation analysis and numerical simulations based on the above extended model including the nonzero $A_{15}$ terms, and found qualitatively similar results. This indicates that the approximation $A_{15}=0$ and the corresponding model simplification do not affect the essential physics of the system.

It is also instructive to consider the particle-density-only model \cref{dotrhoclosed}. Inserting Eqs.~\eqref{b1}--\eqref{b21} into \cref{dotrhoclosed}, which does not depend on the choice of torque $M$ as this reduced model does not incorporate the effect of alignment interactions, we have
\begin{equation}
\begin{split}
\dot{\rho}&= \frac{2D_{\mathrm{T}}^2}{D_{\mathrm{R}}}\partial_i \left \{\rho \left [\tilde{T} \partial_i \ln\rho+ A_{1}^2\partial_i\rho^2 +\tilde{B}_7 \partial_i \rho \right.\right.\\
& \left.\left. +4A_{12} A_{1}(\partial_j \rho) \partial_i \partial_j\rho+2A_{12} A_{1}(\partial_i\rho)\partial_j^2 \rho+ \tilde{B}_{14} \partial_i \partial_j^2\rho \right ] \right \}.
\end{split}  
\label{dotrhoclosedsimple}
\end{equation}
As to be discussed in Appendix \ref{relationamb}, \cref{dotrhoclosedsimple} is a special case of active model B+ \cite{TjhungNC2018} with non-constant mobility.

To ensure that the coefficients for Active Model N are numbered sequentially, we redefine the notations $B_1 \equiv A_1$, $B_2 \equiv A_7$, $B_3 \equiv A_{12}$, $B_4 \equiv A_{14}$, and $B_5 \equiv A_{20}$, i.e.,
\begin{align}
B_{1}&=a^2\pi\beta \left (F_0-2\alpha^2 F_1 \right )\sin\vartheta,\label{eq:b1m}\\
B_2&= \frac{a^3}{2}\pi\beta \left (\sqrt{\pi}F_0-8\alpha^3F_1 \right )\vartheta,\label{eq:b2m}\\
B_3&= \frac{a^4}{8}\pi\beta \left (F_0-12 \alpha^4 F_1 \right )\sin\vartheta,\label{eq:b3m}\\
B_4&=\frac{3a^5}{32}\pi\beta \left (\sqrt{\pi}F_0-64\alpha^5 F_1 \right )\vartheta,\label{eq:b4m}\\
B_5&=-2\pi \beta bc^2\vartheta, \label{eq:b5m}
\end{align}
as used in all the calculations of this work and in Appendix \ref{sec:bifurcation} of bifurcation analysis.

\section{Relation to Active Model B+}
\label{relationamb}

If the polarization field evolves significantly slower than the particle density field and if the system does not exhibit global polarization, we can reduce Active Model N using a quasi-stationary approximation for $\mathbf{P}$ \cite{teVrugtBW2022} (see Appendix \ref{sec:simplified}). For this purpose, we set $\partial{\mathbf{P}}/\partial t=\mathbf{0}$ in Eq.~\eqref{eq:P}, solve the resulting equation for $\mathbf{P}$, and then drop terms of higher than first order in spatial gradients and higher than second order in $\rho$. This yields
\begin{equation}
\mathbf{P} = \frac{2B_1}{\tilde{D}_\mathrm{R}}\bm{\nabla}\rho^2.\label{eq:pstatic}
\end{equation}
Substituting \cref{eq:pstatic} into \cref{eq:rho} gives
\begin{equation}
\begin{split}
 \frac{\partial \rho}{\partial t} & = \bm{\nabla} \cdot \bigg[ \rho \bigg(\bm{\nabla}\ln\rho+ \frac{2B_1^2}{\tilde{D}_\mathrm{R}} \bm{\nabla}\rho^2+ B_2 \bm{\nabla} \rho \\
    &+\frac{4B_1B_3}{\tilde{D}_\mathrm{R}}(\bm{\nabla}\rho) {\nabla}^2 \rho  +  \frac{4B_1B_3}{\tilde{D}_\mathrm{R}} \bm{\nabla}(\bm{\nabla}\rho)^2
    + B_4 \bm{\nabla} \nabla^2 \rho \bigg)\bigg],\label{eq:rhoclosed}
\end{split}
\end{equation}
which is the rescaled form of \cref{dotrhoclosedsimple}. Finally, expanding the logarithm in \cref{eq:rhoclosed} around $\rho = \bar{\rho}$ and defining $a=3/\bar{\rho} + B_2$, $b= - 3/(2\bar{\rho}^2) +2B_1^2/\tilde{D}_\mathrm{R}$, $c= 1/(3\bar{\rho}^3)$, $\lambda=\xi = 4B_1 B_3/\tilde{D}_\mathrm{R}$, and $\kappa = - B_4$, we find
\begin{align}
 \frac{\partial \rho}{\partial t} =&~ \bm{\nabla} \cdot \left\{\rho \left (\bm{\nabla} \left [a\rho + b \rho^2 + c \rho^3 + \lambda(\bm{\nabla}\rho)^2 -\kappa {\nabla}^2\rho \right ] \right.\right. \nonumber\\
 &\left.\left. + \xi (\bm{\nabla}\rho) {\nabla}^2 \rho \right ) \right \},\label{eq:amb+}
\end{align}
which is a special case of Active Model B+ \cite{TjhungNC2018}. (In the general case one would have $\lambda \neq \xi$.) This is a remarkable observation since Active Model B+ is usually derived and interpreted as a theory for phase separation in systems of self-propelled particles \cite{TjhungNC2018,teVrugtBW2022} from several microscopic descriptions \cite{KalzSM2023,teVrugtBW2022}. The fact that Active Model B+ also arises as a limiting case of Active Model N (which does not incorporate any explicit self-propulsion source) shows that it applies more generally, and suggests that the phase separation dynamics captured by Active Model B+ -- usually interpreted as motility-induced phase separation \cite{CatesT2015} -- can be also observed in particles with nonreciprocal interactions that do not have any motility. In addition, \cref{eq:amb+} reduces to passive model B in the reciprocal case where $B_1 = B_3 = 0$, as expected. Note that, if one derives Active Model B+ for a system of self-propelled particles \cite{teVrugtBW2022}, a different mobility would be found since the quasi-stationary approximation then gives, at lowest order in gradients and fields, $\mathbf{P}\propto  \bm{\nabla} \rho$ rather than $\mathbf{P} \propto \bm{\nabla} \rho^2$ as in \cref{eq:pstatic}.

\section{Bifurcation analysis}
\label{sec:bifurcation}

From the rescaled dynamical equations \eqref{eq:rho} and \eqref{eq:P} governing a particle density field $\rho$ and a polarization density field $\mathbf{P} = (P_x, P_y)$ in Active Model N, we get
\begin{equation}
\begin{split}
  & \frac{\partial \rho}{\partial t} = 0, \\
  & \frac{\partial \mathbf{P}}{\partial t} = - \tilde{D}_R \left [ \mathbf{P} + 2B_5 \mathbf{P}
    \left ( \rho + \frac{1}{4} B_5 |\mathbf{P}|^2 \right ) \right ] = \mathbf{0}, 
\end{split}
\label{eq:steadystate}
\end{equation}
in the uniform steady state with $\rho = {\rm const.}$ and $\mathbf{P} = {\rm const.}$
The solutions of \cref{eq:steadystate} correspond to a homogeneous state with constant density $\rho = \bar{\rho}$ and constant polarization
\begin{equation}
  \mathbf{P} = \bar{\mathbf P} = \mathbf{0}, \qquad \bar{P}_x = \bar{P}_y = 0,
\end{equation}
leading to a fully disordered state of both density and polarization fields, or
\begin{equation}
\begin{split}
  & |\mathbf{P}|^2 = |\bar{\mathbf P}|^2 = \bar{P}_x^2 + \bar{P}_y^2
  = -\frac{4}{B_5} \left ( \bar{\rho} + \frac{1}{2B_5} \right ), \\
  & \qquad \textrm{if } \bar{\rho} > -\frac{1}{2B_5} = \frac{1}{4\pi\beta bc^2\vartheta}
  \textrm{ or } \vartheta > \frac{1}{4\pi\beta bc^2\bar{\rho}}, 
\end{split}
\label{eq:Pbar^2}
\end{equation}
for a disordered flocking state with nonzero polarization of fixed magnitude.

\subsection{Primary instability/bifurcation: Flocking transition}
\label{sec:primary_inst}

To conduct the primary instability analysis, we choose the base state as the fully disordered state of $\rho = \bar{\rho} > 0$ and $\bar{P}_x = \bar{P}_y = 0$, and expand
  $\rho = \bar{\rho} + \widehat{\rho}$, $P_x = \bar{P}_x + \widehat{P}_x = \widehat{P}_x$, and $P_y = \bar{P}_y + \widehat{P}_y = \widehat{P}_y$.
In Fourier space, the linearized model equations become
\begin{widetext}
\begin{align}
  & \frac{\partial}{\partial t} \left (
  \begin{array}{c}
    \widehat{\rho}_{\mathbf{q}} \\ \widehat{P}_{x\mathbf{q}} \\ \widehat{P}_{y\mathbf{q}}
  \end{array}
  \right )
  = \mathcal{L}_1 \left (
  \begin{array}{c}
    \widehat{\rho}_{\mathbf{q}} \\ \widehat{P}_{x\mathbf{q}} \\ \widehat{P}_{y\mathbf{q}}
  \end{array}
  \right ) \nonumber\\
  & = \left (
  \begin{array}{ccc}
    B_4 \bar{\rho} q^4 - (1+B_2\bar{\rho})q^2 & iB_1 \bar{\rho} q_x &  iB_1 \bar{\rho} q_y \\
    4iB_1 \bar{\rho} q_x &
    -\tilde{D}_R(1+2B_5\bar{\rho}) - \left ( 1+\frac{B_1^2}{4\tilde{D}_R}\bar{\rho}^2 \right ) q^2 & 0 \\
    4iB_1 \bar{\rho} q_y & 0 &
    -\tilde{D}_R(1+2B_5\bar{\rho}) - \left ( 1+\frac{B_1^2}{4\tilde{D}_R}\bar{\rho}^2 \right ) q^2
  \end{array}
  \right )
  \left (
  \begin{array}{c}
    \widehat{\rho}_{\mathbf{q}} \\ \widehat{P}_{x\mathbf{q}} \\ \widehat{P}_{y\mathbf{q}}
  \end{array}
  \right ). \label{eq:linearized}
\end{align}
\end{widetext}
Assuming $\widehat{\rho}_{\mathbf{q}} = \tilde{\rho}_{\mathbf{q}} e^{\sigma t}$ and $\widehat{\mathbf P}_\mathbf{q} = \tilde{\mathbf P}_\mathbf{q} e^{\sigma t}$, the perturbation growth rate $\sigma = \sigma(q_x,q_y)$ can be obtained from the eigenvalues of $\mathcal{L}_1$, i.e.,
\begin{equation}
  \left | \mathcal{L}_1 - \sigma \mathds{1} \right | = 0 \label{eq:L1_sigma}
\end{equation}
with the unit matrix $\mathds{1}$.

\subsubsection{Primary instability: The reciprocal limit \texorpdfstring{$\vartheta = \pi$}{}}

In the limit of reciprocal interactions, $\vartheta = \pi$ and hence $\sin\vartheta = 0$ and $B_1 = B_3 = 0$. From \cref{eq:linearized}, the first-order dynamics of $\widehat{\rho}_{\mathbf{q}}$, $\widehat{P}_{x\mathbf{q}}$, and $\widehat{P}_{y\mathbf{q}}$ are decoupled, with separate perturbation growth rates
\begin{equation}
\begin{split}
  & \sigma_\rho = B_4 \bar{\rho} q^4 - (1+B_2\bar{\rho})q^2, \\
  & \sigma_\mathbf{P} = \sigma_{P_x} = \sigma_{P_y} = -\tilde{D}_R(1+2B_5\bar{\rho})-q^2. 
\end{split}
\label{eq:sig1_reciprocal}
\end{equation}
From the condition of $\sigma(q \rightarrow \infty) \leq 0$ (to avoid small length scale instability), we need $B_4 < 0$. Equation (\ref{eq:sig1_reciprocal}) shows that in this reciprocal limit there are no oscillatory instabilities since all growth rates are real.

When $B_2\bar{\rho} > -1$ and $B_4 < 0$, we have the maximum perturbation growth rate $\sigma_\rho^{\rm max} = 0$ at $q_m = 0$, with $\sigma_\rho < 0$ for all the finite wave numbers, indicating that the particle density field remains homogeneous with $\rho = \bar{\rho} > 0$. On the other hand, when $B_2\bar{\rho} < -1$ and $B_4 < 0$ we obtain a nonzero value of the most
unstable wave number that is given by $q_m^2 = (1+B_2\bar{\rho})/(2B_4\bar{\rho}) > 0$ for maximum linear instability of $\sigma_\rho^{\rm max} = -(1+B_2\bar{\rho})^2/(4B_4\bar{\rho}) > 0$, corresponding to the occurrence of phase separation.

For the instability of the polarization field, we always have the maximum perturbation growth rate $\sigma_\mathbf{P}^{\rm max} = -\tilde{D}_{\mathrm{R}}(1+2B_5\bar{\rho})$ at $q_m = 0$. If $\bar{\rho} > -{1}/(2B_5) = {1}/(4\pi^2\beta bc^2)$, $\sigma_\mathbf{P}^{\rm max} > 0$ and the system would evolve to another uniform but anisotropic state with constant nonzero polarization $\mathbf{P} = \bar{\mathbf P}$, corresponding to an alignment or flocking state with constant $\bar{\mathbf P} \neq 0$ at high enough particle density $\bar{\rho}$ (note that this polarization instability or flocking/alignment transition is driven by the nonzero torque with $b \neq 0, c \neq 0$); otherwise $\sigma_\mathbf{P}^{\rm max} < 0$ and the system would remain in the orientationally disordered state of $\bar{\mathbf P} = \mathbf{0}$.

\subsubsection{Primary instability: The nonreciprocal cases \texorpdfstring{$0 < \vartheta < \pi$}{}}

In the general case of nonreciprocal interaction, $\vartheta \neq \pi$ (with $0 < \vartheta < \pi$); from Eq.~(\ref{eq:L1_sigma}) the characteristic equation for the perturbation growth rate $\sigma$ becomes
\begin{align}
  & \left ( \sigma - \sigma_1 \right ) \left \{ \left ( \sigma - \sigma_1 \right )
  \left [ \sigma + \left ( 1 + B_2\bar{\rho} \right ) q^2 - B_4 \bar{\rho} q^4 \right ] \right. \nonumber\\
  & \qquad\qquad \left. + 4B_1^2 \bar{\rho}^2 q^2 \right \} = 0.
\end{align}
The three solutions are
\begin{align}
  \sigma_1 & = - \tilde{D}_R \left ( 1 + 2B_5\bar{\rho} \right )
   - \left ( 1 + \frac{B_1^2}{4\tilde{D}_R}\bar{\rho}^2 \right ) q^2, \label{eq:sigma1} \\
  \sigma_{2,3} & = \frac{1}{2} \Big [ - \tilde{D}_R \left ( 1 + 2B_5\bar{\rho} \right )
    - \left ( 2 + B_2\bar{\rho} + \frac{B_1^2}{4\tilde{D}_R}\bar{\rho}^2 \right ) q^2 \nonumber\\
    & ~\qquad + B_4 \bar{\rho} q^4 \pm \sqrt{\Delta} \Big ],
  \label{eq:sigma23}
\end{align}
where
\begin{align}
  \Delta =&~ \left [ \tilde{D}_R \left ( 1 + 2B_5\bar{\rho} \right )
    + \left ( \frac{B_1^2}{4\tilde{D}_R}\bar{\rho}^2 - B_2\bar{\rho} \right ) q^2
    + B_4 \bar{\rho} q^4 \right ]^2 \nonumber\\
    & - 16B_1^2 \bar{\rho}^2 q^2.
\end{align}
Given the requirement of system stability at small length scales, i.e., $\sigma(q \rightarrow \infty) \leq 0$, we have
\begin{equation}
  \sigma_{2,3}(q \rightarrow \infty) \rightarrow
  \frac{1}{2} \left ( B_4 \bar{\rho} q^4 \pm \left |B_4 \bar{\rho} q^4 \right | \right ) \leq 0
  \quad \Rightarrow \; B_4 < 0.
\end{equation}
The system stability is determined by the largest real part ${\rm Re}(\sigma)$ of the above three solutions, with the corresponding wave number $q_m$ of the maximum instability. When the maximum ${\rm Re}(\sigma) > 0$ at $q = q_m$, the linear instability or bifurcation occurs, with the emergence of new pattern (if $q_m > 0$) saturated by the nonlinear terms. When the maximum of ${\rm Re}(\sigma)$ at $q_m$ corresponds to one of the complex roots $\sigma_{2,3} = \sigma_R \pm i \sigma_I$ with $\Delta < 0$, we have an oscillatory instability.

From Eq.~(\ref{eq:sigma1}), the maximum of $\sigma_1$ is given by
\begin{equation}
  \sigma_1^{\rm max} = -\tilde{D}_R \left ( 1 + 2B_5\bar{\rho} \right ), \; \textrm{at } q_m = 0,
  \label{eq:sigma1_max}
\end{equation}
so that $\sigma_1^{\rm max} > 0$ when
\begin{equation}
  1 + 2B_5\bar{\rho} < 0, \quad {\rm i.e.,} \quad \bar{\rho} \vartheta > \frac{1}{4\pi \beta b c^2},
  \label{eq:sig1_condition}
\end{equation}
giving a condition of primary instability and bifurcation at large enough $\bar{\rho}$ or $\vartheta$. For $\sigma_{2,3}$ we need to calculate (\ref{eq:sigma23}) numerically.

It is useful to first examine the long-wavelength limit of $q=0$, for which $\widehat{\rho}_{\mathbf{q}}$, $\widehat{P}_{x\mathbf{q}}$, and $\widehat{P}_{y\mathbf{q}}$ are decoupled as seen from Eq.~(\ref{eq:linearized}), yielding
\begin{equation}
\begin{split}
  & \sigma_\rho(q=0) = 0, \\
  &\sigma_{P_x}(q=0) = \sigma_{P_y}(q=0) = -\tilde{D}_R \left ( 1 + 2B_5\bar{\rho} \right ).
\end{split}
\end{equation}
Also from Eqs.~(\ref{eq:sigma1}) and (\ref{eq:sigma23}),
\begin{equation}
\begin{split}
  & \sigma_1(q=0) = -\tilde{D}_R (1 + 2B_5\bar{\rho}), \\
  & \sigma_{2,3}(q=0) = \frac{1}{2} [-\tilde{D}_R (1 + 2B_5\bar{\rho}) \pm |\tilde{D}_R (1 + 2B_5\bar{\rho})|].
\end{split}
\end{equation}
Thus, noting $B_5 = -2\pi \beta b c^2 \vartheta < 0$, we find that when $\bar{\rho} < -{1}/{(2B_5)}$,
\begin{align}
  &\sigma_2(q=0) = \sigma_\rho(q=0) = 0, \nonumber\\
  &\sigma_1(q=0) = \sigma_3(q=0) = \sigma_\mathbf{P}(q=0) \nonumber\\
  &= -\tilde{D}_R \left ( 1 + 2B_5\bar{\rho} \right )
  < 0, 
\end{align}
and when $\bar{\rho} > -{1}/{(2B_5)}$,
\begin{align}
  &\sigma_3(q=0) = \sigma_\rho(q=0) = 0, \nonumber\\
  &\sigma_1(q=0) = \sigma_2(q=0) = \sigma_\mathbf{P}(q=0) \nonumber\\
  &= -\tilde{D}_R \left ( 1 + 2B_5\bar{\rho} \right )
  > 0, 
\end{align}
indicating a flocking transition to an alignment state with $\mathbf{P} \neq \mathbf{0}$ at high enough particle density $\bar{\rho}$ or large enough opening angle $\vartheta$.

At finite values of $q$, since $\sigma_\rho(q=0) = 0$, if $\left [ {\dif{\rm Re}(\sigma_{2,3})}/{\dif(q^2)} \right ]_{q^2 \rightarrow 0} > 0$ ($\sigma_2$ for $\bar{\rho} < -1/(2B_5)$ and $\sigma_3$ for $\bar{\rho} > -1/(2B_5)$), linear instability would occur at a finite wave number with ${\rm Re}(\sigma)(q \neq 0) > 0$, leading to phase separation of the particle density. From $\Delta(q^2 \rightarrow 0) = [\tilde{D}_R \left ( 1 + 2B_5\bar{\rho} \right )]^2 \geq 0$ and then $\sigma_{2,3}(q^2 \rightarrow 0)$ being real when $\bar{\rho} \neq -1/(2B_5)$, or $\Delta(q^2 \rightarrow 0) \leq 0$ when $\bar{\rho} = -1/(2B_5)$, we have
\begin{align}
  & \left. \frac{\dif\sigma_{2,3}}{\dif(q^2)} \right |_{q^2 \rightarrow 0} = \frac{1}{2} \bigg \{
  - \left ( 2 + B_2\bar{\rho} + \frac{B_1^2}{4\tilde{D}_R}\bar{\rho}^2 \right ) \nonumber\\
  & \quad \pm \frac{1}{\left |\tilde{D}_R (1 + 2B_5\bar{\rho}) \right |} \nonumber\\
  & \quad \times \left [ \tilde{D}_R \left ( 1 + 2B_5\bar{\rho} \right )
    \left ( \frac{B_1^2}{4\tilde{D}_R}\bar{\rho}^2 -B_2 \bar{\rho} \right ) - 8B_1^2 \bar{\rho}^2 \right ] \bigg \} >0,
\end{align}
when $\bar{\rho} \neq -1/(2B_5)$, and
\begin{equation}
  \left. \frac{\dif{\rm Re}(\sigma_{2,3})}{\dif(q^2)} \right |_{q^2 \rightarrow 0}
  = -\frac{1}{2} \left ( 2 + B_2\bar{\rho} + \frac{B_1^2}{4\tilde{D}_R}\bar{\rho}^2 \right ) >0,
\end{equation}
when $\bar{\rho} = -1/(2B_5)$.
Then the conditions for the occurrence of phase separation are given by
\begin{align}
  \textrm{If} ~& \textrm{Re}(\sigma_{2,3}^{\rm max}(q_m > 0)) > \sigma_1^{\rm max}
  = -\tilde{D}_R \left ( 1 + 2B_5\bar{\rho} \right ), \nonumber\\
  & 1 + B_2\bar{\rho} + \frac{4B_1^2\bar{\rho}^2}{\tilde{D}_R (1 + 2B_5\bar{\rho})} < 0, \quad
  \textrm{when } \bar{\rho} \neq -\frac{1}{2B_5}, \nonumber\\
  & 2 + B_2\bar{\rho} + \frac{B_1^2}{4\tilde{D}_R}\bar{\rho}^2 < 0, \quad
  \textrm{when } \bar{\rho} = -\frac{1}{2B_5}.
  \label{eq:instability}
\end{align}
Therefore, (i) when $\bar{\rho} \leq -1/(2B_5)$ or equivalently $\bar{\rho}\vartheta < 1/(4\pi\beta bc^2)$, i.e., for low particle density $\bar{\rho}$ or small vision-cone opening angle $\vartheta$, the system exhibits phase separation without flocking transition if the above conditions are satisfied; otherwise the system remains in the fully disordered state. (ii) At higher particle density $\bar{\rho}$ or large enough opening angle $\vartheta$ when $\bar{\rho} > -1/(2B_5)$ or equivalently $\bar{\rho}\vartheta > 1/(4\pi\beta bc^2)$, both phase separation and flocking occurs if Eq.~(\ref{eq:instability}) is satisfied; otherwise a flocking transition without phase separation (i.e., with homogeneous particle density) occurs.

For completeness, it is also interesting to analytically examine the conditions of oscillatory instability with $\Delta < 0$, if the corresponding maximum perturbation growth rate is larger than that for $\Delta > 0$. In that case we have ${\rm Re}(\sigma_{2,3}) = [- \tilde{D}_R ( 1 + 2B_5\bar{\rho} ) - ( 2 + B_2\bar{\rho} + \frac{B_1^2}{4\tilde{D}_R}\bar{\rho}^2) q^2 + B_4 \bar{\rho} q^4]/2$, and from $d{\rm Re}(\sigma_{2,3}) / d(q^2) = 0$ we get
\begin{equation}
\begin{split}
  & q_m^2 = \frac{1}{2B_4\bar{\rho}} \left ( 2 + B_2\bar{\rho} + \frac{B_1^2}{4\tilde{D}_R}\bar{\rho}^2 \right ), \\
  & {\rm Re}(\sigma_{2,3}^{\rm max}) = -\frac{1}{2} \Bigg [ \tilde{D}_R \left ( 1 + 2B_5\bar{\rho} \right ) \\
  & \qquad\qquad\qquad\quad + \frac{1}{4B_4\bar{\rho}} \left (2 + B_2\bar{\rho} + \frac{B_1^2}{4\tilde{D}_R}\bar{\rho}^2 \right )^2 \Bigg ].
\end{split}
\end{equation}
To satisfy $q_m^2 > 0$, ${\rm Re}(\sigma_{2,3}^{\rm max}) > 0$, ${\rm Re}(\sigma_{2,3}^{\rm max}) > \sigma_1^{\rm max}$, and $\Delta(q=q_m) < 0$, the corresponding conditions are
\begin{align}
  & B_4 < 0, \quad 2 + B_2\bar{\rho} + \frac{B_1^2}{4\tilde{D}_R}\bar{\rho}^2 < 0, \nonumber\\
  & \left ( 2 + B_2\bar{\rho} + \frac{B_1^2}{4\tilde{D}_R}\bar{\rho}^2 \right )^2 >
  - 4B_4\bar{\rho} \left | \tilde{D}_R \left ( 1 + 2B_5\bar{\rho} \right ) \right |, \nonumber\\
  & \Bigg [ 4B_4\bar{\rho}\tilde{D}_R \left ( 1 + 2B_5\bar{\rho} \right ) \nonumber\\
  & + \left ( 2 - B_2 \bar{\rho} + \frac{3B_1^2}{4\tilde{D}_R}\bar{\rho}^2 \right )
    \left ( 2 + B_2 \bar{\rho} + \frac{B_1^2}{4\tilde{D}_R}\bar{\rho}^2 \right ) \Bigg ]^2 \nonumber\\
  & < 128B_1^2B_4\bar{\rho}^3 \left ( 2 + B_2\bar{\rho} + \frac{B_1^2}{4\tilde{D}_R}\bar{\rho}^2 \right ).
\end{align}
If these conditions are satisfied and ${\rm Re}(\sigma_{2,3}^{\rm max})(\Delta < 0) > \sigma_{2,3}^{\rm max}(\Delta > 0)$, when $\bar{\rho} \leq -1/(2B_5)$ or $\bar{\rho}\vartheta \leq 1/(4\pi\beta bc^2)$ we have phase separation with oscillatory instability of frequency $\omega = \sigma_{\rm I} = \sqrt{-\Delta(q_m)}$ but without flocking; when $\bar{\rho} > -1/(2B_5)$ or $\bar{\rho}\vartheta > 1/(4\pi\beta bc^2)$, i.e, at large enough particle density or high enough vision-cone opening angle, both phase separation with oscillatory instability and flocking occur.

For more specific results of system bifurcation and transitions, it would be more straightforward to numerically calculate the perturbation growth rate as a function of wave number $q$ by directly evaluating Eqs.~(\ref{eq:sigma1}) and (\ref{eq:sigma23}) across various values of model parameters and then identify the maximum growth rate $\sigma_{\rm max}$ and the corresponding most unstable wave number $q_m$. In principle there could be six possible phases, including (i) the fully disordered state with $\rho=\bar{\rho}$ and $\mathbf{P}=\bar{\mathbf P}=\mathbf{0}$, (ii) homogeneous or uniform flocking phase with polarization alignment and homogeneous particle density, i.e., $\rho=\bar{\rho}$ and $\mathbf{P} = \bar{\mathbf P} \neq \mathbf{0}$, for which $q_m=0$ and $\sigma_{\rm max} = \sigma_1^{\rm max} = \sigma_{\mathbf P}(q=0) > 0$ (i.e., Eq.~(\ref{eq:sigma1_max})) satisfying the condition of Eq.~(\ref{eq:sig1_condition}), (iii) phase separation of particle density $\rho$ without flocking ($\mathbf{P} = \bar{\mathbf P} = \mathbf{0}$), for which $q_m > 0$, $\sigma_{\rm max} = \sigma_2^{\rm max} > 0$ with $\Delta > 0$, and $\sigma_1^{\rm max} < 0$, (iv) both phase separation and flocking, with $q_m > 0$, $\sigma_{\rm max} = \sigma_3^{\rm max} > 0$ with $\Delta > 0$, and $\sigma_1^{\rm max} > 0$, (v) phase separation with oscillatory instability but without flocking, with the conditions of (iii) other than $\sigma_{\rm max} = \textrm{Re}(\sigma_{2,3}^{\rm max}) > 0$ with $\Delta < 0$, and (vi) both phase separation and flocking, with oscillatory instability, corresponding to the conditions of (iv) other than $\sigma_{\rm max} = \textrm{Re}(\sigma_{2,3}^{\rm max}) > 0$ with $\Delta < 0$.

For the model parameters used in this study, as given in Eq.~\eqref{eq:para_g_tau} for a soft nonreciprocal interaction, our calculations show that only phase (i), i.e., the fully disordered base state, or phase (ii) of homogeneous flocking state, could occur for primary bifurcation, while the conditions for cases (iii)--(vi) cannot be satisfied across all the parameter ranges of $\vartheta$, $\bar{\rho}$, and $\tilde{D}_R$. The choices of other types of interaction functions of force and torque, or other parameter combinations that might lead to the emergence of any other phases while still maintaining the numerical convergence of the full nonlinear model equations, are beyond the scope of this work.

\subsection{Secondary instability/bifurcation}
\label{sec:secondary_inst}

When the system is in the homogeneous flocking state, as developed from the above primary bifurcation and characterized by uniform particle density $\rho=\bar{\rho}$ and a specific polarization alignment $P_x=\bar{P}_x$ and $P_y =\bar{P}_y$ with nonzero polarization magnitude $|\bar{\mathbf P}|^2 = -4(\bar{\rho}+1/2B_5)/B_5 > 0$ (i.e., Eq.~(\ref{eq:Pbar^2}), satisfying the condition of Eq.~(\ref{eq:sig1_condition})), the corresponding secondary instability analysis can be conducted via expanding
\begin{equation}
  \rho = \bar{\rho} + \widehat{\rho}, \quad P_x = \bar{P}_x + \widehat{P}_x, \quad
  P_y = \bar{P}_y + \widehat{P}_y, \label{eq:expand}
\end{equation}
to get the first-order equations for $\widehat{\rho}_{\mathbf{q}}$ and $\widehat{\mathbf P}_{\mathbf{q}}$ in Fourier space, i.e.,
\begin{equation}
\begin{split}
  \frac{\partial}{\partial t} \left (
  \begin{array}{c}
    \widehat{\rho}_{\mathbf{q}} \\ \widehat{P}_{x\mathbf{q}} \\ \widehat{P}_{y\mathbf{q}}
  \end{array}
  \right )
  &= \mathcal{L} \left (
  \begin{array}{c}
    \widehat{\rho}_{\mathbf{q}} \\ \widehat{P}_{x\mathbf{q}} \\ \widehat{P}_{y\mathbf{q}}
  \end{array}
  \right ) \\
  &= \left ( \begin{array}{ccc}
    \alpha_{11} & iB_1 \bar{\rho} q_x &  iB_1 \bar{\rho} q_y \\
    \alpha_{21} & \alpha_{22} & \alpha_{23} \\
    \alpha_{31} & \alpha_{32} & \alpha_{33}
  \end{array} \right )
  \left ( \begin{array}{c}
    \widehat{\rho}_{\mathbf{q}} \\ \widehat{P}_{x\mathbf{q}} \\ \widehat{P}_{y\mathbf{q}}
  \end{array} \right ), 
\end{split}
\label{eq:secondary_linearized}
\end{equation}
where (after using Eq.~(\ref{eq:Pbar^2}) for $|\bar{\mathbf P}|^2$) 
\begin{align}
  \alpha_{11} =& -(1+B_2\bar{\rho})q^2 + B_4\bar{\rho}q^4 \nonumber\\
  & + i\left ( B_1-3B_3q^2 \right ) \left ( q_x\bar{P}_x+q_y\bar{P}_y \right ), \nonumber\\
  \alpha_{21} =& -\left ( 2\tilde{D}_RB_5 + \frac{B_1^2}{4\tilde{D}_R}\bar{\rho}q^2 \right ) \bar{P}_x \nonumber\\
  & + iq_x \left [ 4B_1\bar{\rho} - \frac{1}{4}B_1B_5\left ( \bar{P}_x^2-3\bar{P}_y^2 \right ) \right ] \nonumber\\
  & - iq_yB_1B_5\bar{P}_x\bar{P}_y, \nonumber\\
  \alpha_{22} =& - \tilde{D}_RB_5^2\bar{P}_x^2 - \left ( 1 + \frac{B_1^2}{4\tilde{D}_R}\bar{\rho}^2
  \right ) q^2 \nonumber\\
  & - i \frac{3}{4} B_1B_5\bar{\rho}\left ( q_x\bar{P}_x+q_y\bar{P}_y \right ), \nonumber\\
  \alpha_{23} =& - \tilde{D}_RB_5^2\bar{P}_x\bar{P}_y - \frac{5}{4} iB_1B_5\bar{\rho}
  \left ( q_y\bar{P}_x-q_x\bar{P}_y \right ), \nonumber\\
  \alpha_{31} =& -\left ( 2\tilde{D}_RB_5 + \frac{B_1^2}{4\tilde{D}_R}\bar{\rho}q^2 \right ) \bar{P}_y
  - iq_xB_1B_5\bar{P}_x\bar{P}_y \nonumber\\
  & + iq_y \left [ 4B_1\bar{\rho} + \frac{1}{4}B_1B_5\left ( 3\bar{P}_x^2-\bar{P}_y^2 \right ) \right ]
  = \left. \alpha_{21} \right |_{x \leftrightarrow y}, \nonumber\\
  \alpha_{32} =& \alpha_{23}^* = \left. \alpha_{23} \right |_{x \leftrightarrow y}, \nonumber\\
  \alpha_{33} =& - \tilde{D}_RB_5^2\bar{P}_y^2 - \left ( 1 + \frac{B_1^2}{4\tilde{D}_R}\bar{\rho}^2 \right ) q^2 \nonumber\\
  & - i \frac{3}{4} B_1B_5\bar{\rho}\left ( q_x\bar{P}_x+q_y\bar{P}_y \right )
  = \left. \alpha_{22} \right |_{x \leftrightarrow y},\label{eq:dynamicalmatrix}
\end{align}
indicating the non-Hermiticity of the dynamical matrix as induced by force nonreciprocity, i.e., $\mathcal{L}^\dagger \neq \mathcal{L}$ when $\vartheta \neq \pi$. Similarly, the perturbation growth rate $\sigma = \sigma(q_x,q_y)$ of the secondary instability is determined by the eigenvalues of $\mathcal{L}$, with
\begin{equation}
  \left | \mathcal{L} - \sigma \mathds{1} \right | = 0. \label{eq:L_sigma}
\end{equation}

\subsubsection{Secondary instability: The reciprocal limit \texorpdfstring{$\vartheta = \pi$}{}}

In the reciprocal limit with $\vartheta=\pi$ and hence $B_1=B_3=0$, the dynamics of $\widehat{\rho}_{\mathbf{q}}$ is decoupled from that of $\widehat{P}_{x\mathbf{q}}$ and $\widehat{P}_{y\mathbf{q}}$, and the dynamical matrix for $\widehat{P}_{x\mathbf{q}}$ and $\widehat{P}_{y\mathbf{q}}$ is Hermitian. The perturbation growth rate for $\rho$ is given by
\begin{equation}
  \sigma_\rho = - (1+B_2\bar{\rho})q^2 + B_4 \bar{\rho} q^4, \label{eq:sig_rho_reci}
\end{equation}
which is the same as that of primary instability. When $B_2\bar{\rho} < -1$, phase separation of particle density $\rho$ occurs, with the most unstable wave number $q_m = \sqrt{(1+B_2\bar{\rho}) /(2B_4\bar{\rho})}$ (noting $B_4<0$ to satisfy the condition of small length scale stability); otherwise, $\rho$ remains homogeneous when $B_2\bar{\rho} > -1$ (as for the model parameters used in this study).

For the polarization field,
\begin{align}
  \sigma_{P1} &= -\tilde{D}_R \left ( 1 + 2B_5\bar{\rho} + \frac{1}{2} B_5^2 \bar{P}^2 \right )
  - q^2 = -q^2, \nonumber\\
  \sigma_{P2} &= -\tilde{D}_R \left ( 1 + 2B_5\bar{\rho} + \frac{3}{2} B_5^2 \bar{P}^2 \right ) - q^2 \nonumber\\
  &= -\tilde{D}_R B_5^2 \bar{P}^2 -q^2, \label{eq:sig_P_reci}
\end{align}
where Eq.~(\ref{eq:Pbar^2}) has been used for $\bar{P}$. Thus we always have $\sigma_P \leq 0$, yielding the stable uniform flocking phase with stationary polarization alignment in the reciprocal limit.

\subsubsection{Secondary instability: The nonreciprocal cases \texorpdfstring{$0 < \vartheta < \pi$}{}}

In the general case of nonreciprocal interaction with $\vartheta \neq \pi$, the corresponding characteristic equation (\ref{eq:L_sigma}) is a cubic equation for the perturbation growth rate $\sigma$, i.e.,
\begin{align}
  & \sigma^3 - \left ( \alpha_{11} + \alpha_{22} + \alpha_{33} \right ) \sigma^2
  + \big [ \alpha_{11} \left ( \alpha_{22} + \alpha_{33} \right ) + \alpha_{22}\alpha_{33} \nonumber\\
  & \quad - \alpha_{23}\alpha_{32} - iB_1\bar{\rho} \left ( q_x \alpha_{21} + q_y \alpha_{31} \right )
    \big ] \sigma \nonumber\\
  & + \alpha_{11} \left ( \alpha_{23}\alpha_{32} - \alpha_{22}\alpha_{33} \right ) \nonumber\\
  & + iB_1\bar{\rho}
  \left [ q_x \left ( \alpha_{21}\alpha_{33} - \alpha_{23}\alpha_{31} \right )
    + q_y \left ( \alpha_{22}\alpha_{31} - \alpha_{21}\alpha_{32} \right ) \right ] \nonumber\\
    & = 0,
  \label{eq:sigma}
\end{align}
giving three solutions $\sigma_1$, $\sigma_2$, and $\sigma_3$. At $q=0$, i.e., $q_x=q_y=0$ in the long-wavelength limit, the exact solutions are
\begin{align}
  &\sigma_1(q=0) = \sigma_\rho(q=0) = 0, \nonumber\\
  &\sigma_2(q=0) = \sigma_\mathbf{P}(q=0) \nonumber\\
  &= -\tilde{D}_R \left ( 1 + 2B_5\bar{\rho}
  + \frac{1}{2} B_5^2 \bar{P}^2 \right ) = 0, \label{eq:sig_q=0}\\
  &\sigma_3(q=0) = \sigma_\mathbf{P}(q=0) \nonumber\\
  &= -\tilde{D}_R \left ( 1 + 2B_5\bar{\rho}
  + \frac{3}{2} B_5^2 \bar{P}^2 \right )= -\tilde{D}_R B_5^2 \bar{P}^2 < 0, \nonumber  
\end{align}
with the use of Eq.~(\ref{eq:Pbar^2}). For $\mathbf{q} \neq \mathbf{0}$, when $B_1 \neq 0$ or $B_3 \neq 0$ (for nonreciprocal interactions with $\vartheta \neq \pi$) the coefficients of the characteristic equation (\ref{eq:sigma}) are complex. If the maximum of the real part of perturbation growth rate Re$(\sigma_j) > 0$ at a finite wave vector $\mathbf{q}_m$ when the corresponding $\sigma_j$ solution is complex, an oscillatory periodic instability would occur, while our numerical results show that either real or complex values of $\sigma_j$ for maximum Re$(\sigma_j)$ could be obtained, depending on the model parameter values and $\mathbf{q}_m$. The exact solutions (Cardano's solution) of this cubic characteristic equation (\ref{eq:sigma}) are evaluated across a range of wave vector $\mathbf{q}=(q_x,q_y)$ for various values of model parameters $\vartheta$ and $\bar{\rho}$. The corresponding results of this bifurcation analysis are given in the $\bar{\rho}$ vs $\vartheta$ phase diagram of Fig.~\ref{fig:phase_diagram}(f), and well agree with numerical simulations of the full nonlinear equations of active Model N in terms of the corresponding phase transitions.

\section{Model parameters and numerical simulations}
\label{sec:parameters}

In this study, we use the soft-interaction form of \cref{eq:g_tau} for the force and torque functions and choose the model parameters
\begin{equation}
  \beta a F_0 = 0.1, \;\frac{F_1}{F_0} = 0.001, \; \beta b = 0.03, \; \alpha=2, \; \frac{c}{a} = 2.
  \label{eq:para_g_tau}
\end{equation}
From Eqs.~\eqref{eq:b1m}--\eqref{eq:b5m}, the coupling coefficients $B_i$ in Active Model N are then found to be
\begin{align}
  B_1 & = \pi \beta a F_0 \left ( 1 - 2\alpha^2 \frac{F_1}{F_0} \right ) \sin\vartheta = 0.0992\pi \sin\vartheta, \nonumber\\
  B_2 & = \frac{\pi}{2} \beta a F_0 \left ( \sqrt{\pi} - 8\alpha^3 \frac{F_1}{F_0} \right ) \vartheta = 0.05\pi (\sqrt{\pi} - 0.064) \vartheta, \nonumber\\
  B_3 & = \frac{\pi}{8} \beta a F_0 \left ( 1 - 12\alpha^4 \frac{F_1}{F_0} \right ) \sin\vartheta = 0.0101\pi \sin\vartheta, \nonumber\\
  B_4 & = \frac{3}{32} \pi \beta a F_0 \left ( \sqrt{\pi} - 64\alpha^5 \frac{F_1}{F_0} \right ) \vartheta \nonumber\\
  & = -(3\pi/320) (2.048 - \sqrt{\pi}) \vartheta, \nonumber\\
  B_5 & = -2\pi \beta b \left ( \frac{c}{a} \right )^2 \vartheta = -0.24\pi \vartheta,
\end{align}
where we have rescaled $B_1 \rightarrow B_1/a$, $B_2 \rightarrow B_2/a^2$, $B_3 \rightarrow B_3/a^3$, $B_4 \rightarrow B_4/a^4$, and $B_5 \rightarrow B_5/a^2$ in Eqs.~(\ref{eq:b1m})--(\ref{eq:b5m}) to make all the parameters dimensionless. In addition, $\tilde{D}_\mathrm{R} = a^2 D_\mathrm{R} / D_\mathrm{T} = 0.1$ is used.

Equations \eqref{eq:rho}, \eqref{eq:P}, and \eqref{eq:Qij} of Active Model N are solved numerically using a pseudospectral method with periodic boundary conditions. The simulated system size ranges from $256 \times 256$ to $2048 \times 2048$ grid points, with grid spacing $\Delta x = \Delta y = 1$. The time step size $\Delta t$ is fixed within each simulation, but chosen from $0.001$ to $0.1$ depending on the values of $\bar{\rho}$ and $\vartheta$ to ensure numerical convergence and accuracy. The simulation snapshots shown in Figs.~\ref{fig:phase_diagram}(a)--\ref{fig:phase_diagram}(e) and Fig.~\ref{fig:pattern_evolution} and results of Figs.~\ref{fig:activeyarn_varying}--\ref{fig:activeyarn_width} are for a system size $512 \times 512$. For most simulation results presented here (all except for Fig.~\ref{fig:phase_diagram}(e)), the initial conditions are chosen as a homogeneous or uniform flocking state with homogeneous particle density $\rho=\bar{\rho}$ and polarization $\mathbf{P}=\mathbf{\bar{P}}$ imposed by random initial fluctuations. Similar and consistent results are obtained if using different values of initial polarization alignment $\mathbf{\bar{P}} = (\bar{P}_x, \bar{P}_y)$. A different initial setup is used for the reciprocal limit of $\vartheta = \pi$ (Fig.~\ref{fig:phase_diagram}(e)), where random initial conditions are adopted for both particle density and polarization fields in the fully disordered state with $\rho=\bar{\rho}$ and $\mathbf{P}=\mathbf{0}$.

All the results of active droplet growth dynamics given in Fig.~\ref{fig:droplet_growth} are obtained from simulations with a system size $2048 \times 2048$. Initially both the droplet nucleus and outside medium are in the homogeneous flocking phase with $\rho = \bar{\rho}$ and $\mathbf{P} = \bar{\mathbf{P}} = (\bar{P}, 0)$ with $\bar{P} = |\bar{\mathbf{P}}|$ determined by \cref{eq:Pbar}, such that the droplet self-travels along the negative $x$ direction. Initial noise fluctuations are set inside the small circular nucleus of radius $r_0=5$, but not in the outside medium, which allows the droplet to expand and grow with time, with the development of patterns inside.

\bibliography{refs}

\begin{thebibliography}{65}%
\makeatletter
\providecommand \@ifxundefined [1]{%
 \@ifx{#1\undefined}
}%
\providecommand \@ifnum [1]{%
 \ifnum #1\expandafter \@firstoftwo
 \else \expandafter \@secondoftwo
 \fi
}%
\providecommand \@ifx [1]{%
 \ifx #1\expandafter \@firstoftwo
 \else \expandafter \@secondoftwo
 \fi
}%
\providecommand \natexlab [1]{#1}%
\providecommand \enquote  [1]{``#1''}%
\providecommand \bibnamefont  [1]{#1}%
\providecommand \bibfnamefont [1]{#1}%
\providecommand \citenamefont [1]{#1}%
\providecommand \href@noop [0]{\@secondoftwo}%
\providecommand \href [0]{\begingroup \@sanitize@url \@href}%
\providecommand \@href[1]{\@@startlink{#1}\@@href}%
\providecommand \@@href[1]{\endgroup#1\@@endlink}%
\providecommand \@sanitize@url [0]{\catcode `\\12\catcode `\$12\catcode `\&12\catcode `\#12\catcode `\^12\catcode `\_12\catcode `\%12\relax}%
\providecommand \@@startlink[1]{}%
\providecommand \@@endlink[0]{}%
\providecommand \url  [0]{\begingroup\@sanitize@url \@url }%
\providecommand \@url [1]{\endgroup\@href {#1}{\urlprefix }}%
\providecommand \urlprefix  [0]{URL }%
\providecommand \Eprint [0]{\href }%
\providecommand \doibase [0]{https://doi.org/}%
\providecommand \selectlanguage [0]{\@gobble}%
\providecommand \bibinfo  [0]{\@secondoftwo}%
\providecommand \bibfield  [0]{\@secondoftwo}%
\providecommand \translation [1]{[#1]}%
\providecommand \BibitemOpen [0]{}%
\providecommand \bibitemStop [0]{}%
\providecommand \bibitemNoStop [0]{.\EOS\space}%
\providecommand \EOS [0]{\spacefactor3000\relax}%
\providecommand \BibitemShut  [1]{\csname bibitem#1\endcsname}%
\let\auto@bib@innerbib\@empty
\bibitem [{\citenamefont {Petroff}\ \emph {et~al.}(2015)\citenamefont {Petroff}, \citenamefont {Wu},\ and\ \citenamefont {Libchaber}}]{PetroffPRL15}%
  \BibitemOpen
  \bibfield  {author} {\bibinfo {author} {\bibfnamefont {A.~P.}\ \bibnamefont {Petroff}}, \bibinfo {author} {\bibfnamefont {X.-L.}\ \bibnamefont {Wu}},\ and\ \bibinfo {author} {\bibfnamefont {A.}~\bibnamefont {Libchaber}},\ }\bibfield  {title} {\bibinfo {title} {Fast-moving bacteria self-organize into active two-dimensional crystals of rotating cells},\ }\href@noop {} {\bibfield  {journal} {\bibinfo  {journal} {Phys. Rev. Lett.}\ }\textbf {\bibinfo {volume} {114}},\ \bibinfo {pages} {158102} (\bibinfo {year} {2015})}\BibitemShut {NoStop}%
\bibitem [{\citenamefont {Tan}\ \emph {et~al.}(2022)\citenamefont {Tan}, \citenamefont {Mietke}, \citenamefont {Li}, \citenamefont {Chen}, \citenamefont {Higinbotham}, \citenamefont {Foster}, \citenamefont {Gokhale}, \citenamefont {Dunkel},\ and\ \citenamefont {Fakhri}}]{TanNature22}%
  \BibitemOpen
  \bibfield  {author} {\bibinfo {author} {\bibfnamefont {T.~H.}\ \bibnamefont {Tan}}, \bibinfo {author} {\bibfnamefont {A.}~\bibnamefont {Mietke}}, \bibinfo {author} {\bibfnamefont {J.}~\bibnamefont {Li}}, \bibinfo {author} {\bibfnamefont {Y.}~\bibnamefont {Chen}}, \bibinfo {author} {\bibfnamefont {H.}~\bibnamefont {Higinbotham}}, \bibinfo {author} {\bibfnamefont {P.~J.}\ \bibnamefont {Foster}}, \bibinfo {author} {\bibfnamefont {S.}~\bibnamefont {Gokhale}}, \bibinfo {author} {\bibfnamefont {J.}~\bibnamefont {Dunkel}},\ and\ \bibinfo {author} {\bibfnamefont {N.}~\bibnamefont {Fakhri}},\ }\bibfield  {title} {\bibinfo {title} {Odd dynamics of living chiral crystals},\ }\href@noop {} {\bibfield  {journal} {\bibinfo  {journal} {Nature}\ }\textbf {\bibinfo {volume} {607}},\ \bibinfo {pages} {287} (\bibinfo {year} {2022})}\BibitemShut {NoStop}%
\bibitem [{\citenamefont {Lavergne}\ \emph {et~al.}(2019)\citenamefont {Lavergne}, \citenamefont {Wendehenne}, \citenamefont {B{\"a}uerle},\ and\ \citenamefont {Bechinger}}]{LavergneWBB2019}%
  \BibitemOpen
  \bibfield  {author} {\bibinfo {author} {\bibfnamefont {F.~A.}\ \bibnamefont {Lavergne}}, \bibinfo {author} {\bibfnamefont {H.}~\bibnamefont {Wendehenne}}, \bibinfo {author} {\bibfnamefont {T.}~\bibnamefont {B{\"a}uerle}},\ and\ \bibinfo {author} {\bibfnamefont {C.}~\bibnamefont {Bechinger}},\ }\bibfield  {title} {\bibinfo {title} {Group formation and cohesion of active particles with visual perception--dependent motility},\ }\href@noop {} {\bibfield  {journal} {\bibinfo  {journal} {Science}\ }\textbf {\bibinfo {volume} {364}},\ \bibinfo {pages} {70} (\bibinfo {year} {2019})}\BibitemShut {NoStop}%
\bibitem [{\citenamefont {Barberis}\ and\ \citenamefont {Peruani}(2016)}]{BarberisPRL16}%
  \BibitemOpen
  \bibfield  {author} {\bibinfo {author} {\bibfnamefont {L.}~\bibnamefont {Barberis}}\ and\ \bibinfo {author} {\bibfnamefont {F.}~\bibnamefont {Peruani}},\ }\bibfield  {title} {\bibinfo {title} {Large-scale patterns in a minimal cognitive flocking model: Incidental leaders, nematic patterns, and aggregates},\ }\href@noop {} {\bibfield  {journal} {\bibinfo  {journal} {Phys. Rev. Lett.}\ }\textbf {\bibinfo {volume} {117}},\ \bibinfo {pages} {248001} (\bibinfo {year} {2016})}\BibitemShut {NoStop}%
\bibitem [{\citenamefont {Bastien}\ and\ \citenamefont {Romanczuk}(2020)}]{Bastien2020}%
  \BibitemOpen
  \bibfield  {author} {\bibinfo {author} {\bibfnamefont {R.}~\bibnamefont {Bastien}}\ and\ \bibinfo {author} {\bibfnamefont {P.}~\bibnamefont {Romanczuk}},\ }\bibfield  {title} {\bibinfo {title} {A model of collective behavior based purely on vision},\ }\href@noop {} {\bibfield  {journal} {\bibinfo  {journal} {Sci. Adv.}\ }\textbf {\bibinfo {volume} {6}},\ \bibinfo {pages} {eaay0792} (\bibinfo {year} {2020})}\BibitemShut {NoStop}%
\bibitem [{\citenamefont {Negi}\ \emph {et~al.}(2022)\citenamefont {Negi}, \citenamefont {Winkler},\ and\ \citenamefont {Gompper}}]{NegiWG2022}%
  \BibitemOpen
  \bibfield  {author} {\bibinfo {author} {\bibfnamefont {R.~S.}\ \bibnamefont {Negi}}, \bibinfo {author} {\bibfnamefont {R.~G.}\ \bibnamefont {Winkler}},\ and\ \bibinfo {author} {\bibfnamefont {G.}~\bibnamefont {Gompper}},\ }\bibfield  {title} {\bibinfo {title} {Emergent collective behavior of active {B}rownian particles with visual perception},\ }\href@noop {} {\bibfield  {journal} {\bibinfo  {journal} {Soft Matter}\ }\textbf {\bibinfo {volume} {18}},\ \bibinfo {pages} {6167} (\bibinfo {year} {2022})}\BibitemShut {NoStop}%
\bibitem [{\citenamefont {Negi}\ \emph {et~al.}(2024{\natexlab{a}})\citenamefont {Negi}, \citenamefont {Winkler},\ and\ \citenamefont {Gompper}}]{NegiWG2024}%
  \BibitemOpen
  \bibfield  {author} {\bibinfo {author} {\bibfnamefont {R.~S.}\ \bibnamefont {Negi}}, \bibinfo {author} {\bibfnamefont {R.~G.}\ \bibnamefont {Winkler}},\ and\ \bibinfo {author} {\bibfnamefont {G.}~\bibnamefont {Gompper}},\ }\bibfield  {title} {\bibinfo {title} {Collective behavior of self-steering active particles with velocity alignment and visual perception},\ }\href@noop {} {\bibfield  {journal} {\bibinfo  {journal} {Phys. Rev. Res.}\ }\textbf {\bibinfo {volume} {6}},\ \bibinfo {pages} {013118} (\bibinfo {year} {2024}{\natexlab{a}})}\BibitemShut {NoStop}%
\bibitem [{\citenamefont {Negi}\ \emph {et~al.}(2024{\natexlab{b}})\citenamefont {Negi}, \citenamefont {Iyer},\ and\ \citenamefont {Gompper}}]{NegiIG2024}%
  \BibitemOpen
  \bibfield  {author} {\bibinfo {author} {\bibfnamefont {R.~S.}\ \bibnamefont {Negi}}, \bibinfo {author} {\bibfnamefont {P.}~\bibnamefont {Iyer}},\ and\ \bibinfo {author} {\bibfnamefont {G.}~\bibnamefont {Gompper}},\ }\bibfield  {title} {\bibinfo {title} {Controlling inter-particle distances in crowds of motile, cognitive, active particles},\ }\href@noop {} {\bibfield  {journal} {\bibinfo  {journal} {Sci. Rep.}\ }\textbf {\bibinfo {volume} {14}},\ \bibinfo {pages} {9443} (\bibinfo {year} {2024}{\natexlab{b}})}\BibitemShut {NoStop}%
\bibitem [{\citenamefont {Loos}\ \emph {et~al.}(2023)\citenamefont {Loos}, \citenamefont {Klapp},\ and\ \citenamefont {Martynec}}]{LoosKM2023}%
  \BibitemOpen
  \bibfield  {author} {\bibinfo {author} {\bibfnamefont {S.~A.~M.}\ \bibnamefont {Loos}}, \bibinfo {author} {\bibfnamefont {S.~H.~L.}\ \bibnamefont {Klapp}},\ and\ \bibinfo {author} {\bibfnamefont {T.}~\bibnamefont {Martynec}},\ }\bibfield  {title} {\bibinfo {title} {Long-range order and directional defect propagation in the nonreciprocal {XY} model with vision cone interactions},\ }\href@noop {} {\bibfield  {journal} {\bibinfo  {journal} {Phys. Rev. Lett.}\ }\textbf {\bibinfo {volume} {130}},\ \bibinfo {pages} {198301} (\bibinfo {year} {2023})}\BibitemShut {NoStop}%
\bibitem [{\citenamefont {Rouzaire}\ \emph {et~al.}(2024)\citenamefont {Rouzaire}, \citenamefont {Levis},\ and\ \citenamefont {Pagonabarraga}}]{RouzaireLP2024}%
  \BibitemOpen
  \bibfield  {author} {\bibinfo {author} {\bibfnamefont {Y.}~\bibnamefont {Rouzaire}}, \bibinfo {author} {\bibfnamefont {D.}~\bibnamefont {Levis}},\ and\ \bibinfo {author} {\bibfnamefont {I.}~\bibnamefont {Pagonabarraga}},\ }\bibfield  {title} {\bibinfo {title} {Non-reciprocal interactions reshape topological defect annihilation},\ }\href@noop {} {\bibfield  {journal} {\bibinfo  {journal} {arXiv:2401.12637}\ } (\bibinfo {year} {2024})}\BibitemShut {NoStop}%
\bibitem [{\citenamefont {Bililign}\ \emph {et~al.}(2022)\citenamefont {Bililign}, \citenamefont {Usabiaga}, \citenamefont {Ganan}, \citenamefont {Poncet}, \citenamefont {Soni}, \citenamefont {Magkiriadou}, \citenamefont {Shelley}, \citenamefont {Bartolo},\ and\ \citenamefont {Irvine}}]{BililignNatPhys22}%
  \BibitemOpen
  \bibfield  {author} {\bibinfo {author} {\bibfnamefont {E.~S.}\ \bibnamefont {Bililign}}, \bibinfo {author} {\bibfnamefont {F.~B.}\ \bibnamefont {Usabiaga}}, \bibinfo {author} {\bibfnamefont {Y.~A.}\ \bibnamefont {Ganan}}, \bibinfo {author} {\bibfnamefont {A.}~\bibnamefont {Poncet}}, \bibinfo {author} {\bibfnamefont {V.}~\bibnamefont {Soni}}, \bibinfo {author} {\bibfnamefont {S.}~\bibnamefont {Magkiriadou}}, \bibinfo {author} {\bibfnamefont {M.~J.}\ \bibnamefont {Shelley}}, \bibinfo {author} {\bibfnamefont {D.}~\bibnamefont {Bartolo}},\ and\ \bibinfo {author} {\bibfnamefont {W.~T.~M.}\ \bibnamefont {Irvine}},\ }\bibfield  {title} {\bibinfo {title} {Motile dislocations knead odd crystals into whorls},\ }\href@noop {} {\bibfield  {journal} {\bibinfo  {journal} {Nat. Phys.}\ }\textbf {\bibinfo {volume} {18}},\ \bibinfo {pages} {212} (\bibinfo {year} {2022})}\BibitemShut {NoStop}%
\bibitem [{\citenamefont {Poncet}\ and\ \citenamefont {Bartolo}(2022)}]{PoncetPRL22}%
  \BibitemOpen
  \bibfield  {author} {\bibinfo {author} {\bibfnamefont {A.}~\bibnamefont {Poncet}}\ and\ \bibinfo {author} {\bibfnamefont {D.}~\bibnamefont {Bartolo}},\ }\bibfield  {title} {\bibinfo {title} {When soft crystals defy {Newton's} third law: Nonreciprocal mechanics and dislocation motility},\ }\href@noop {} {\bibfield  {journal} {\bibinfo  {journal} {Phys. Rev. Lett.}\ }\textbf {\bibinfo {volume} {128}},\ \bibinfo {pages} {048002} (\bibinfo {year} {2022})}\BibitemShut {NoStop}%
\bibitem [{\citenamefont {Veenstra}\ \emph {et~al.}(2024)\citenamefont {Veenstra}, \citenamefont {Gamayun}, \citenamefont {Guo}, \citenamefont {Sarvi}, \citenamefont {Meinersen},\ and\ \citenamefont {Coulais}}]{VeenstraGGSMC2024}%
  \BibitemOpen
  \bibfield  {author} {\bibinfo {author} {\bibfnamefont {J.}~\bibnamefont {Veenstra}}, \bibinfo {author} {\bibfnamefont {O.}~\bibnamefont {Gamayun}}, \bibinfo {author} {\bibfnamefont {X.}~\bibnamefont {Guo}}, \bibinfo {author} {\bibfnamefont {A.}~\bibnamefont {Sarvi}}, \bibinfo {author} {\bibfnamefont {C.~V.}\ \bibnamefont {Meinersen}},\ and\ \bibinfo {author} {\bibfnamefont {C.}~\bibnamefont {Coulais}},\ }\bibfield  {title} {\bibinfo {title} {Non-reciprocal topological solitons in active metamaterials},\ }\href@noop {} {\bibfield  {journal} {\bibinfo  {journal} {Nature}\ }\textbf {\bibinfo {volume} {627}},\ \bibinfo {pages} {528} (\bibinfo {year} {2024})}\BibitemShut {NoStop}%
\bibitem [{\citenamefont {Brandenbourger}\ \emph {et~al.}(2019)\citenamefont {Brandenbourger}, \citenamefont {Locsin}, \citenamefont {Lerner},\ and\ \citenamefont {Coulais}}]{BrandenbourgerLLC2019}%
  \BibitemOpen
  \bibfield  {author} {\bibinfo {author} {\bibfnamefont {M.}~\bibnamefont {Brandenbourger}}, \bibinfo {author} {\bibfnamefont {X.}~\bibnamefont {Locsin}}, \bibinfo {author} {\bibfnamefont {E.}~\bibnamefont {Lerner}},\ and\ \bibinfo {author} {\bibfnamefont {C.}~\bibnamefont {Coulais}},\ }\bibfield  {title} {\bibinfo {title} {Non-reciprocal robotic metamaterials},\ }\href@noop {} {\bibfield  {journal} {\bibinfo  {journal} {Nat. Commun.}\ }\textbf {\bibinfo {volume} {10}},\ \bibinfo {pages} {4608} (\bibinfo {year} {2019})}\BibitemShut {NoStop}%
\bibitem [{\citenamefont {Vossel}\ \emph {et~al.}(2023)\citenamefont {Vossel}, \citenamefont {de~Groot},\ and\ \citenamefont {Godec}}]{VosseldGG2023}%
  \BibitemOpen
  \bibfield  {author} {\bibinfo {author} {\bibfnamefont {M.}~\bibnamefont {Vossel}}, \bibinfo {author} {\bibfnamefont {B.~L.}\ \bibnamefont {de~Groot}},\ and\ \bibinfo {author} {\bibfnamefont {A.}~\bibnamefont {Godec}},\ }\bibfield  {title} {\bibinfo {title} {The allosteric lever: Towards a principle of specific allosteric response},\ }\href@noop {} {\bibfield  {journal} {\bibinfo  {journal} {arXiv:2311.12025}\ } (\bibinfo {year} {2023})}\BibitemShut {NoStop}%
\bibitem [{\citenamefont {Ivlev}\ \emph {et~al.}(2015)\citenamefont {Ivlev}, \citenamefont {Bartnick}, \citenamefont {Heinen}, \citenamefont {Du}, \citenamefont {Nosenko},\ and\ \citenamefont {L{\"o}wen}}]{IvlevEtAl2015}%
  \BibitemOpen
  \bibfield  {author} {\bibinfo {author} {\bibfnamefont {A.~V.}\ \bibnamefont {Ivlev}}, \bibinfo {author} {\bibfnamefont {J.}~\bibnamefont {Bartnick}}, \bibinfo {author} {\bibfnamefont {M.}~\bibnamefont {Heinen}}, \bibinfo {author} {\bibfnamefont {C.-R.}\ \bibnamefont {Du}}, \bibinfo {author} {\bibfnamefont {V.}~\bibnamefont {Nosenko}},\ and\ \bibinfo {author} {\bibfnamefont {H.}~\bibnamefont {L{\"o}wen}},\ }\bibfield  {title} {\bibinfo {title} {Statistical mechanics where {N}ewton's third law is broken},\ }\href@noop {} {\bibfield  {journal} {\bibinfo  {journal} {Phys. Rev. X}\ }\textbf {\bibinfo {volume} {5}},\ \bibinfo {pages} {011035} (\bibinfo {year} {2015})}\BibitemShut {NoStop}%
\bibitem [{\citenamefont {Bartnick}\ \emph {et~al.}(2016)\citenamefont {Bartnick}, \citenamefont {Kaiser}, \citenamefont {L{\"o}wen},\ and\ \citenamefont {Ivlev}}]{BartnickEtAl2016}%
  \BibitemOpen
  \bibfield  {author} {\bibinfo {author} {\bibfnamefont {J.}~\bibnamefont {Bartnick}}, \bibinfo {author} {\bibfnamefont {A.}~\bibnamefont {Kaiser}}, \bibinfo {author} {\bibfnamefont {H.}~\bibnamefont {L{\"o}wen}},\ and\ \bibinfo {author} {\bibfnamefont {A.~V.}\ \bibnamefont {Ivlev}},\ }\bibfield  {title} {\bibinfo {title} {Emerging activity in bilayered dispersions with wake-mediated interactions},\ }\href@noop {} {\bibfield  {journal} {\bibinfo  {journal} {J. Chem. Phys.}\ }\textbf {\bibinfo {volume} {144}},\ \bibinfo {pages} {224901} (\bibinfo {year} {2016})}\BibitemShut {NoStop}%
\bibitem [{\citenamefont {You}\ \emph {et~al.}(2020)\citenamefont {You}, \citenamefont {Baskaran},\ and\ \citenamefont {Marchetti}}]{YouPNAS20}%
  \BibitemOpen
  \bibfield  {author} {\bibinfo {author} {\bibfnamefont {Z.}~\bibnamefont {You}}, \bibinfo {author} {\bibfnamefont {A.}~\bibnamefont {Baskaran}},\ and\ \bibinfo {author} {\bibfnamefont {M.~C.}\ \bibnamefont {Marchetti}},\ }\bibfield  {title} {\bibinfo {title} {Nonreciprocity as a generic route to traveling states},\ }\href@noop {} {\bibfield  {journal} {\bibinfo  {journal} {Proc. Natl. Acad. Sci. U.S.A.}\ }\textbf {\bibinfo {volume} {117}},\ \bibinfo {pages} {19767} (\bibinfo {year} {2020})}\BibitemShut {NoStop}%
\bibitem [{\citenamefont {Saha}\ \emph {et~al.}(2020)\citenamefont {Saha}, \citenamefont {Agudo-Canalejo},\ and\ \citenamefont {Golestanian}}]{SahaAG2020}%
  \BibitemOpen
  \bibfield  {author} {\bibinfo {author} {\bibfnamefont {S.}~\bibnamefont {Saha}}, \bibinfo {author} {\bibfnamefont {J.}~\bibnamefont {Agudo-Canalejo}},\ and\ \bibinfo {author} {\bibfnamefont {R.}~\bibnamefont {Golestanian}},\ }\bibfield  {title} {\bibinfo {title} {Scalar active mixtures: The nonreciprocal {C}ahn-{H}illiard model},\ }\href@noop {} {\bibfield  {journal} {\bibinfo  {journal} {Phys. Rev. X}\ }\textbf {\bibinfo {volume} {10}},\ \bibinfo {pages} {041009} (\bibinfo {year} {2020})}\BibitemShut {NoStop}%
\bibitem [{\citenamefont {Fruchart}\ \emph {et~al.}(2021)\citenamefont {Fruchart}, \citenamefont {Hanai}, \citenamefont {Littlewood},\ and\ \citenamefont {Vitelli}}]{FruchartNature21}%
  \BibitemOpen
  \bibfield  {author} {\bibinfo {author} {\bibfnamefont {M.}~\bibnamefont {Fruchart}}, \bibinfo {author} {\bibfnamefont {R.}~\bibnamefont {Hanai}}, \bibinfo {author} {\bibfnamefont {P.~B.}\ \bibnamefont {Littlewood}},\ and\ \bibinfo {author} {\bibfnamefont {V.}~\bibnamefont {Vitelli}},\ }\bibfield  {title} {\bibinfo {title} {Non-reciprocal phase transitions},\ }\href@noop {} {\bibfield  {journal} {\bibinfo  {journal} {Nature}\ }\textbf {\bibinfo {volume} {592}},\ \bibinfo {pages} {363} (\bibinfo {year} {2021})}\BibitemShut {NoStop}%
\bibitem [{\citenamefont {Kreienkamp}\ and\ \citenamefont {Klapp}(2022)}]{KreienkampK2022}%
  \BibitemOpen
  \bibfield  {author} {\bibinfo {author} {\bibfnamefont {K.~L.}\ \bibnamefont {Kreienkamp}}\ and\ \bibinfo {author} {\bibfnamefont {S.~H.~L.}\ \bibnamefont {Klapp}},\ }\bibfield  {title} {\bibinfo {title} {Clustering and flocking of repulsive chiral active particles with non-reciprocal couplings},\ }\href@noop {} {\bibfield  {journal} {\bibinfo  {journal} {New J. Phys.}\ }\textbf {\bibinfo {volume} {24}},\ \bibinfo {pages} {123009} (\bibinfo {year} {2022})}\BibitemShut {NoStop}%
\bibitem [{\citenamefont {Frohoff-H{\"u}lsmann}\ and\ \citenamefont {Thiele}(2023)}]{FrohoffT2023}%
  \BibitemOpen
  \bibfield  {author} {\bibinfo {author} {\bibfnamefont {T.}~\bibnamefont {Frohoff-H{\"u}lsmann}}\ and\ \bibinfo {author} {\bibfnamefont {U.}~\bibnamefont {Thiele}},\ }\bibfield  {title} {\bibinfo {title} {Nonreciprocal {C}ahn-{H}illiard model emerges as a universal amplitude equation},\ }\href@noop {} {\bibfield  {journal} {\bibinfo  {journal} {Phys. Rev. Lett.}\ }\textbf {\bibinfo {volume} {131}},\ \bibinfo {pages} {107201} (\bibinfo {year} {2023})}\BibitemShut {NoStop}%
\bibitem [{\citenamefont {Suchanek}\ \emph {et~al.}(2023{\natexlab{a}})\citenamefont {Suchanek}, \citenamefont {Kroy},\ and\ \citenamefont {Loos}}]{SuchanekKL2023}%
  \BibitemOpen
  \bibfield  {author} {\bibinfo {author} {\bibfnamefont {T.}~\bibnamefont {Suchanek}}, \bibinfo {author} {\bibfnamefont {K.}~\bibnamefont {Kroy}},\ and\ \bibinfo {author} {\bibfnamefont {S.~A.~M.}\ \bibnamefont {Loos}},\ }\bibfield  {title} {\bibinfo {title} {Time-reversal and parity-time symmetry breaking in non-{H}ermitian field theories},\ }\href@noop {} {\bibfield  {journal} {\bibinfo  {journal} {Phys. Rev. E}\ }\textbf {\bibinfo {volume} {108}},\ \bibinfo {pages} {064123} (\bibinfo {year} {2023}{\natexlab{a}})}\BibitemShut {NoStop}%
\bibitem [{\citenamefont {Frohoff-H{\"u}lsmann}\ \emph {et~al.}(2023)\citenamefont {Frohoff-H{\"u}lsmann}, \citenamefont {Thiele},\ and\ \citenamefont {Pismen}}]{FrohoffTP2023}%
  \BibitemOpen
  \bibfield  {author} {\bibinfo {author} {\bibfnamefont {T.}~\bibnamefont {Frohoff-H{\"u}lsmann}}, \bibinfo {author} {\bibfnamefont {U.}~\bibnamefont {Thiele}},\ and\ \bibinfo {author} {\bibfnamefont {L.~M.}\ \bibnamefont {Pismen}},\ }\bibfield  {title} {\bibinfo {title} {Non-reciprocity induces resonances in a two-field {C}ahn--{H}illiard model},\ }\href@noop {} {\bibfield  {journal} {\bibinfo  {journal} {Philos. Trans. R. Soc. A}\ }\textbf {\bibinfo {volume} {381}},\ \bibinfo {pages} {20220087} (\bibinfo {year} {2023})}\BibitemShut {NoStop}%
\bibitem [{\citenamefont {Brauns}\ and\ \citenamefont {Marchetti}(2024)}]{BraunsPRX24}%
  \BibitemOpen
  \bibfield  {author} {\bibinfo {author} {\bibfnamefont {F.}~\bibnamefont {Brauns}}\ and\ \bibinfo {author} {\bibfnamefont {M.~C.}\ \bibnamefont {Marchetti}},\ }\bibfield  {title} {\bibinfo {title} {Nonreciprocal pattern formation of conserved fields},\ }\href@noop {} {\bibfield  {journal} {\bibinfo  {journal} {Phys. Rev. X}\ }\textbf {\bibinfo {volume} {14}},\ \bibinfo {pages} {021014} (\bibinfo {year} {2024})}\BibitemShut {NoStop}%
\bibitem [{\citenamefont {Cates}(2022)}]{Cates2019b}%
  \BibitemOpen
  \bibfield  {author} {\bibinfo {author} {\bibfnamefont {M.~E.}\ \bibnamefont {Cates}},\ }\bibfield  {title} {\bibinfo {title} {Active field theories},\ }in\ \href@noop {} {\emph {\bibinfo {booktitle} {Active Matter and Nonequilibrium Statistical Physics, Lecture Notes of the Les Houches Summer School: Volume 112, September 2018}}},\ \bibinfo {editor} {edited by\ \bibinfo {editor} {\bibfnamefont {J.}~\bibnamefont {Tailleur}}, \bibinfo {editor} {\bibfnamefont {G.}~\bibnamefont {Gompper}}, \bibinfo {editor} {\bibfnamefont {M.~C.}\ \bibnamefont {Marchetti}}, \bibinfo {editor} {\bibfnamefont {J.~M.}\ \bibnamefont {Yeomans}},\ and\ \bibinfo {editor} {\bibfnamefont {C.}~\bibnamefont {Salomon}}}\ (\bibinfo  {publisher} {Oxford University Press},\ \bibinfo {address} {Oxford},\ \bibinfo {year} {2022})\ pp.\ \bibinfo {pages} {180--216}\BibitemShut {NoStop}%
\bibitem [{\citenamefont {{te Vrugt}}\ \emph {et~al.}(2023{\natexlab{a}})\citenamefont {{te Vrugt}}, \citenamefont {Bickmann},\ and\ \citenamefont {Wittkowski}}]{teVrugtBW2022}%
  \BibitemOpen
  \bibfield  {author} {\bibinfo {author} {\bibfnamefont {M.}~\bibnamefont {{te Vrugt}}}, \bibinfo {author} {\bibfnamefont {J.}~\bibnamefont {Bickmann}},\ and\ \bibinfo {author} {\bibfnamefont {R.}~\bibnamefont {Wittkowski}},\ }\bibfield  {title} {\bibinfo {title} {How to derive a predictive field theory for active {B}rownian particles: a step-by-step tutorial},\ }\href@noop {} {\bibfield  {journal} {\bibinfo  {journal} {J. Phys. Condens. Matter}\ }\textbf {\bibinfo {volume} {35}},\ \bibinfo {pages} {313001} (\bibinfo {year} {2023}{\natexlab{a}})}\BibitemShut {NoStop}%
\bibitem [{\citenamefont {Doostmohammadi}\ \emph {et~al.}(2018)\citenamefont {Doostmohammadi}, \citenamefont {Ign{\'e}s-Mullol}, \citenamefont {Yeomans},\ and\ \citenamefont {Sagu{\'e}s}}]{DoostmohammadiIYS2018}%
  \BibitemOpen
  \bibfield  {author} {\bibinfo {author} {\bibfnamefont {A.}~\bibnamefont {Doostmohammadi}}, \bibinfo {author} {\bibfnamefont {J.}~\bibnamefont {Ign{\'e}s-Mullol}}, \bibinfo {author} {\bibfnamefont {J.~M.}\ \bibnamefont {Yeomans}},\ and\ \bibinfo {author} {\bibfnamefont {F.}~\bibnamefont {Sagu{\'e}s}},\ }\bibfield  {title} {\bibinfo {title} {Active nematics},\ }\href@noop {} {\bibfield  {journal} {\bibinfo  {journal} {Nat. Commun.}\ }\textbf {\bibinfo {volume} {9}},\ \bibinfo {pages} {3246} (\bibinfo {year} {2018})}\BibitemShut {NoStop}%
\bibitem [{\citenamefont {Bickmann}\ and\ \citenamefont {Wittkowski}(2020)}]{BickmannW2020}%
  \BibitemOpen
  \bibfield  {author} {\bibinfo {author} {\bibfnamefont {J.}~\bibnamefont {Bickmann}}\ and\ \bibinfo {author} {\bibfnamefont {R.}~\bibnamefont {Wittkowski}},\ }\bibfield  {title} {\bibinfo {title} {Predictive local field theory for interacting active {B}rownian spheres in two spatial dimensions},\ }\href@noop {} {\bibfield  {journal} {\bibinfo  {journal} {J. Phys. Condens. Matter}\ }\textbf {\bibinfo {volume} {32}},\ \bibinfo {pages} {214001} (\bibinfo {year} {2020})}\BibitemShut {NoStop}%
\bibitem [{\citenamefont {Wittkowski}\ \emph {et~al.}(2014)\citenamefont {Wittkowski}, \citenamefont {Tiribocchi}, \citenamefont {Stenhammar}, \citenamefont {Allen}, \citenamefont {Marenduzzo},\ and\ \citenamefont {Cates}}]{WittkowskiTSAMC2014}%
  \BibitemOpen
  \bibfield  {author} {\bibinfo {author} {\bibfnamefont {R.}~\bibnamefont {Wittkowski}}, \bibinfo {author} {\bibfnamefont {A.}~\bibnamefont {Tiribocchi}}, \bibinfo {author} {\bibfnamefont {J.}~\bibnamefont {Stenhammar}}, \bibinfo {author} {\bibfnamefont {R.~J.}\ \bibnamefont {Allen}}, \bibinfo {author} {\bibfnamefont {D.}~\bibnamefont {Marenduzzo}},\ and\ \bibinfo {author} {\bibfnamefont {M.~E.}\ \bibnamefont {Cates}},\ }\bibfield  {title} {\bibinfo {title} {Scalar $\phi^4$ field theory for active-particle phase separation},\ }\href@noop {} {\bibfield  {journal} {\bibinfo  {journal} {Nat. Commun.}\ }\textbf {\bibinfo {volume} {5}},\ \bibinfo {pages} {4351} (\bibinfo {year} {2014})}\BibitemShut {NoStop}%
\bibitem [{\citenamefont {Tiribocchi}\ \emph {et~al.}(2015)\citenamefont {Tiribocchi}, \citenamefont {Wittkowski}, \citenamefont {Marenduzzo},\ and\ \citenamefont {Cates}}]{TiribocchiPRL15}%
  \BibitemOpen
  \bibfield  {author} {\bibinfo {author} {\bibfnamefont {A.}~\bibnamefont {Tiribocchi}}, \bibinfo {author} {\bibfnamefont {R.}~\bibnamefont {Wittkowski}}, \bibinfo {author} {\bibfnamefont {D.}~\bibnamefont {Marenduzzo}},\ and\ \bibinfo {author} {\bibfnamefont {M.~E.}\ \bibnamefont {Cates}},\ }\bibfield  {title} {\bibinfo {title} {Active model {H}: Scalar active matter in a momentum-conserving fluid},\ }\href@noop {} {\bibfield  {journal} {\bibinfo  {journal} {Phys. Rev. Lett.}\ }\textbf {\bibinfo {volume} {115}},\ \bibinfo {pages} {188302} (\bibinfo {year} {2015})}\BibitemShut {NoStop}%
\bibitem [{\citenamefont {Tjhung}\ \emph {et~al.}(2018)\citenamefont {Tjhung}, \citenamefont {Nardini},\ and\ \citenamefont {Cates}}]{TjhungNC2018}%
  \BibitemOpen
  \bibfield  {author} {\bibinfo {author} {\bibfnamefont {E.}~\bibnamefont {Tjhung}}, \bibinfo {author} {\bibfnamefont {C.}~\bibnamefont {Nardini}},\ and\ \bibinfo {author} {\bibfnamefont {M.~E.}\ \bibnamefont {Cates}},\ }\bibfield  {title} {\bibinfo {title} {Cluster phases and bubbly phase separation in active fluids: reversal of the {O}stwald process},\ }\href@noop {} {\bibfield  {journal} {\bibinfo  {journal} {Phys. Rev. X}\ }\textbf {\bibinfo {volume} {8}},\ \bibinfo {pages} {031080} (\bibinfo {year} {2018})}\BibitemShut {NoStop}%
\bibitem [{\citenamefont {{te Vrugt}}\ \emph {et~al.}(2023{\natexlab{b}})\citenamefont {{te Vrugt}}, \citenamefont {Frohoff-H{\"u}lsmann}, \citenamefont {Heifetz}, \citenamefont {Thiele},\ and\ \citenamefont {Wittkowski}}]{teVrugtFHHTW2023}%
  \BibitemOpen
  \bibfield  {author} {\bibinfo {author} {\bibfnamefont {M.}~\bibnamefont {{te Vrugt}}}, \bibinfo {author} {\bibfnamefont {T.}~\bibnamefont {Frohoff-H{\"u}lsmann}}, \bibinfo {author} {\bibfnamefont {E.}~\bibnamefont {Heifetz}}, \bibinfo {author} {\bibfnamefont {U.}~\bibnamefont {Thiele}},\ and\ \bibinfo {author} {\bibfnamefont {R.}~\bibnamefont {Wittkowski}},\ }\bibfield  {title} {\bibinfo {title} {From a microscopic inertial active matter model to the {S}chr\"odinger equation},\ }\href@noop {} {\bibfield  {journal} {\bibinfo  {journal} {Nat. Commun.}\ }\textbf {\bibinfo {volume} {14}},\ \bibinfo {pages} {1302} (\bibinfo {year} {2023}{\natexlab{b}})}\BibitemShut {NoStop}%
\bibitem [{\citenamefont {Menzel}\ and\ \citenamefont {L{\"o}wen}(2013)}]{MenzelPRL13}%
  \BibitemOpen
  \bibfield  {author} {\bibinfo {author} {\bibfnamefont {A.~M.}\ \bibnamefont {Menzel}}\ and\ \bibinfo {author} {\bibfnamefont {H.}~\bibnamefont {L{\"o}wen}},\ }\bibfield  {title} {\bibinfo {title} {Traveling and resting crystals in active systems},\ }\href@noop {} {\bibfield  {journal} {\bibinfo  {journal} {Phys. Rev. Lett.}\ }\textbf {\bibinfo {volume} {110}},\ \bibinfo {pages} {055702} (\bibinfo {year} {2013})}\BibitemShut {NoStop}%
\bibitem [{\citenamefont {Menzel}\ \emph {et~al.}(2014)\citenamefont {Menzel}, \citenamefont {Ohta},\ and\ \citenamefont {L{\"o}wen}}]{MenzelPRE14}%
  \BibitemOpen
  \bibfield  {author} {\bibinfo {author} {\bibfnamefont {A.~M.}\ \bibnamefont {Menzel}}, \bibinfo {author} {\bibfnamefont {T.}~\bibnamefont {Ohta}},\ and\ \bibinfo {author} {\bibfnamefont {H.}~\bibnamefont {L{\"o}wen}},\ }\bibfield  {title} {\bibinfo {title} {Active crystals and their stability},\ }\href@noop {} {\bibfield  {journal} {\bibinfo  {journal} {Phys. Rev. E}\ }\textbf {\bibinfo {volume} {89}},\ \bibinfo {pages} {022301} (\bibinfo {year} {2014})}\BibitemShut {NoStop}%
\bibitem [{\citenamefont {Huang}\ \emph {et~al.}(2020)\citenamefont {Huang}, \citenamefont {Menzel},\ and\ \citenamefont {L{\"o}wen}}]{HuangPRL20}%
  \BibitemOpen
  \bibfield  {author} {\bibinfo {author} {\bibfnamefont {Z.-F.}\ \bibnamefont {Huang}}, \bibinfo {author} {\bibfnamefont {A.~M.}\ \bibnamefont {Menzel}},\ and\ \bibinfo {author} {\bibfnamefont {H.}~\bibnamefont {L{\"o}wen}},\ }\bibfield  {title} {\bibinfo {title} {Dynamical crystallites of active chiral particles},\ }\href@noop {} {\bibfield  {journal} {\bibinfo  {journal} {Phys. Rev. Lett.}\ }\textbf {\bibinfo {volume} {125}},\ \bibinfo {pages} {218002} (\bibinfo {year} {2020})}\BibitemShut {NoStop}%
\bibitem [{\citenamefont {Huang}\ \emph {et~al.}(2022)\citenamefont {Huang}, \citenamefont {L{\"o}wen},\ and\ \citenamefont {Voigt}}]{HuangCP22}%
  \BibitemOpen
  \bibfield  {author} {\bibinfo {author} {\bibfnamefont {Z.-F.}\ \bibnamefont {Huang}}, \bibinfo {author} {\bibfnamefont {H.}~\bibnamefont {L{\"o}wen}},\ and\ \bibinfo {author} {\bibfnamefont {A.}~\bibnamefont {Voigt}},\ }\bibfield  {title} {\bibinfo {title} {Defect dynamics in active smectics induced by confining geometry and topology},\ }\href@noop {} {\bibfield  {journal} {\bibinfo  {journal} {Commun. Phys.}\ }\textbf {\bibinfo {volume} {5}},\ \bibinfo {pages} {294} (\bibinfo {year} {2022})}\BibitemShut {NoStop}%
\bibitem [{\citenamefont {Nestler}\ \emph {et~al.}(2024)\citenamefont {Nestler}, \citenamefont {Praetorius}, \citenamefont {Huang}, \citenamefont {L{\"o}wen},\ and\ \citenamefont {Voigt}}]{NestlerJPCM24}%
  \BibitemOpen
  \bibfield  {author} {\bibinfo {author} {\bibfnamefont {M.}~\bibnamefont {Nestler}}, \bibinfo {author} {\bibfnamefont {S.}~\bibnamefont {Praetorius}}, \bibinfo {author} {\bibfnamefont {Z.-F.}\ \bibnamefont {Huang}}, \bibinfo {author} {\bibfnamefont {H.}~\bibnamefont {L{\"o}wen}},\ and\ \bibinfo {author} {\bibfnamefont {A.}~\bibnamefont {Voigt}},\ }\bibfield  {title} {\bibinfo {title} {Active smectics on a sphere},\ }\href@noop {} {\bibfield  {journal} {\bibinfo  {journal} {J. Phys. Condens. Matter}\ }\textbf {\bibinfo {volume} {36}},\ \bibinfo {pages} {185001} (\bibinfo {year} {2024})}\BibitemShut {NoStop}%
\bibitem [{\citenamefont {Suchanek}\ \emph {et~al.}(2023{\natexlab{b}})\citenamefont {Suchanek}, \citenamefont {Kroy},\ and\ \citenamefont {Loos}}]{SuchanekKL2023b}%
  \BibitemOpen
  \bibfield  {author} {\bibinfo {author} {\bibfnamefont {T.}~\bibnamefont {Suchanek}}, \bibinfo {author} {\bibfnamefont {K.}~\bibnamefont {Kroy}},\ and\ \bibinfo {author} {\bibfnamefont {S.~A.~M.}\ \bibnamefont {Loos}},\ }\bibfield  {title} {\bibinfo {title} {Entropy production in the nonreciprocal {C}ahn-{H}illiard model},\ }\href@noop {} {\bibfield  {journal} {\bibinfo  {journal} {Phys. Rev. E}\ }\textbf {\bibinfo {volume} {108}},\ \bibinfo {pages} {064610} (\bibinfo {year} {2023}{\natexlab{b}})}\BibitemShut {NoStop}%
\bibitem [{\citenamefont {Hansen}\ and\ \citenamefont {McDonald}(2009)}]{HansenMD2009}%
  \BibitemOpen
  \bibfield  {author} {\bibinfo {author} {\bibfnamefont {J.-P.}\ \bibnamefont {Hansen}}\ and\ \bibinfo {author} {\bibfnamefont {I.~R.}\ \bibnamefont {McDonald}},\ }\href@noop {} {\emph {\bibinfo {title} {Theory of Simple Liquids: with Applications to Soft Matter}}},\ \bibinfo {edition} {4th}\ ed.\ (\bibinfo  {publisher} {Elsevier Academic Press},\ \bibinfo {address} {Oxford},\ \bibinfo {year} {2009})\BibitemShut {NoStop}%
\bibitem [{\citenamefont {{te Vrugt}}\ \emph {et~al.}(2020)\citenamefont {{te Vrugt}}, \citenamefont {L{\"o}wen},\ and\ \citenamefont {Wittkowski}}]{teVrugtLW2020}%
  \BibitemOpen
  \bibfield  {author} {\bibinfo {author} {\bibfnamefont {M.}~\bibnamefont {{te Vrugt}}}, \bibinfo {author} {\bibfnamefont {H.}~\bibnamefont {L{\"o}wen}},\ and\ \bibinfo {author} {\bibfnamefont {R.}~\bibnamefont {Wittkowski}},\ }\bibfield  {title} {\bibinfo {title} {Classical dynamical density functional theory: from fundamentals to applications},\ }\href@noop {} {\bibfield  {journal} {\bibinfo  {journal} {Adv. Phys.}\ }\textbf {\bibinfo {volume} {69}},\ \bibinfo {pages} {121} (\bibinfo {year} {2020})}\BibitemShut {NoStop}%
\bibitem [{\citenamefont {Jeggle}\ \emph {et~al.}(2020)\citenamefont {Jeggle}, \citenamefont {Stenhammar},\ and\ \citenamefont {Wittkowski}}]{JeggleSW2020}%
  \BibitemOpen
  \bibfield  {author} {\bibinfo {author} {\bibfnamefont {J.}~\bibnamefont {Jeggle}}, \bibinfo {author} {\bibfnamefont {J.}~\bibnamefont {Stenhammar}},\ and\ \bibinfo {author} {\bibfnamefont {R.}~\bibnamefont {Wittkowski}},\ }\bibfield  {title} {\bibinfo {title} {Pair-distribution function of active {B}rownian spheres in two spatial dimensions: Simulation results and analytic representation},\ }\href@noop {} {\bibfield  {journal} {\bibinfo  {journal} {J. Chem. Phys.}\ }\textbf {\bibinfo {volume} {152}},\ \bibinfo {pages} {194903} (\bibinfo {year} {2020})}\BibitemShut {NoStop}%
\bibitem [{\citenamefont {Br{\"o}ker}\ \emph {et~al.}(2024)\citenamefont {Br{\"o}ker}, \citenamefont {te~Vrugt}, \citenamefont {Jeggle}, \citenamefont {Stenhammar},\ and\ \citenamefont {Wittkowski}}]{BrokertVJSW2024}%
  \BibitemOpen
  \bibfield  {author} {\bibinfo {author} {\bibfnamefont {S.}~\bibnamefont {Br{\"o}ker}}, \bibinfo {author} {\bibfnamefont {M.}~\bibnamefont {te~Vrugt}}, \bibinfo {author} {\bibfnamefont {J.}~\bibnamefont {Jeggle}}, \bibinfo {author} {\bibfnamefont {J.}~\bibnamefont {Stenhammar}},\ and\ \bibinfo {author} {\bibfnamefont {R.}~\bibnamefont {Wittkowski}},\ }\bibfield  {title} {\bibinfo {title} {Pair-distribution function of active {B}rownian spheres in three spatial dimensions: simulation results and analytical representation},\ }\href@noop {} {\bibfield  {journal} {\bibinfo  {journal} {Soft Matter}\ }\textbf {\bibinfo {volume} {20}},\ \bibinfo {pages} {224} (\bibinfo {year} {2024})}\BibitemShut {NoStop}%
\bibitem [{\citenamefont {Louis}\ \emph {et~al.}(2000)\citenamefont {Louis}, \citenamefont {Bolhuis},\ and\ \citenamefont {Hansen}}]{LouisBG2000}%
  \BibitemOpen
  \bibfield  {author} {\bibinfo {author} {\bibfnamefont {A.~A.}\ \bibnamefont {Louis}}, \bibinfo {author} {\bibfnamefont {P.~G.}\ \bibnamefont {Bolhuis}},\ and\ \bibinfo {author} {\bibfnamefont {J.~P.}\ \bibnamefont {Hansen}},\ }\bibfield  {title} {\bibinfo {title} {Mean-field fluid behavior of the {G}aussian core model},\ }\href@noop {} {\bibfield  {journal} {\bibinfo  {journal} {Phys. Rev. E}\ }\textbf {\bibinfo {volume} {62}},\ \bibinfo {pages} {7961} (\bibinfo {year} {2000})}\BibitemShut {NoStop}%
\bibitem [{\citenamefont {Reinken}\ \emph {et~al.}(2018)\citenamefont {Reinken}, \citenamefont {Klapp}, \citenamefont {B{\"a}r},\ and\ \citenamefont {Heidenreich}}]{ReinkenKBH2018}%
  \BibitemOpen
  \bibfield  {author} {\bibinfo {author} {\bibfnamefont {H.}~\bibnamefont {Reinken}}, \bibinfo {author} {\bibfnamefont {S.~H.~L.}\ \bibnamefont {Klapp}}, \bibinfo {author} {\bibfnamefont {M.}~\bibnamefont {B{\"a}r}},\ and\ \bibinfo {author} {\bibfnamefont {S.}~\bibnamefont {Heidenreich}},\ }\bibfield  {title} {\bibinfo {title} {Derivation of a hydrodynamic theory for mesoscale dynamics in microswimmer suspensions},\ }\href@noop {} {\bibfield  {journal} {\bibinfo  {journal} {Phys. Rev. E}\ }\textbf {\bibinfo {volume} {97}},\ \bibinfo {pages} {022613} (\bibinfo {year} {2018})}\BibitemShut {NoStop}%
\bibitem [{\citenamefont {{te Vrugt}}\ and\ \citenamefont {Wittkowski}(2020)}]{teVrugtW2020b}%
  \BibitemOpen
  \bibfield  {author} {\bibinfo {author} {\bibfnamefont {M.}~\bibnamefont {{te Vrugt}}}\ and\ \bibinfo {author} {\bibfnamefont {R.}~\bibnamefont {Wittkowski}},\ }\bibfield  {title} {\bibinfo {title} {Relations between angular and {C}artesian orientational expansions},\ }\href@noop {} {\bibfield  {journal} {\bibinfo  {journal} {AIP Adv.}\ }\textbf {\bibinfo {volume} {10}},\ \bibinfo {pages} {035106} (\bibinfo {year} {2020})}\BibitemShut {NoStop}%
\bibitem [{\citenamefont {Archer}\ \emph {et~al.}(2019)\citenamefont {Archer}, \citenamefont {Ratliff}, \citenamefont {Rucklidge},\ and\ \citenamefont {Subramanian}}]{ArcherRRS2019}%
  \BibitemOpen
  \bibfield  {author} {\bibinfo {author} {\bibfnamefont {A.~J.}\ \bibnamefont {Archer}}, \bibinfo {author} {\bibfnamefont {D.~J.}\ \bibnamefont {Ratliff}}, \bibinfo {author} {\bibfnamefont {A.~M.}\ \bibnamefont {Rucklidge}},\ and\ \bibinfo {author} {\bibfnamefont {P.}~\bibnamefont {Subramanian}},\ }\bibfield  {title} {\bibinfo {title} {Deriving phase field crystal theory from dynamical density functional theory: consequences of the approximations},\ }\href@noop {} {\bibfield  {journal} {\bibinfo  {journal} {Phys. Rev. E}\ }\textbf {\bibinfo {volume} {100}},\ \bibinfo {pages} {022140} (\bibinfo {year} {2019})}\BibitemShut {NoStop}%
\bibitem [{\citenamefont {Robinson}\ \emph {et~al.}(2024)\citenamefont {Robinson}, \citenamefont {Machon},\ and\ \citenamefont {Speck}}]{RobinsonMS2024}%
  \BibitemOpen
  \bibfield  {author} {\bibinfo {author} {\bibfnamefont {J.~F.}\ \bibnamefont {Robinson}}, \bibinfo {author} {\bibfnamefont {T.}~\bibnamefont {Machon}},\ and\ \bibinfo {author} {\bibfnamefont {T.}~\bibnamefont {Speck}},\ }\bibfield  {title} {\bibinfo {title} {Universal limiting behaviour of reaction-diffusion systems with conservation laws},\ }\href@noop {} {\bibfield  {journal} {\bibinfo  {journal} {arXiv:2406.02409}\ } (\bibinfo {year} {2024})}\BibitemShut {NoStop}%
\bibitem [{\citenamefont {De~Luca}\ \emph {et~al.}(2024)\citenamefont {De~Luca}, \citenamefont {Maryshev},\ and\ \citenamefont {Frey}}]{DeLucaMF2024}%
  \BibitemOpen
  \bibfield  {author} {\bibinfo {author} {\bibfnamefont {F.}~\bibnamefont {De~Luca}}, \bibinfo {author} {\bibfnamefont {I.}~\bibnamefont {Maryshev}},\ and\ \bibinfo {author} {\bibfnamefont {E.}~\bibnamefont {Frey}},\ }\bibfield  {title} {\bibinfo {title} {Supramolecular assemblies in active motor-filament systems: micelles, bilayers, and foams},\ }\href@noop {} {\bibfield  {journal} {\bibinfo  {journal} {arXiv:2401.05070}\ } (\bibinfo {year} {2024})}\BibitemShut {NoStop}%
\bibitem [{\citenamefont {Zhang}\ \emph {et~al.}(2021)\citenamefont {Zhang}, \citenamefont {Alert}, \citenamefont {Yan}, \citenamefont {Wingreen},\ and\ \citenamefont {Granick}}]{ZhangNatPhys21}%
  \BibitemOpen
  \bibfield  {author} {\bibinfo {author} {\bibfnamefont {J.}~\bibnamefont {Zhang}}, \bibinfo {author} {\bibfnamefont {R.}~\bibnamefont {Alert}}, \bibinfo {author} {\bibfnamefont {J.}~\bibnamefont {Yan}}, \bibinfo {author} {\bibfnamefont {N.~S.}\ \bibnamefont {Wingreen}},\ and\ \bibinfo {author} {\bibfnamefont {S.}~\bibnamefont {Granick}},\ }\bibfield  {title} {\bibinfo {title} {Active phase separation by turning towards regions of higher density},\ }\href@noop {} {\bibfield  {journal} {\bibinfo  {journal} {Nat. Phys.}\ }\textbf {\bibinfo {volume} {17}},\ \bibinfo {pages} {961} (\bibinfo {year} {2021})}\BibitemShut {NoStop}%
\bibitem [{\citenamefont {Dauchot}(2021)}]{DauchotNatPhys21}%
  \BibitemOpen
  \bibfield  {author} {\bibinfo {author} {\bibfnamefont {O.}~\bibnamefont {Dauchot}},\ }\bibfield  {title} {\bibinfo {title} {Turn towards the crowd},\ }\href@noop {} {\bibfield  {journal} {\bibinfo  {journal} {Nat. Phys.}\ }\textbf {\bibinfo {volume} {17}},\ \bibinfo {pages} {875} (\bibinfo {year} {2021})}\BibitemShut {NoStop}%
\bibitem [{\citenamefont {{Cates}}\ and\ \citenamefont {{Tailleur}}(2015)}]{CatesT2015}%
  \BibitemOpen
  \bibfield  {author} {\bibinfo {author} {\bibfnamefont {M.~E.}\ \bibnamefont {{Cates}}}\ and\ \bibinfo {author} {\bibfnamefont {J.}~\bibnamefont {{Tailleur}}},\ }\bibfield  {title} {\bibinfo {title} {Motility-induced phase separation},\ }\href@noop {} {\bibfield  {journal} {\bibinfo  {journal} {Annu. Rev. Condens. Matter Phys.}\ }\textbf {\bibinfo {volume} {6}},\ \bibinfo {pages} {219} (\bibinfo {year} {2015})}\BibitemShut {NoStop}%
\bibitem [{\citenamefont {Bray}(1994)}]{BrayAP94}%
  \BibitemOpen
  \bibfield  {author} {\bibinfo {author} {\bibfnamefont {A.~J.}\ \bibnamefont {Bray}},\ }\bibfield  {title} {\bibinfo {title} {Theory of phase ordering kinetics},\ }\href@noop {} {\bibfield  {journal} {\bibinfo  {journal} {Adv. Phys.}\ }\textbf {\bibinfo {volume} {43}},\ \bibinfo {pages} {357} (\bibinfo {year} {1994})}\BibitemShut {NoStop}%
\bibitem [{\citenamefont {Caballero}\ and\ \citenamefont {Marchetti}(2022)}]{CaballeroPRL22}%
  \BibitemOpen
  \bibfield  {author} {\bibinfo {author} {\bibfnamefont {F.}~\bibnamefont {Caballero}}\ and\ \bibinfo {author} {\bibfnamefont {M.~C.}\ \bibnamefont {Marchetti}},\ }\bibfield  {title} {\bibinfo {title} {Activity-suppressed phase separation},\ }\href@noop {} {\bibfield  {journal} {\bibinfo  {journal} {Phys. Rev. Lett.}\ }\textbf {\bibinfo {volume} {129}},\ \bibinfo {pages} {268002} (\bibinfo {year} {2022})}\BibitemShut {NoStop}%
\bibitem [{\citenamefont {Caprini}\ \emph {et~al.}(2020)\citenamefont {Caprini}, \citenamefont {Marini Bettolo~Marconi},\ and\ \citenamefont {Puglisi}}]{CapriniMP2020}%
  \BibitemOpen
  \bibfield  {author} {\bibinfo {author} {\bibfnamefont {L.}~\bibnamefont {Caprini}}, \bibinfo {author} {\bibfnamefont {U.}~\bibnamefont {Marini Bettolo~Marconi}},\ and\ \bibinfo {author} {\bibfnamefont {A.}~\bibnamefont {Puglisi}},\ }\bibfield  {title} {\bibinfo {title} {Spontaneous velocity alignment in motility-induced phase separation},\ }\href@noop {} {\bibfield  {journal} {\bibinfo  {journal} {Phys. Rev. Lett.}\ }\textbf {\bibinfo {volume} {124}},\ \bibinfo {pages} {078001} (\bibinfo {year} {2020})}\BibitemShut {NoStop}%
\bibitem [{\citenamefont {Das}\ \emph {et~al.}(2024)\citenamefont {Das}, \citenamefont {Ciarchi}, \citenamefont {Zhou}, \citenamefont {Yan}, \citenamefont {Zhang},\ and\ \citenamefont {Alert}}]{Das2024}%
  \BibitemOpen
  \bibfield  {author} {\bibinfo {author} {\bibfnamefont {S.}~\bibnamefont {Das}}, \bibinfo {author} {\bibfnamefont {M.}~\bibnamefont {Ciarchi}}, \bibinfo {author} {\bibfnamefont {Z.}~\bibnamefont {Zhou}}, \bibinfo {author} {\bibfnamefont {J.}~\bibnamefont {Yan}}, \bibinfo {author} {\bibfnamefont {J.}~\bibnamefont {Zhang}},\ and\ \bibinfo {author} {\bibfnamefont {R.}~\bibnamefont {Alert}},\ }\bibfield  {title} {\bibinfo {title} {Flocking by turning away},\ }\href@noop {} {\bibfield  {journal} {\bibinfo  {journal} {Phys. Rev. X (in press, arXiv:2401.17153)}\ } (\bibinfo {year} {2024})}\BibitemShut {NoStop}%
\bibitem [{\citenamefont {Heckenthaler}\ \emph {et~al.}(2023)\citenamefont {Heckenthaler}, \citenamefont {Holder}, \citenamefont {Amir}, \citenamefont {Feinerman},\ and\ \citenamefont {Fonio}}]{HeckenthalerPRXLife23}%
  \BibitemOpen
  \bibfield  {author} {\bibinfo {author} {\bibfnamefont {T.}~\bibnamefont {Heckenthaler}}, \bibinfo {author} {\bibfnamefont {T.}~\bibnamefont {Holder}}, \bibinfo {author} {\bibfnamefont {A.}~\bibnamefont {Amir}}, \bibinfo {author} {\bibfnamefont {O.}~\bibnamefont {Feinerman}},\ and\ \bibinfo {author} {\bibfnamefont {E.}~\bibnamefont {Fonio}},\ }\bibfield  {title} {\bibinfo {title} {Connecting cooperative transport by ants with the physics of self-propelled particles},\ }\href@noop {} {\bibfield  {journal} {\bibinfo  {journal} {PRX Life}\ }\textbf {\bibinfo {volume} {1}},\ \bibinfo {pages} {023001} (\bibinfo {year} {2023})}\BibitemShut {NoStop}%
\bibitem [{\citenamefont {Zitterbart}\ \emph {et~al.}(2011)\citenamefont {Zitterbart}, \citenamefont {Wienecke}, \citenamefont {Butler},\ and\ \citenamefont {Fabry}}]{ZitterbartWBF2011}%
  \BibitemOpen
  \bibfield  {author} {\bibinfo {author} {\bibfnamefont {D.~P.}\ \bibnamefont {Zitterbart}}, \bibinfo {author} {\bibfnamefont {B.}~\bibnamefont {Wienecke}}, \bibinfo {author} {\bibfnamefont {J.~P.}\ \bibnamefont {Butler}},\ and\ \bibinfo {author} {\bibfnamefont {B.}~\bibnamefont {Fabry}},\ }\bibfield  {title} {\bibinfo {title} {Coordinated movements prevent jamming in an emperor penguin huddle},\ }\href@noop {} {\bibfield  {journal} {\bibinfo  {journal} {PLoS One}\ }\textbf {\bibinfo {volume} {6}},\ \bibinfo {pages} {e20260} (\bibinfo {year} {2011})}\BibitemShut {NoStop}%
\bibitem [{\citenamefont {Schadschneider}\ and\ \citenamefont {Seyfried}(2011)}]{SchadschneiderS2011}%
  \BibitemOpen
  \bibfield  {author} {\bibinfo {author} {\bibfnamefont {A.}~\bibnamefont {Schadschneider}}\ and\ \bibinfo {author} {\bibfnamefont {A.}~\bibnamefont {Seyfried}},\ }\bibfield  {title} {\bibinfo {title} {Empirical results for pedestrian dynamics and their implications for modeling},\ }\href@noop {} {\bibfield  {journal} {\bibinfo  {journal} {Netw. Heterog. Media}\ }\textbf {\bibinfo {volume} {6}},\ \bibinfo {pages} {545} (\bibinfo {year} {2011})}\BibitemShut {NoStop}%
\bibitem [{\citenamefont {Kaspar}\ \emph {et~al.}(2021)\citenamefont {Kaspar}, \citenamefont {Ravoo}, \citenamefont {{van der Wiel}}, \citenamefont {Wegner},\ and\ \citenamefont {Pernice}}]{KasparRvdWWP2021}%
  \BibitemOpen
  \bibfield  {author} {\bibinfo {author} {\bibfnamefont {C.}~\bibnamefont {Kaspar}}, \bibinfo {author} {\bibfnamefont {B.~J.}\ \bibnamefont {Ravoo}}, \bibinfo {author} {\bibfnamefont {W.~G.}\ \bibnamefont {{van der Wiel}}}, \bibinfo {author} {\bibfnamefont {S.~V.}\ \bibnamefont {Wegner}},\ and\ \bibinfo {author} {\bibfnamefont {W.~H.~P.}\ \bibnamefont {Pernice}},\ }\bibfield  {title} {\bibinfo {title} {The rise of intelligent matter},\ }\href@noop {} {\bibfield  {journal} {\bibinfo  {journal} {Nature}\ }\textbf {\bibinfo {volume} {594}},\ \bibinfo {pages} {345} (\bibinfo {year} {2021})}\BibitemShut {NoStop}%
\bibitem [{\citenamefont {Walther}(2020)}]{Walther2020}%
  \BibitemOpen
  \bibfield  {author} {\bibinfo {author} {\bibfnamefont {A.}~\bibnamefont {Walther}},\ }\bibfield  {title} {\bibinfo {title} {From responsive to adaptive and interactive materials and materials systems: A roadmap},\ }\href@noop {} {\bibfield  {journal} {\bibinfo  {journal} {Adv. Mater.}\ }\textbf {\bibinfo {volume} {32}},\ \bibinfo {pages} {1905111} (\bibinfo {year} {2020})}\BibitemShut {NoStop}%
\bibitem [{\citenamefont {Dunajova}\ \emph {et~al.}(2023)\citenamefont {Dunajova}, \citenamefont {Mateu}, \citenamefont {Radler}, \citenamefont {Lim}, \citenamefont {Brandis}, \citenamefont {Velicky}, \citenamefont {Danzl}, \citenamefont {Wong}, \citenamefont {Elgeti}, \citenamefont {Hannezo},\ and\ \citenamefont {Loose}}]{DunajovaEtAl2023}%
  \BibitemOpen
  \bibfield  {author} {\bibinfo {author} {\bibfnamefont {Z.}~\bibnamefont {Dunajova}}, \bibinfo {author} {\bibfnamefont {B.~P.}\ \bibnamefont {Mateu}}, \bibinfo {author} {\bibfnamefont {P.}~\bibnamefont {Radler}}, \bibinfo {author} {\bibfnamefont {K.}~\bibnamefont {Lim}}, \bibinfo {author} {\bibfnamefont {D.}~\bibnamefont {Brandis}}, \bibinfo {author} {\bibfnamefont {P.}~\bibnamefont {Velicky}}, \bibinfo {author} {\bibfnamefont {J.~G.}\ \bibnamefont {Danzl}}, \bibinfo {author} {\bibfnamefont {R.~W.}\ \bibnamefont {Wong}}, \bibinfo {author} {\bibfnamefont {J.}~\bibnamefont {Elgeti}}, \bibinfo {author} {\bibfnamefont {E.}~\bibnamefont {Hannezo}},\ and\ \bibinfo {author} {\bibfnamefont {M.}~\bibnamefont {Loose}},\ }\bibfield  {title} {\bibinfo {title} {Chiral and nematic phases of flexible active filaments},\ }\href@noop {} {\bibfield  {journal} {\bibinfo  {journal} {Nat. Phys.}\ }\textbf {\bibinfo {volume} {19}},\ \bibinfo {pages} {1916} (\bibinfo {year} {2023})}\BibitemShut {NoStop}%
\bibitem [{\citenamefont {Kole}\ \emph {et~al.}(2021)\citenamefont {Kole}, \citenamefont {Alexander}, \citenamefont {Ramaswamy},\ and\ \citenamefont {Maitra}}]{KoleARM2021}%
  \BibitemOpen
  \bibfield  {author} {\bibinfo {author} {\bibfnamefont {S.~J.}\ \bibnamefont {Kole}}, \bibinfo {author} {\bibfnamefont {G.~P.}\ \bibnamefont {Alexander}}, \bibinfo {author} {\bibfnamefont {S.}~\bibnamefont {Ramaswamy}},\ and\ \bibinfo {author} {\bibfnamefont {A.}~\bibnamefont {Maitra}},\ }\bibfield  {title} {\bibinfo {title} {Layered chiral active matter: beyond odd elasticity},\ }\href@noop {} {\bibfield  {journal} {\bibinfo  {journal} {Phys. Rev. Lett.}\ }\textbf {\bibinfo {volume} {126}},\ \bibinfo {pages} {248001} (\bibinfo {year} {2021})}\BibitemShut {NoStop}%
\bibitem [{\citenamefont {Liebchen}\ and\ \citenamefont {Levis}(2022)}]{LiebchenL2022b}%
  \BibitemOpen
  \bibfield  {author} {\bibinfo {author} {\bibfnamefont {B.}~\bibnamefont {Liebchen}}\ and\ \bibinfo {author} {\bibfnamefont {D.}~\bibnamefont {Levis}},\ }\bibfield  {title} {\bibinfo {title} {Chiral active matter},\ }\href@noop {} {\bibfield  {journal} {\bibinfo  {journal} {EPL}\ }\textbf {\bibinfo {volume} {139}},\ \bibinfo {pages} {67001} (\bibinfo {year} {2022})}\BibitemShut {NoStop}%
\bibitem [{\citenamefont {Kalz}\ \emph {et~al.}(2024)\citenamefont {Kalz}, \citenamefont {Sharma},\ and\ \citenamefont {Metzler}}]{KalzSM2023}%
  \BibitemOpen
  \bibfield  {author} {\bibinfo {author} {\bibfnamefont {E.}~\bibnamefont {Kalz}}, \bibinfo {author} {\bibfnamefont {A.}~\bibnamefont {Sharma}},\ and\ \bibinfo {author} {\bibfnamefont {R.}~\bibnamefont {Metzler}},\ }\bibfield  {title} {\bibinfo {title} {Field theory of active chiral hard disks: a first-principles approach to steric interactions},\ }\href@noop {} {\bibfield  {journal} {\bibinfo  {journal} {J. Phys. A Math. Theor.}\ }\textbf {\bibinfo {volume} {57}},\ \bibinfo {pages} {265002} (\bibinfo {year} {2024})}\BibitemShut {NoStop}%
\end{thebibliography}%

\end{document}